\pdfoutput=1
\documentclass[pdftex,12pt]{JHEP3}
\usepackage{amsfonts,amsbsy,latexsym,amssymb,amscd,amstext}
\usepackage[pdftex]{epsfig}

\newcommand{\AmS}{{\protect\the\textfont2
  A\kern-.1667em\lower.5ex\hbox{M}\kern-.125emS}}
\newcommand{\infl}{\chi} 

\newcommand{\mh}{m_{_{\rm H}}}
\newcommand{\mw}{m_{_{\rm W}}}
\newcommand{\gw}{g_{_{\rm W}}}
\newcommand{\gy}{g_{_{\rm Y}}}

\newcommand{\be}{\begin{equation}}
\newcommand{\ee}{\end{equation}}
\newcommand{\ba}{\begin{array}}
\newcommand{\ea}{\end{array}}
\newcommand{\baa}{\begin{array}}
\newcommand{\eaa}{\end{array}}
\newcommand{\bea}{\begin{eqnarray}}
\newcommand{\eea}{\end{eqnarray}}
\newcommand{\half}{{1\over2}}

\newcommand{\pmin}{p_{\mbox{\tiny min}}}

\newcommand{\Y}{{\rm Y}}
\newcommand{\Z}{{\rm Z}}

\newcommand{\hm}{{\hspace*{2mm}}}

\newcommand{\cV}{{\cal V}}

\newcommand{\cA}{{\cal A}}   
\newcommand{\cB}{{\cal B}}   
\newcommand{\cF}{{\cal F}}   
\newcommand{\cG}{{\cal G}}   
\newcommand{\cD}{{\cal D}}   
\newcommand{\cZ}{{\cal Z}}   

\date{} \preprint{IFT-UAM/CSIC-08-02, hep-ph/yymmnnn}
\title{Primordial magnetic fields from preheating at the electroweak
scale} \author{Andr\'es D\'{\i}az-Gil $^{a}$, Juan Garc{\'\i}a-Bellido
$^{a,b}$, Margarita Garc{\'\i}a P\'erez $^a$ and Antonio
Gonz\'alez-Arroyo $^a$ \\ $^a$ Instituto de F\'{\i}sica Te\'orica
UAM/CSIC and Departamento de F\'{\i}sica Te\'orica \\ Universidad
Aut\'onoma de Madrid, E-28049--Madrid, Spain \\ $^b$ Kavli Institute
for Theoretical Physics, University of California Santa Barbara, CA
93106-4030,\\ E-mail: \email{andres.diazgil@uam.es,
juan.garciabellido@uam.es, margarita.garcia@uam.es,
antonio.gonzalez-arroyo@uam.es }}
\abstract{
We analyze the generation of helical magnetic fields during preheating
in a model of low-scale electroweak (EW) hybrid inflation.  We show
how the inhomogeneities in the Higgs field, resulting from tachyonic
preheating after inflation, seed the magnetic fields in a way
analogous to that predicted by Vachaspati and Cornwall in the context
of the EW symmetry breaking. At this stage, the helical nature of the
generated magnetic fields is linked to the non-trivial winding of the
Higgs-field. We analyze non-perturbatively the evolution of
these helical seeds through the highly non-linear stages of symmetry
breaking (SB) and beyond. Electroweak SB occurs via the 
nucleation and growth of Higgs bubbles which squeeze the magnetic 
fields into string-like structures. The $W$-boson charge density 
clusters in lumps  around the magnetic strings. After symmetry breaking, 
a detailed analysis of the magnetic field Fourier spectrum shows two 
well differentiated components: a UV radiation
tail at a temperature $T\sim 0.23\, \mh$, slowly growing with time,
and an IR peak associated to the helical magnetic fields, which seems to
follow inverse cascade. The system enters a
regime in which we observe that both the amplitude ($\rho_B/\rho_{\rm
EW}\sim 10^{-2}$) and the correlation length of the magnetic field
grow linearly with time. During this stage of evolution we also observe
a power-law growth in the helical susceptibility.
These properties support the possibility that our scenario could provide
the seeds eventually evolving into the microgauss fields observed today in
galaxies and clusters of galaxies.
}

\begin{document}

{\vskip 1cm}


\section{Introduction}\label{intro}

The origin of magnetic fields (MF) is one of the remaining mysteries
in relativistic astrophysics and cosmology (for reviews see the list
of references~\cite{Parker1979}-\cite{Barrow2007}). Magnetic fields
play an important role in the evolution of the primordial plasma in
the early universe (possibly also in cosmic phase transitions), in the
propagation of cosmic rays in our galaxy, as well as in clusters of
galaxies. They may influence galaxy formation and large scale
structures, and they may generate a stochastic background of
gravitational waves.  The connection between magnetic fields and
gravitational waves is particularly intriguing. Since MF induce an
anisotropic stress tensor, this can act as a source of gravitational
waves (see \cite{DurrerCaprini2003}).  Large amplitude magnetic fields
from primordial turbulence could induce a significant stochastic
background of gravitational waves which could be seen by LIGO or BBO,
with a specific spectral signature.

Magnetic fields have been found on the scale of galaxies and clusters 
of galaxies with a magnitude of order the microgauss. There is even some 
evidence of their existence on the scale of superclusters (for a review on 
observational results see \cite{Vallee1997}). 
Summarizing the measured MF values on all scales $L$ :

\begin{itemize}

\item galaxies: $B \simeq 50~\mu$G at $L < $ 1 kpc; 
$B \simeq 5-10~\mu$G at $L \sim $ 10 kpc.

\item clusters: $B \simeq 1~\mu$G at $L \sim $ 1 Mpc.

\item superclusters:  $B < 10^{-2} - 10^{-3}~\mu$G at $L \sim  1 - 50$ Mpc.

\item CMB: $B < 10^{-3} - 10^{-5}~\mu$G at $L > $ 100 Mpc.

\item Primordial nucleosynthesis: $B < 10^{11}$ G at $T = 10^9$ K.

\end{itemize}

\noindent
where the last bound (BBN) comes from the modification that such a
background would imply for the expansion rate of the universe at
primordial nucleosynthesis, which would change the observed Helium 
abundance.

The main difficulty in understanding the origin of magnetic fields
is not in their amplitude (i.e. magnitude) but in its correlation
scale, from galaxies to clusters to superclusters. The microgauss
order of magnitude of present galactic MF could be explained easily
from an amplification via a dynamo mechanism initiated by a tiny seed,
with $B \sim 10^{-23} - 10^{-30}$ G (when taking into account
gravitational collapse in a flat $\Lambda$CDM model).
The explanation of the scale of the magnetic seed in this case 
is rather straightforward. The dynamo mechanism is
an exponential mechanism which makes the MF amplitude increase a
factor $e$ at every turn of the object (typically a galaxy) with free
charge and thus large electrical conductivity. Since the typical
galaxy has made around 30 turns in their lifetime, the growth factor
is $e^{30} = 10^{13}$. Since we observe microgauss, we just need a seed
$B_{\rm seed} \sim 10^{-19}$ G over a scale of 30 kpc. This is the MF
{\it after} gravitational collapse. Typically a galaxy forms by
gravitational collapse of a lump of matter the size of about a Mpc
with density of order the critical density, and ends collapsing to a
size of order 30 kpc and density $\rho_{\rm gal} \sim 10^6\rho_c$.
By flux conservation, the gravitational collapse amplification gives
an extra factor
$$(\rho_{\rm gal}/\rho_c)^{2/3} \sim 10^4\,,$$
which gives a seed $B_{\rm seed} \sim 10^{-23}$ G over a scale of 1 Mpc.
This calculation was done assuming matter domination. If we consider
a $\Lambda$CDM universe, then gravitational collapse amplification
is greater and the seed can start with $B_{\rm seed} \sim 10^{-30}$ G
over a scale of 1 Mpc. This is the {\it minimal} value required for
a typical galaxy.  

The microgauss amplitude at cluster scales is more difficult to
explain via a dynamo mechanism because it did not have as much time
since its formation to build up from such a tiny seed, and the order
of fractions of microgauss amplitude at supercluster scales is simply
impossible to explain by dynamo mechanisms or gravitational collapse.
In any case, even in the presence of dynamo amplification, an initial
magnetic seed is required which is not provided by the dynamo
mechanism itself. Theoretical models trying to account for the origin
of the primordial seeds can be classified in two groups:

\begin{itemize}

\item Astrophysical: Biermann battery in intergalactic shocks, stellar
magnetic winds (like in our Sun), supernova explosions, galactic
outflows in the inter-galactic medium (IGM), quasar outflows of
magnetized plasma into the intra-cluster medium 
(ICM), see Refs.~\cite{GrassoRubinstein2001,Widrow2002,Beck2005}, 
and a recently suggested proposal in conjunction with high energy 
cosmic rays~\cite{DarDeRujula2005}.

\item Cosmological: Early universe phase 
transitions~\cite{Quashnock:1988}-\cite{Banerjee}, 
magnetic helicity together with the baryon asymmetry of the universe
(BAU) at the electroweak (EW) 
transition~\cite{Vachaspati1991}-\cite{GrassoRiotto1998}, 
via hypercharge and hypermagnetic field generation before
EW transition~\cite{Joyce1997,Giovannini1998a}, from second order cosmological 
perturbations from inflation~\cite{TurnerWidrow1988}-\cite{Durrer2006}, 
from preheating after 
inflation~\cite{CalzettaKandus2002}-\cite{Boyanovsky2005}, etc.

\end{itemize}

Moreover, MF have also been observed in quasars at redshift $z\sim2$,
again with a magnitude of order the microgauss. This indicates not
only ubiquity but also invariance (within an order of magnitude) with
time. Such features cry for a cosmological, rather than astrophysical,
origin of MF. Could it be that some yet unknown mechanism directly
generated microgauss MF on all scales? The first reaction is to ask
about the dynamo mechanism in galaxies, would it not amplify this
microgauss MF to even larger amplitudes, as can be seen in neutron
stars, and even our Sun? The surprising answer is no, because a few
microgauss is the {\it maximum} magnetic field possible on galactic
scales, due to the existence of relativistic cosmic rays and ionized
gas moving at large speeds. If one computes the total energy density
in cosmic rays (integrating the measured flux spectrum over all
energies), one finds
$$\half \rho_{\rm CR} v^2/c^2 = 0.5~{\rm eV/cm}^3\,,$$
and a similar number for the energy density in the ionized gas
moving with rotation speeds of order 200 km/s,
$$\half \rho_{\rm gas} v^2/c^2 = 0.3~{\rm eV/cm}^3\,.$$
If we assume that magnetic fields are in equilibrium, due to their
interaction with the cosmic rays and the gas, and furthermore we 
suppose equipartition, then their energy density (using 1 G $= 
1.95\times10^{-20}$ GeV$^2$) becomes
$$\rho_{\rm B}=  B^2/(8\pi) = 0.5~{\rm eV/cm}^3 =
(5~\mu{\rm G})^2/(8\pi)\,,$$
which corresponds to a few microgauss, in surprising agreement with
observations. Some people suggest that this argument may also explain
the cluster MF value.

The ubiquity of MF with similar amplitude on all scales reminds us of
the issue of Helium abundance in the universe. Early measurements in
the fourties indicated that the Helium mass fraction to Hydrogen in
the Universe was about a quarter, very nearly {\it everywhere}. This
observation was correctly interpreted by Gamow and collaborators as
indicating a primordial origin. Simple order of magnitude computation
of nuclear interaction rates (mainly those of deuterium, a necessary
step in the reactions from H to He) and comparison with the rate of
expansion in the early universe at temperatures of order the nuclear
transitions (i.e. MeV), together with the then largely unknown neutron
decay rate, suggested that the present abundance of Helium could have
been produced from Hydrogen in the early universe and thus be present
everywhere. The other light elements seemed to require further synthesis 
in stars and thus depended on location, but the Helium was ubiquitous
because it was there from the very beginning.

Something similar may have happened with magnetic fields, if they were
generated in the early universe by some unknown mechanism and then
redshifted until today. The question is what is the typical energy
density which today gives the order microgauss fields? These fields
(if homogeneous) redshift as radiation, i.e. $\rho_{\rm B}(a) =
\rho_{\rm B}({\rm today}) (a_0/a)^4$. Like with Helium, we have to ask what 
was the energy scale of interactions responsible for the generation of
primordial magnetic fields?  Photons are massless so in principle any
scale, as long as there are charged particles, is sufficient to
generate magnetic fields, and this is the reason why there is still so
much debate as to their origin.  However, was the universe always
permeated with electromagnetic waves?  The answer is no, the
electromagnetic interaction as we know it came into being at a very
precise time, when the electroweak (EW) force broke into the weak
interactions plus electromagnetism. Before we could not talk about
photons and magnetic fields. This occurred when the typical energy (or
temperature) in the universe was around $T_{\rm EW} \sim$ 100 GeV. If
we construct an energy density with this scale we get $\rho_{\rm
EW}\sim 10^8$ GeV$^4$. At that time the universe was (or became)
radiation dominated.  If we now redshift this MF energy density until
today ($T_0 = 2.725$ K) we get
$$\rho_{\rm B}({\rm today}) = (T_0/T_{\rm EW})^4 \rho_{\rm EW} \sim
3.04\times10^{-53}\ {\rm GeV}^4 = 0.4\ {\rm eV/cm}^3$$ which is {\it
precisely} the order of magnitude of the present MF energy
density.\footnote{We could be even more conservative and suppose that
the fraction of magnetic field energy density to radiation at the time
of the EW transition was given by $f=\rho_{\rm B}/\rho_{\rm rad}<1$.
In this case, the present MF magnitude would be $B_0 \sim
5\,f^{1/2}~\mu$G.}  This would be enough to explain the cluster and
supercluster values, and would perhaps require a mild dynamo mechanism
to grow to galactic values (if the fraction $f\ll1$). The question is
whether this is just a coincidence or it is hinting directly at its
origin.\footnote{Some authors suppose that the
generation occurred {\em earlier} in the form of hypermagnetic fields
and was then converted into ordinary magnetic fields at the EW
scale \cite{Giovannini1998a}.}
While other mechanisms require a seed with an arbitrary scale
(typically $B\sim10^{-23}$ G, so that today we observe microgauss MF
on galactic scales via the dynamo mechanism), there is no physical
reason behind this scale.  On the other hand, the EW scale is a 
{\it natural} scale for the generation on magnetic fields since it is the
scale at which electromagnetism arises for the first time as a
fundamental interaction.

Whether this is sufficient reason to assign the EW energy scale to the
origin of magnetic fields is another issue. In particular, it is not
clear how to obtain the large correlation length of magnetic fields
observed at galactic and cluster scales. Any physical mechanism that
creates magnetic fields must be necessarily causal, but at high
temperatures in the early universe there is also a natural coherence
scale given by the particle horizon. At the electroweak scale the
physical horizon is $10^{-10}$ light-seconds ($\sim 3$ cm), which today
corresponds to a co-moving scale of 0.3 mpc ($\sim 1$ AU), clearly
insufficient when compared even with the irregular (turbulent)
component of the galactic magnetic field ($L \sim 100$ pc), not to
mention the regular (uniform) component, which has correlations $L
\sim 10$ kpc. It thus seems impossible to explain the coherent magnetic
fields observed on galaxy clusters and supercluster scales (of order 
10 Mpc) with intensities of order $\mu$G to $n$G.

There is however a second coincidence, which makes things even more
intriguing. If we assume that the plasma after the electroweak
transition is sufficiently turbulent to maintain magnetic fields
of the largest possible coherence scales via inverse cascade 
\cite{Brandenburg1996}-\cite{Banerjee}, then we could reach cosmological 
scales today.  Let us follow the argument.
The largest coherence scale at the electroweak transition is the
physical horizon, of order 3cm. If a strong inverse cascade is active,
then the coherence length of the magnetic fields will grow as fast
as the horizon (it cannot grow faster). This means that it grows like
the scale factor {\em squared} during the radiation dominated era.
This ideal situation could only last while there is a plasma and
thus it is bound to stop acting at photon decoupling, when the universe
becomes neutral. Since then, the correlation length can only grow
with the expansion of the universe, as the scale factor. If we take 
this effect into account from the electroweak scale until today we
find, using the adiabatic expansion relation $T\propto a^{-1}$,
\begin{equation}
\xi_0 = \xi_{\rm EW}\,\left({a_{\rm dec}\over a_{\rm EW}}\right)^2
{a_0\over a_{\rm dec}} = 3\,{\rm cm} \left({T_{\rm EW}\over 
T_{\rm eq}}\right)^2{T_{\rm eq}\over T_0} \sim 6\times 10^{25}\,{\rm cm}
= 20\ {\rm Mpc}\,,
\end{equation}
where we have made the approximation that equality and decoupling
occurred more or less simultaneously (a careful computation gives only
a minor correction). The surprising thing is that this simple
calculation gives {\em precisely} the order of magnitude for the
largest correlation length of cosmic MF ever observed (i.e. cluster
scales). If the agreement in the magnitude of the primordial MF seed
seemed peculiar, the fact that an inverse cascade could also be
responsible for the observed correlation length becomes a surprising
coincidence, probably hinting at an underlying mechanism. It is
therefore worthwhile exploring the conditions that could have taken
place at the electroweak transition which could give rise to a
significant fraction of energy density in magnetic fields, and be
responsible for a sustained period of inverse cascade until photon
decoupling. It has been shown in
Refs.~\cite{ChristenssonHindmarsh2001}-\cite{Banerjee} that one
important ingredient is the generation of magnetic fields with a non
trivial helical component, which guarantees an optimal amplification
of the magnetic correlation length through inverse cascade. A very
good account of the large number of works investigating these issues,
with a complete list of references is given in
Ref.~\cite{GrassoRubinstein2001} (see also
\cite{BaymBodeker1996}-\cite{Giovannini1998a}).

In this paper we propose a scenario in which the electroweak
transition takes place at the end of a brief period of hybrid
inflation\footnote{Note that we do not need a 60 e-fold period of
  inflation, just a few ($\sim5$) e-folds of low scale thermal
  inflation~\cite{thermalinflation,Easther2008} to cool down the
  universe.  The amplitude of CMB temperature fluctuations would be
  determined by the usual 55 e-folds of high-scale (e.g. GUT)
  inflation.}  It has been conjectured that preheating and early
reheating in this model could provide an alternative mechanism to
generate the baryon asymmetry in the universe~\cite{CEWB}-\cite{smit}
and a way to source gravitational waves~\cite{GW}. In this paper we
analyze whether it could also give rise to primordial magnetic fields
with the required amplitude and correlation length. This issue has
been partially addressed in a recent letter~\cite{ajmtPRL}. Here we
will present a complete account of the results obtained and a detailed
description of the approach employed in the analysis. Our set up
provides a specific realization of some of the proposals described
above.  In particular, we will see how helical magnetic fields arise
from the inhomogeneities in the spatial distribution of the Higgs
field, along the lines conjectured by
Vachaspati~\cite{Vachaspati1991,Vachaspati2001} and
Cornwall~\cite{Cornwall1997} some years ago.

The paper is organized as follows. In section \ref{model} we describe
the hybrid inflation model that we will be using and revise, following
Ref.~\cite{jmt}, how to solve the quantum evolution of the system from
the end of inflation until non-linearities start to become important.
Beyond this time, a fully non-perturbative approach is
required. Fortunately, the time evolution at this stage can be
described within the classical approximation as demonstrated in
Ref.~\cite{jmt}. Details on the methodological set up and the lattice
implementation are presented in section~\ref{numer}. 
Section~\ref{generation} analyzes the mechanism leading to helical 
magnetic fields following from the inhomogeneities in the Higgs field, 
which are seeded by the Higgs quantum
fluctuations that arise from the period of linear quantum evolution.
Strings of magnetic flux, carrying non-vanishing helicity, are clearly
observed.  They persist and are even enhanced as the system progresses
towards the true vacuum. We also analyze here the structure of the 
plasma of $W-$charges which accompany the magnetic fields during
this period.  The fate of these magnetic fields at later times is
discussed in section~\ref{latet}, 
where we present a detailed study of the
spectrum of the magnetic field. We will show that there is a
significant helical magnetic field remnant whose amplitude and
correlation length are amplified linearly in time.  In section
\ref{arte} we will discuss lattice and finite volume independence of
our results, as well as the dependence of magnetic field production on the
Higgs- to W-mass ratio. Conclusions and prospects for further work are
presented in section~\ref{conclusions}.  Finally, a few technical
points about the lattice discretization of the classical equations of
motion, the Maxwell equations and electromagnetic radiation are
described in the Appendices \ref{app1}, \ref{app2} and
\ref{app3}. Appendix \ref{app4} is devoted to an analysis of the
Gaussian random fields that provide the initial Higgs field
distribution.


\section{The model}\label{model}

The scenario we will be considering is that of preheating after a
period of hybrid inflation which ends at the EW scale.  This was first
introduced in Ref.~\cite{CEWB} to provide a new mechanism for the
generation of baryon asymmetry in the Universe (BAU). It has been
extensively studied since then both in connection with
BAU \cite{CEWB}-\cite{smit} and in relation with the production of
gravitational waves~\cite{GW}. In this paper and in
Ref. \cite{ajmtPRL} we include for the first time the Hypercharge
field in order to study the generation of electromagnetic fields
during preheating (preliminary results can be found in~\cite{ajmt}).
In this section we will introduce the model and describe the first
stages of evolution after inflation ends which provide the initial
conditions for the non-linear approach addressed in
section~\ref{evolution}.

The Hybrid inflation model is attained by extending the Standard Model
with the addition of a scalar field, the inflaton, singlet under the
gauge group.  The scalar sector thus includes the Higgs field:
$\Phi=\half(\phi^0\,1\!{\rm l}+i\phi^a\tau_a)$ ($\tau_a$ are the Pauli
matrices) and the singlet inflaton $\infl$ which couples only to the
Higgs via the scalar potential:

\be
{\rm V} (\Phi,\infl) =
{\rm V}_0 + \half(g^2\infl^2-m^2)\,|\phi|^2 + \frac{\lambda}{4}
|\phi|^4 + \half \mu^2 \infl^2 \,,
\ee
where $|\phi|^2\!\equiv\!2{\rm Tr}\, \Phi^\dag\Phi$, $\mu$ is the
inflaton mass in the false vacuum and $\mh = \sqrt{2}\,m \!\equiv\!
\sqrt{2\lambda}\,v$ the Higgs mass, with $\,v\!=\!246$ GeV the 
Higgs vacuum expectation value at the electroweak scale. The gauge 
sector contains both the SU(2) and the hypercharge U(1) fields with
\be
\cG^a_{\mu\nu}=\partial_\mu \cA_\nu^a - \partial_\nu \cA_\mu^a +
\gw \epsilon^{abc} \cA_\mu^b \cA_\nu^c 
\ee
and
\be
\cF^\Y_{\mu\nu}=\partial_\mu \cB_\nu- \partial_\nu \cB_\mu ,
\ee
their respective field strengths.
The covariant derivative is:
\be
\cD_\mu = \partial_\mu - {i\over2} \gw \cA_\mu^a\tau_a - {i\over2} \gy \cB_\mu,
\ee
with $\gw$ the SU(2) gauge coupling and $\gy$ the hypercharge coupling.
In this work we can safely ignore fermionic fields since the time scales 
involved in the perturbative decay of the Higgs field into fermions 
are much larger than the ones considered here.

With all these definitions the Lagrangian density of the model becomes:
\be
{\cal L} = - {1\over4}\cG^a_{\mu\nu}\cG^{\mu\nu}_a -
{1\over4}\cF^\Y_{\mu\nu}\cF_\Y^{\mu\nu}+
  {\rm Tr}\Big [(\cD_\mu\Phi)^\dag \cD^\mu\Phi\Big ]
  + \half \partial_\mu\infl \partial^\mu\infl - V(\Phi,\infl) \,.
\ee

For our analysis we have fixed the $W$ mass and the $Z$ to $W$ mass
ratio to the experimental values~\cite{PDG}. We have analyzed three
different values of the Higgs to $W$ mass ratio: $\mh/\mw = 
2\sqrt{2\lambda}/\gw =$ 2, 3 and 4.65. 
The Higgs-inflaton coupling has been fixed to $ g^2 = 2 \lambda$
as in super-symmetric models
\cite{GBKLT,GBL} and we have taken the inflaton bare mass 
$\mu = 10^{-5} g v \approx 0$.

The extraction of the electromagnetic content of the SU(2)$\times$U(1)
fields in the Lagrangian proceeds in the usual way. Fixing the unitary
gauge, $\Phi(x) = \rho(x) \,1\!{\rm l}$, 
the $Z$-boson field and the electromagnetic field are extracted
from appropriate orthogonal combinations of the SU(2) and hypercharge
vector potentials:
\bea
{\cal Z}_{\mu}(x) &=& \cos\theta_W \, {\cal A}_{\mu}^3(x) + \sin\theta_W \, {\cal B}_{\mu}(x) \,,  \\
{\cal A}_{\mu}^{\gamma}(x) &=&  \sin\theta_W \, {\cal A}_{\mu}^3(x) - \cos \theta_W \, {\cal B}_{\mu}(x) \,. 
\eea   
with $\varphi(x) = \Phi(x) (1,0)^{\rm T}$ the Higgs doublet. 
This separation can only be done unambiguously when the Higgs field 
is on the true vacuum, i.e. in the broken symmetry phase. However even in 
that phase there can be points where the Higgs field vanishes and the
symmetry is locally restored (a typical example of a configuration exhibiting
such behavior is the sphaleron). At those points there is no unique
way  to define the electromagnetic fields. In fact
't Hooft was the first to point out in Ref.~\cite{hooft}
the consequences of this ambiguity in the Georgi-Glashow model,
tying it to the appearance of non-trivial configurations like monopoles
or strings, acting as sources of magnetic fields. In Ref.~\cite{Vachaspati1991}
Vachaspati pointed out that a similar mechanism was at work in the electroweak
model where the sources for magnetic field generation are tied to the presence
of non-homogeneous phases in the Higgs field. In the following sections we will
analyze in detail how this mechanism is realized during the period of preheating 
after inflation.

\subsection{Linear quantum evolution}

Following Refs. \cite{jmt,jmt2}, we will address here the
first stages of evolution starting at the end of inflation. 

 The period of inflation is characterized by the fact that the Higgs
and inflaton fields are displaced from the true minimum of the
potential. In this case, inflation is driven by the false vacuum
energy, $V_0 = \lambda v^4 / 4$.  During this time the inflaton
homogeneous mode, $\chi_0\equiv \langle \chi \rangle$, dominates the
dynamics. After only about 5-10 e-folds the Universe has cooled down
and all other particle species have been diluted, remaining in the de
Sitter vacuum.\footnote{For electroweak-scale inflation and the range
of momenta we will be considering, de Sitter vacuum is
indistinguishable from the Minkowski vacuum.}

The interaction between the Higgs and inflaton fields drives the end of
inflation and triggers EW symmetry breaking. The way this proceeds is
as follows.  Close to the time when inflation ends, denoted by $t_c$,
the time evolution of the inflaton zero mode can be approximated by:
\be
\chi_0 (t)= \chi_c (1  - V m(t-t_c))
\ee
where $\chi_c = \chi_0 (t_c) \equiv m/g$. Here $V$ denotes the inflaton
dimensionless velocity, defined through this equation and fixed to
$V=0.024$ in our analysis~\cite{jmt}. The variation of $\chi_0(t)$ 
induces, via
the Higgs-inflaton coupling, a time dependence of the effective Higgs
mass parameter, $m_\phi^2 = - M^3(t-t_c) \equiv -2 V m^3 (t-t_c)$,
which changes from positive to negative, triggering electroweak
symmetry breaking.  Accordingly, the time when inflation ends, $t_c$,
is characterized as the critical point where the Higgs field
becomes massless.

As described in detail in Refs.~\cite{jmt,jmt2}, it is possible to
solve exactly the quantum evolution of the system around $t_c$ if
non-linearities in the Higgs field and the interaction with the gauge
fields are neglected. As we will see below, this is a reasonable
assumption at this stage.  In this approximation the Higgs field is
effectively described as a free scalar field with a time dependent
mass $m_\phi(t)$.  Its quantum evolution can be solved in terms of
Airy functions~\cite{jmt}. After $t_c$, low momentum modes of the
Higgs field grow exponentially in a process known as ``tachyonic
preheating''~\cite{GBKLT}.  Due to the tachyonic growth, low momentum
Higgs field modes acquire large occupation numbers and, soon after
$t_c$, they evolve as classical modes.  This is a very fast process so
that all other modes can be taken to remain in the quantum vacuum
(ground) state, justifying thus the linear approximation. These modes
will be later populated through the interaction with the Higgs field
tachyonic modes. Once non-linearities start to become relevant the
approximation ceases to be valid and a full non-linear treatment is
required.  Our strategy for dealing with the later stages of evolution
will be presented in the next section.


\section{Methodological set up}\label{numer}

Beyond the quantum linear evolution described in the previous  section
we have to deal with the non-linear dynamics of the Higgs field and
its coupling to the gauge fields. Our approach is based upon the 
classical approximation (details can be found in~\cite{jmt,jmt2} -
see also~\cite{smit}). The validity of this approximation relies on the
fast growth of tachyonic modes as explained previously. In what follows 
we will describe several aspects of our procedure. 

\subsection{Initial conditions for the  the non-linear evolution}
\label{evolution}

As mentioned previously the initial stages after the end of inflation
($t=t_c$) lead to a rapid growth of the tachyonic modes which tend to
behave classically.  The correlation functions of the Higgs field
resulting from the initial quantum evolution can be computed. Our
approach is to use these results as initial conditions for the
classical evolution of the system.  The quantum fluctuations translate
into stochastic initial conditions for the Higgs field, whose
correlations are designed to match the Weyl-ordered quantum
expectation values.  The matching of the two methods is done at an
initial time $t=t_i>t_c$ that must be large enough for classical
behaviour to set in and small enough to make the non-linear terms
small. This leaves a window of possible values of $t_i$. We tested the
robustness of the results with respect to changes in $t_i$ within
these limits, giving confidence on the self-consistency of our
approach.

Given the linear character of the initial quantum evolution, the Higgs
field momentum modes $\phi^{\alpha}_{k}$, at $t=t_i$ behave as
Gaussian random variables  of zero-mean following a Rayleigh distribution:
\be\label{rayleigh}
\exp\Big(\!-{|\phi_k^\alpha|^2\over(\sigma_k^\alpha)^2}\Big)\,
{d|\phi_k^\alpha|^2\over(\sigma_k^\alpha)^2}\,{d\theta_k^\alpha\over2\pi}\,,
\ee
where $\theta_k^\alpha$ is the phase of the complex random variable 
$\phi^{\alpha}_{k}$.  The dispersion 
of the modulus is expressed in terms of the power spectrum 
$P(k,t_i)= k^3 (\sigma_k^\alpha)^2 $, and can be computed analytically 
in terms of Airy functions~\cite{jmt}. For practical purposes it is
better to work with a simple functional  fit  to the power spectrum
(Eq.~(D.1)). Notice that we have introduced a momentum cut-off,
removing modes which have not become tachyonic (classical). As
explained in Ref.~\cite{jmt}, this is  compensated by a renormalization of
the parameters.

The study of the properties of this 4-component Gaussian random field
is collected in Appendix~\ref{app4}.  Its features depend on several
parameters: the Higgs mass, the initial inflaton velocity $V$, the
momentum cut-off and the choice of initial time $t_i$. The first two
appear combined in a new scale $M= (2V)^{1/3} m$ characteristic of the
initial linear evolution. The main conclusion drawn in
Appendix~\ref{app4}, is that to a large extent all these parameter
dependencies translate into setting two main scales: a spatial length
scale, $\xi_0$, and the Higgs dispersion, $\sigma$, which determines
the magnitude of the Higgs field.

More specifically, in Appendix~\ref{app4} we study the distribution of
local maxima in $|\phi(x)|$. These ``peaks'' are the seeds that will
later grow with time and develop into bubbles which start expanding
and colliding among themselves once the Higgs fields enters the
non-linear regime characteristic of symmetry breaking. This process
was described in detail in Ref.~\cite{jmt}. Note that the
multicomponent character of the Higgs field affects the results but,
more importantly, it gives rise to new observables, some of which are
intimately connected to the physical phenomena which are the main goal
of this paper. This will be described in the next section.

To complete the description of the initial conditions, we mention that,
similarly to the high-momentum modes of the Higgs field, all other
non-tachyonic modes are set to zero. These include the non-homogeneous
modes of the inflaton and the vector potentials of the SU(2) and
hypercharge gauge fields. The initial time-derivatives of these
quantities are also set to zero except for the gauge fields which have
to be chosen such that the Gauss constraint is satisfied as an initial
condition. The dynamic equations guarantee that the constraint will
continue to hold at later times. The aforementioned robustness of the
results to the choice of initial time $t_i$ implies that our physical
conclusions do not depend on minor modifications of these initial
conditions.

\subsection{Numerical procedure}\label{electro}

In order to study the non-linear evolution of the system with our
stochastic initial conditions we have made use of the lattice
approach.  This has the advantage that classical equations of motion
are discretized preserving full gauge invariance of the
system. Generally speaking the procedure is standard.  Details on the
lattice Lagrangian and the lattice form of the equations of motion are
presented in Appendix~\ref{app1}.  At early times the errors
associated to discretization are very small due to the cut-off form of
the initial spectrum.  This shows up in the very mild dependence of
the results on the spatial $a$ and temporal $a_t$ lattice
spacings. This contrasts with other situations in which lattice
techniques have been used. As time evolves higher momenta of the
fields grow and start to play a role, eventually leading to a
breakdown of the approximation. We have explicitly analysed that this
does not occur for the range of times covered in this paper. A
different approach needs to be followed if one wishes to reach times
in which full thermalisation has been reached. Notice, however, that
this goal also demands the introduction of fermionic degrees of
freedom which can be safely ignored in our time span. Our present
results can be used as initial conditions for the study of the late
time behaviour of the system.

\TABLE{
\begin{tabular}{||c|c|c|c||}\hline\hline
\hm \hm  $N_s$ \hm \hm  &\hm \hm  $ma$\hm \hm
& \hm \hm $ma_t$\hm  \hm &\hm  \hm $\pmin/m$\hm \hm  \\
\hline\hline
 $64$  & $0.65$ & $1/40$ & $0.150$ \\ \hline
 $80$  & $0.52$ & $1/40$ & $0.150$ \\ \hline
 $100$ & $0.42$ & $1/40$ & $0.150$ \\ \hline
 $100$ & $0.52$ & $1/40$ & $0.125$ \\ \hline
 $100$ & $0.65$ & $1/40$ & $0.100$ \\ \hline
 $120$ & $0.65$ & $1/40$ & $0.080$\\ \hline \hline
\end{tabular}
\caption{List of lattice parameters: $a$ and  $a_t$ are respectively the
spatial and temporal lattice spacings, $N_s$ is the number of lattice
points  and $\pmin= 2\pi /(N_s a)$ is the minimum momentum. The $N_s=120$
lattice has only been used for the study of the initial configuration. The
number of different configurations of each lattice ranges from $80$ to
$200$, depending of the lattice and the choice of parameters.}
\label{lparameters}}

Another approximation  needed for the numerical procedure is to put
the system in a box with periodic boundary conditions.  
The physical volume, ${\cal V} = L^3$ is given in terms of
the minimum momentum: $L =2 \pi /\pmin$. The latter has to be chosen
judiciously to lie well within the tachyonic band of the Higgs field. 
The dependencies of the results can be monitored by using different
values for the parameters of the simulation. In Table~\ref{lparameters} 
we enumerate the different lattice sizes, spacings and physical
volumes that we have used.

Due to its relevance for the goals of our paper, 
we will now explain in detail how 
the electromagnetic and $Z$ fields are defined in our lattice approach. 
This can only be done unambiguously when the Higgs field is in the true vacuum, 
i.e. in the broken symmetry phase. One can compute, in a gauge invariant way,  
the field  associated to the $Z$-boson potential as:
\bea
 Z_\mu (m) &=& { - i{\rm Tr} \Big [ \hat n 
 ( D_\mu  \Phi(m))  \Phi^\dagger (m)\Big ] \over  |\phi(m)| | \phi(m+\mu)|}  
\\
&\equiv& - i {\rm Tr} \Big [ \tau_3 { \Phi^\dagger(m) \over 
| \phi(m)|}  U_\mu (m) { \Phi(m+\mu) \over | \phi(m+\mu)|}
 B_\mu(m) \Big ] 
\stackrel{a\rightarrow 0}{\longrightarrow}  \hspace{5mm}  a_\mu g_Z \cZ _\mu(x)
\,, \label{zeta}
\eea
where we have introduced the adjoint unit vector $\hat n = n_a \tau_a$, with 
components:
\be
n_a(x)  = {\varphi^\dagger (x) \tau_a \varphi(x) \over |\varphi(x)|^2}\,,
\ee
with $\varphi(x) = \Phi(x) (1,0)^{\rm T}$ the Higgs doublet.
The $Z$ boson coupling is denoted by $g_Z$ and $a_i = a$, $a_0=a_t$.
$D_\mu$ is the lattice covariant derivative operator defined in 
Eq.~(\ref{eq:deriv1}) of Appendix~\ref{app1}. $U_\mu(n)$ and $B_\mu(n)$ are,
respectively, the SU(2) and hypercharge link fields introduced in 
Appendix~\ref{app1}.
Notice that continuum quantities are defined  with calligraphic
letters to distinguish them from the lattice quantities. 
Our definition  of the $Z$ boson potential corresponds to the
standard one in the unitary gauge.

We define  the $Z$ boson and hypercharge field strengths through the 
clover averages~\footnote{Corrections to the continuum approach of the 
time-space clover averages are order ${\cal O}(a_0a^2)$.}
\be
 F^\Z_{\mu \nu} (m) \!=\! \langle \Delta_\mu Z_\nu (m)\!-\! \Delta_\nu  Z_\mu (m)  \rangle_{\rm clov}
\ \stackrel{a\rightarrow 0}{\longrightarrow} \ a_\mu a_\nu g_Z \cF
_{\mu \nu}^Z(x) + {\cal O}(a^4)  
\ee
and 
\be
 F^\Y_{\mu \nu} (m) \!=\! \langle \Delta_\mu  \theta_\nu (m)\!-\! \Delta_\nu  \theta_\mu (m)  \rangle_{\rm clov}
\ \stackrel{a\rightarrow 0}{\longrightarrow} \ a_\mu a_\nu \gy \cF
_{\mu \nu}^Y(x) + {\cal O}(a^4)\,,
\ee
where $ B_\mu(m)\equiv \exp (i  \theta_\mu(m) \tau_3/2)$ is the hypercharge
link, $\Delta_\mu$ is the lattice derivative operator introduced in 
Eq. (\ref{eq:derivlat})
and $\langle  O \rangle_{\rm clov}$ denotes the clover averages defined in
Eqs. (\ref{eq:clover})-(\ref{eq:clovers}). 
In terms of them we can compute the lattice electromagnetic field strength  as:
\be
F^{\gamma}_{\mu \nu}(n)  = \sin^2(\theta_W)
 F^{\Z}_{\mu \nu}(n) -   F^{\Y}_{\mu \nu}(n)
\stackrel{a\rightarrow 0}{\longrightarrow} a_\mu a_\nu e \cF _{\mu
\nu}^\gamma(x) + {\cal O}(a^4)\,,
\label{fem}
\ee
where $\cF ^{\gamma}_{\mu \nu}$ is the corresponding continuum
electromagnetic field strength.  This provides a lattice gauge
invariant definition of the electromagnetic field which is equivalent
to the usual definition in the unitary gauge.


\section{The mechanism underlying  magnetic field generation}\label{generation}

In this section we study the production of magnetic fields during the
first stages of our EW preheating scenario. This analysis is performed
in two steps. The first is to investigate the presence, size and
structure of the magnetic fields generated by our Gaussian random
field initial distribution. This complements the results presented in
Appendix~\ref{app4}. Then we will track the evolution of these
magnetic fields through the highly non-linear stages associated to EW
symmetry breaking. This is a crucial period where there are no viable
alternatives to our methodological approach.

\subsection{Initial Magnetic fields}
\FIGURE{
\centerline{
\psfig{file=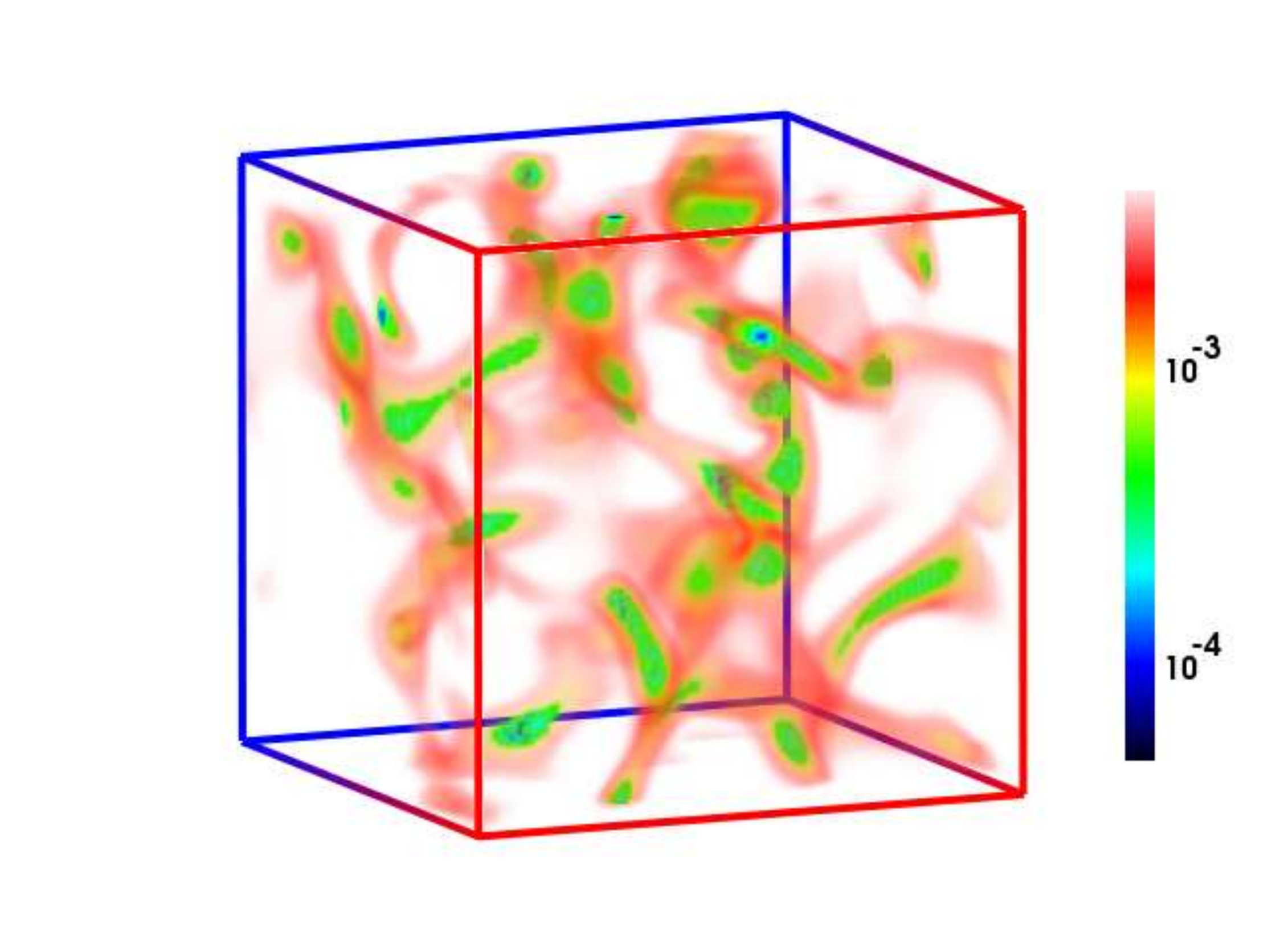,width=9cm}
\psfig{file=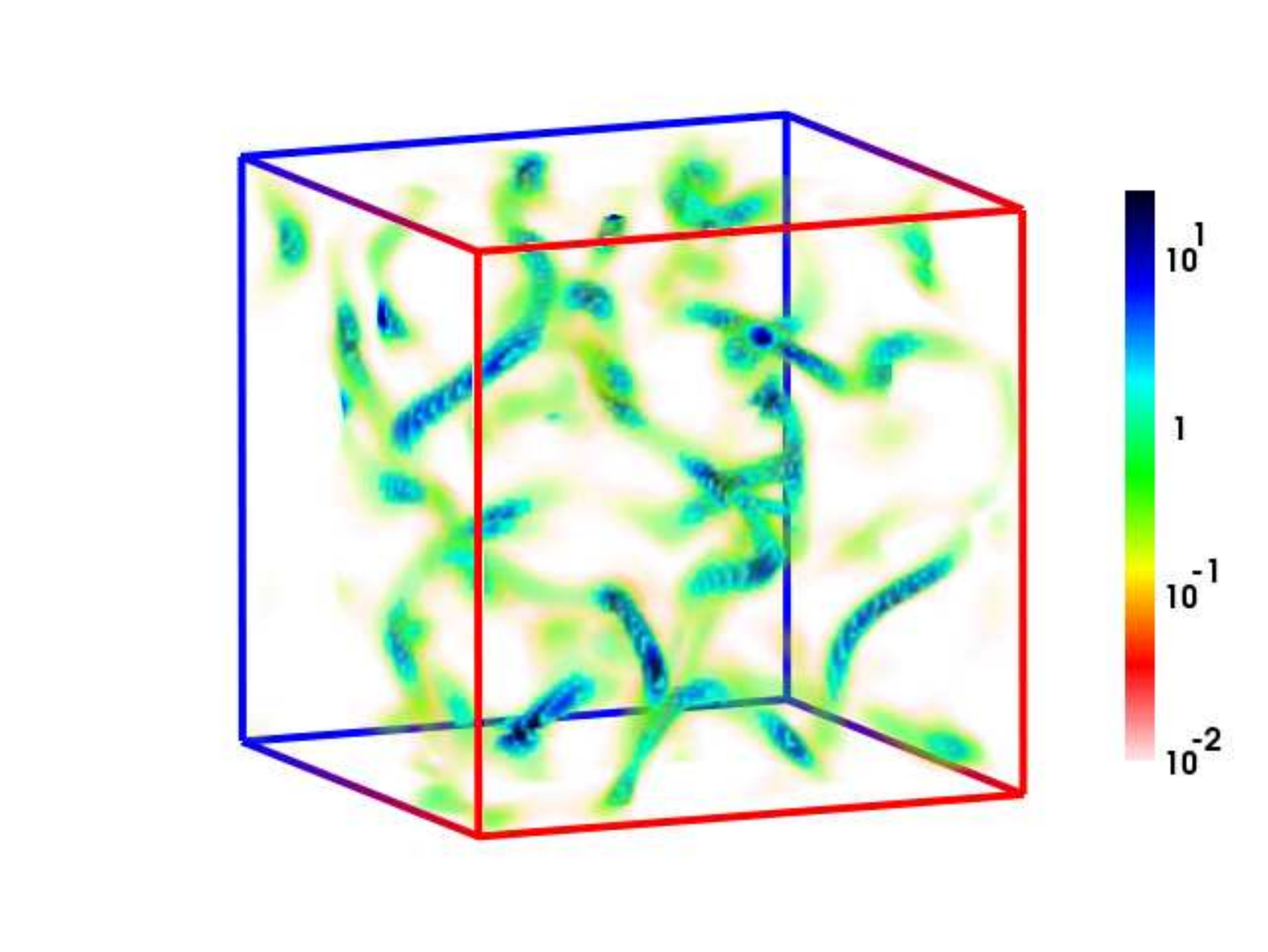,width=9cm}
\caption{(Left) Locus of points where the value of the Higgs field
norm is below $0.03\, v$ . (Right) Locus of points where twice
the magnetic energy density ($|\vec B|^2$) is above  $0.03\, m^4$.
Data correspond to $mt=5$ and $\mh=2\mw$.
}}
\label{fig:mt5}}
A close look at our expression of the photon field reveals that
Abelian electromagnetic fields are present in the first stages of the
evolution. The discussion on how this comes about follows a line of
argument very similar to that developed by Vachaspati in
Ref.~\cite{Vachaspati1991}. The tachyonic preheating phase leads to a
multicomponent Gaussian Higgs field. The SU(2) and hypercharge gauge
fields remain very small. This is incorporated into our initial
conditions by setting the hypercharge and SU(2) magnetic-like fields
to zero and fixing the corresponding electric fields in order to
satisfy the Gauss constraint.  We work in the $A_\mu=0$ gauge, which on
the lattice corresponds to $U_\mu(t=t_i) = B_\mu(t=t_i) = {\rm 1\!l}$. 
Projecting onto the $Z$ and electromagnetic fields we obtain: 
\bea 
\label{eq:initdef}
{\cal Z}_\mu(x) &=& 
{i \over g_z} \, {\rm Tr} \, \Big[ \, \hat n \,\Omega(x)
  \partial_\mu \Omega^\dagger (x) \, \Big ] \label{eq:initial}\\ 
{\cal F}^{\Z}_{\mu \nu}(n) &=& {i \over g_z} \, {\rm Tr} \, \Big [\, 
  \hat n \, \Big (\partial_\nu \Omega(x) \partial_\mu \Omega^\dagger
  (x) - \partial_\mu \Omega(x) \partial_\nu \Omega^\dagger (x) \Big)\,
  \Big ] \nonumber \\ 
{\cal F}^{\gamma}_{\mu \nu}(x) &=& \tan\theta_W
{\cal F}^{\Z}_{\mu \nu}(x) \equiv {i\sin\theta_W \over g_W} \, {\rm Tr} \,
\Big [\, \hat n \, \Big (\partial_\nu \Omega(x) \partial_\mu
  \Omega^\dagger (x) - \partial_\mu \Omega(x) \partial_\nu
  \Omega^\dagger (x) \Big) \, \Big ] \nonumber\, , \nonumber 
\eea
expressed in terms of the SU(2) matrix: 
\be 
\Omega(x) = {\Phi(x) \over |\phi(x)|}\,.  
\ee 
It becomes clear that electromagnetic fields are sourced by the
presence of inhomogeneities in the Higgs field orientation.  This is
one of the essential ingredients in Vachaspati's proposal for
magnetogenesis.

\FIGURE{
\centerline{
\psfig{file=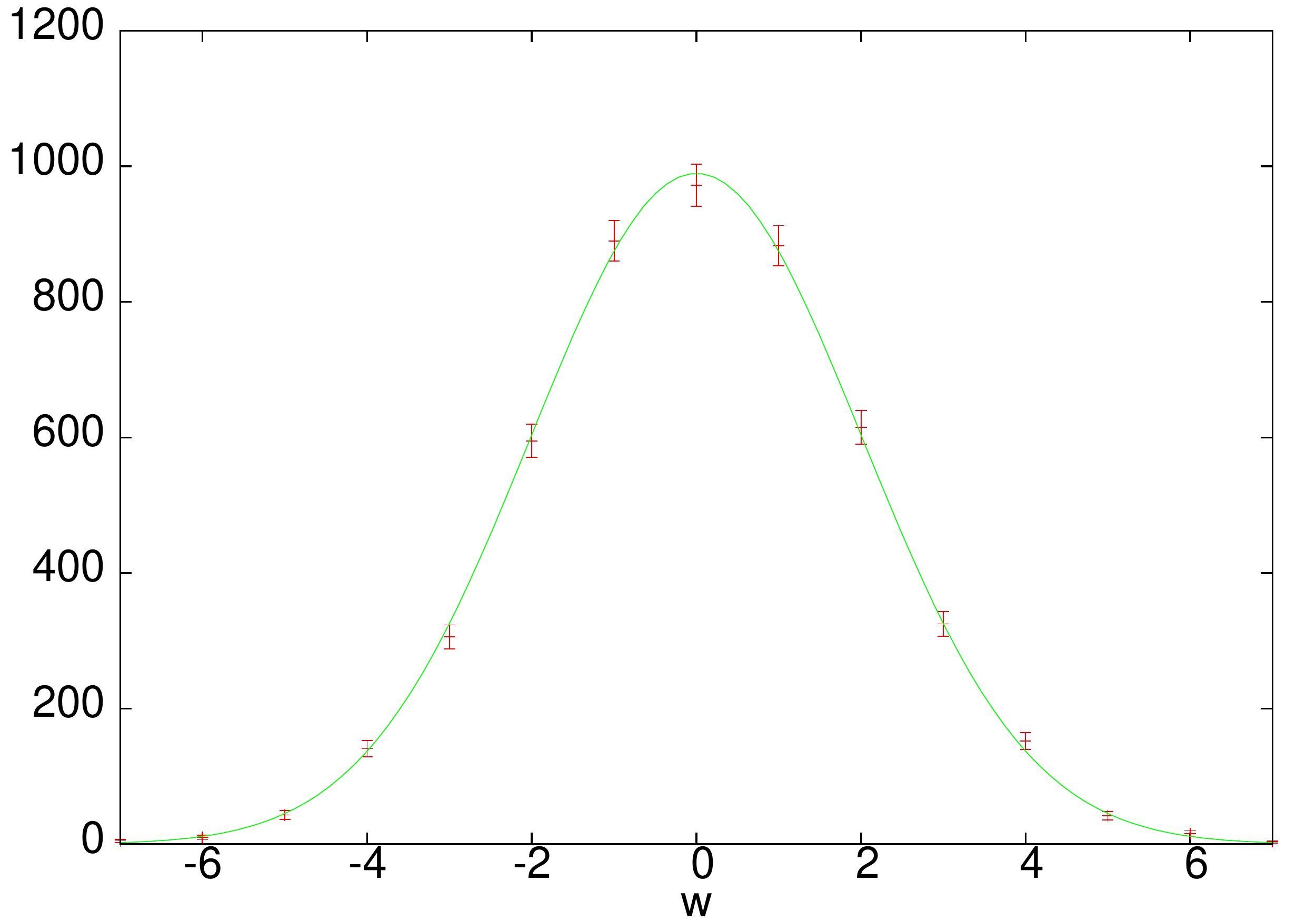,angle=-90,width=9cm}
\caption{Histogram of Higgs winding number
for the initial configuration  $mt=5$ for $\pmin=0.15\, m$.}}
\label{fig:wind}}

The size and spatial distribution of this initial electromagnetic and
$Z$ fields can be obtained from the multicomponent Gaussian random
field. In Appendix~\ref{app4} we displayed the histogram of magnetic
field values. Here we will focus on another aspect which is
particularly interesting for the later evolution. This is the spatial
distribution of points where the magnetic field intensity is larger.
To investigate this, we show in Fig.~\ref{fig:mt5} a 3-dimensional
plot displaying the locus of points where the magnetic energy density
is above 0.03 m$^{4}$ for our initial configuration at $mt_i=5$.
Notice that the regions of higher magnetic energy density exhibit a
string-like geometry. Indeed, this spatial distribution tracks the
location of regions of low Higgs field value, which are also presented
in the figure. Although, the strings seem to end at certain spatial
points, this is simply a reflection of the spreading of magnetic flux
lines. Our electromagnetic field satisfies the Maxwell equations
without magnetic sources or sinks.  According to our formulas the
initial magnetic-like component of the $Z$-boson field strength is
directly proportional to the electromagnetic field and has identical
structure.

\FIGURE{
\centerline{
\psfig{file=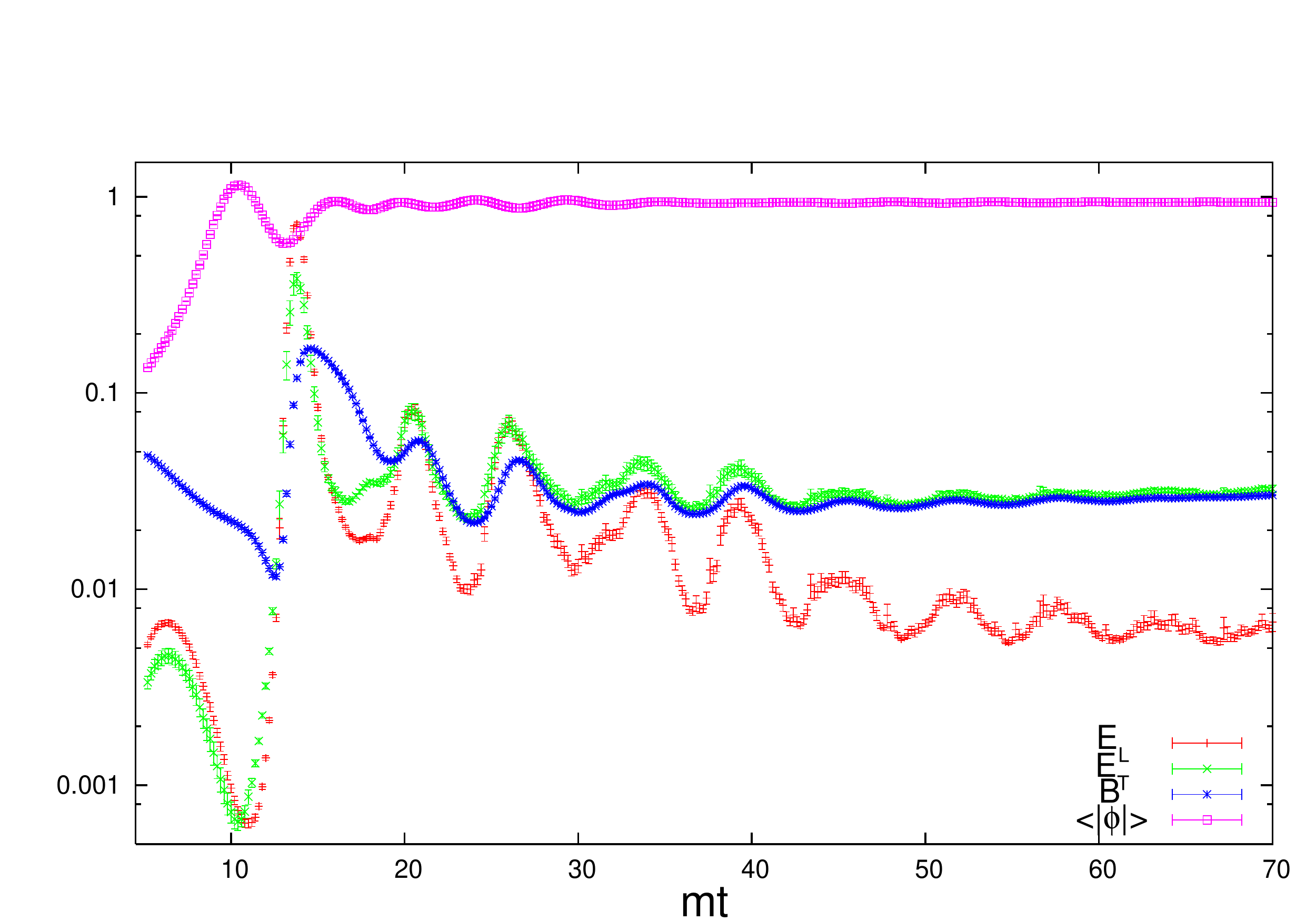,angle=-90,width=11cm}
\caption{ We show the time evolution of the electric (transverse $E_T$
and longitudinal $E_L$) and magnetic energy densities averaged over 150
configurations for $\mh = 3\mw$, $ma=0.42$ and $\pmin=0.15\,m$. }}
\label{fig:energy}}

There is another important feature of magnetic fields which we have
investigated. It corresponds to whether the initial field gives rise
to a sizable helicity. In a finite volume the total magnetic helicity
is defined both in configuration and in momentum space as: 
\be \label{helicity}
H \equiv
\int d^3 x \ h(x) = \int d^3 x \, \vec A \cdot \vec B \equiv {-i\over
  {\cal V} } \sum_k {\vec k \over |\vec k|^2} \cdot (\vec B(\vec k)
\times \vec B^*(\vec k)) \,,
\ee 
where ${\cal V}$ is the volume of space. Notice that this equality
makes use of Maxwell's condition $\vec \nabla \vec B =0$, which is
ensured by our magnetic field definition (\ref{fem}).  At our initial
time, by virtue of Eqs.~(\ref{eq:initial}), this quantity is
proportional to the winding of the Higgs field. This is defined as the
index of the map from the spatial volume to the group SU(2)=$S_3$,
provided by the matrix $\Omega(x)$. A histogram of the winding
obtained for our initial Gaussian random field configurations is
displayed in Fig.~\ref{fig:wind}.  The data are well described by a
Gaussian distribution.  Since we have not included CP violating terms,
the mean value of the winding number is zero.  However, we observe a
non-zero dispersion from which one can obtain a non-zero
volume-independent topological susceptibility $\chi=0.52\times
10^{-4}\, m^3$.  This translates into a corresponding non-vanishing
helical magnetic susceptibility $\chi_H\equiv \langle H^2 \rangle
/{\cal V}=0.38(3)\, m^3$. The $Z$ helical susceptibility at this
initial stage is $\chi_Z\equiv \tan^{-4}\theta_W \,\chi_H$.

In the next subsection we will study the evolution of this helical
magnetic field during the highly non-linear epoch of symmetry
breaking.  This provides a connection between magnetic field helicity,
Z-strings and the occurrence of configurations carrying non-trivial
Chern-Simons number.  This result, which relates baryon number
generation and magnetic helicity, has been proposed, although along
somewhat different lines, by Cornwall~\cite{Cornwall1997}. The
connection has been studied recently by Copi et al.~\cite{Copi2008}. 
They showed that the sphaleron decay indeed gives
rise to helical magnetic fields.
 
At later stages, the Chern-Simons number creation processes stop,
leaving behind a remnant magnetic helicity
component~\cite{Vachaspati2001}, which is preserved in a plasma with
high electrical conductivity. Thus, this could provide a signature of
EW generation of primordial magnetic fields.

\FIGURE{
\centerline{
\psfig{file=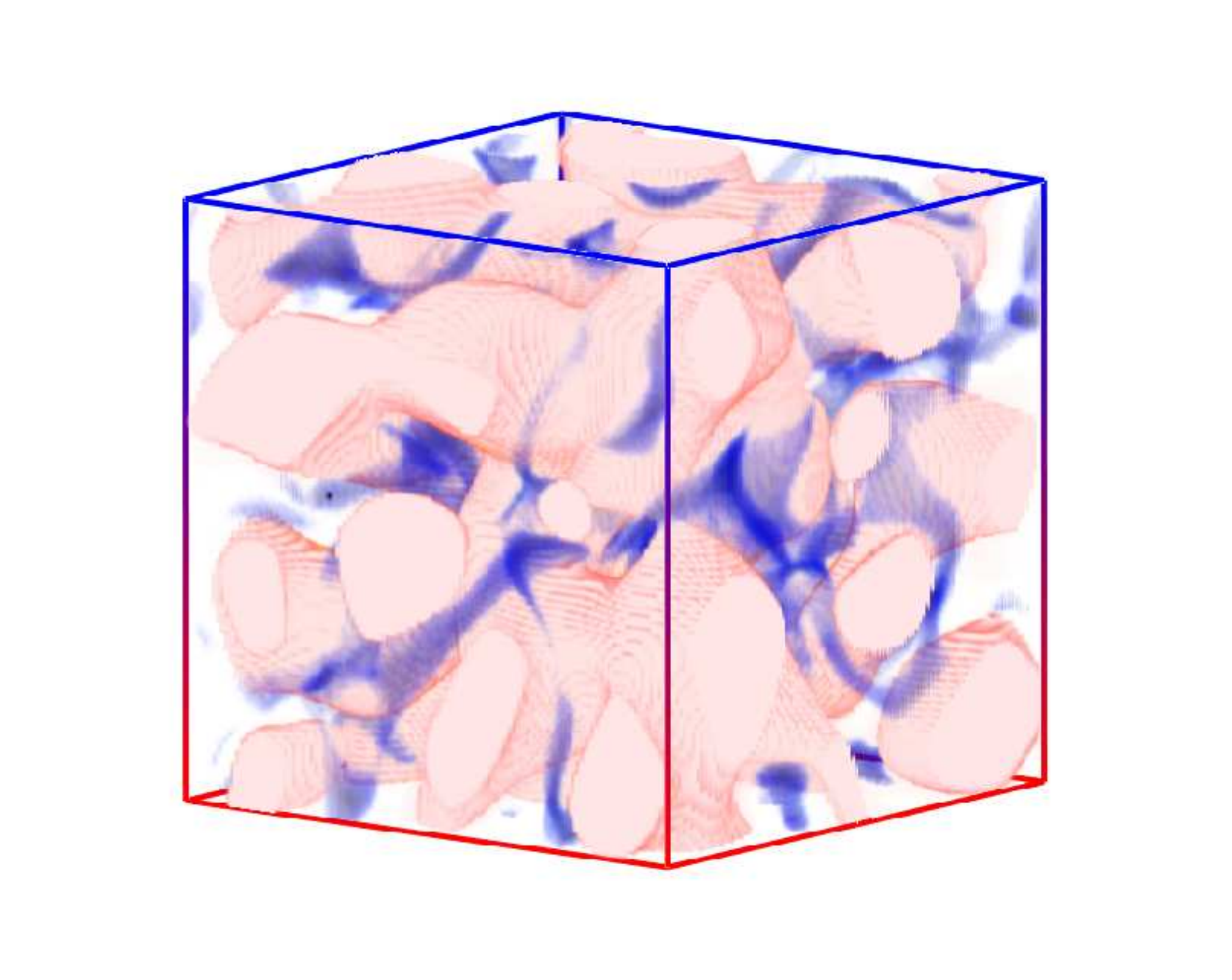,width=9cm}
\psfig{file=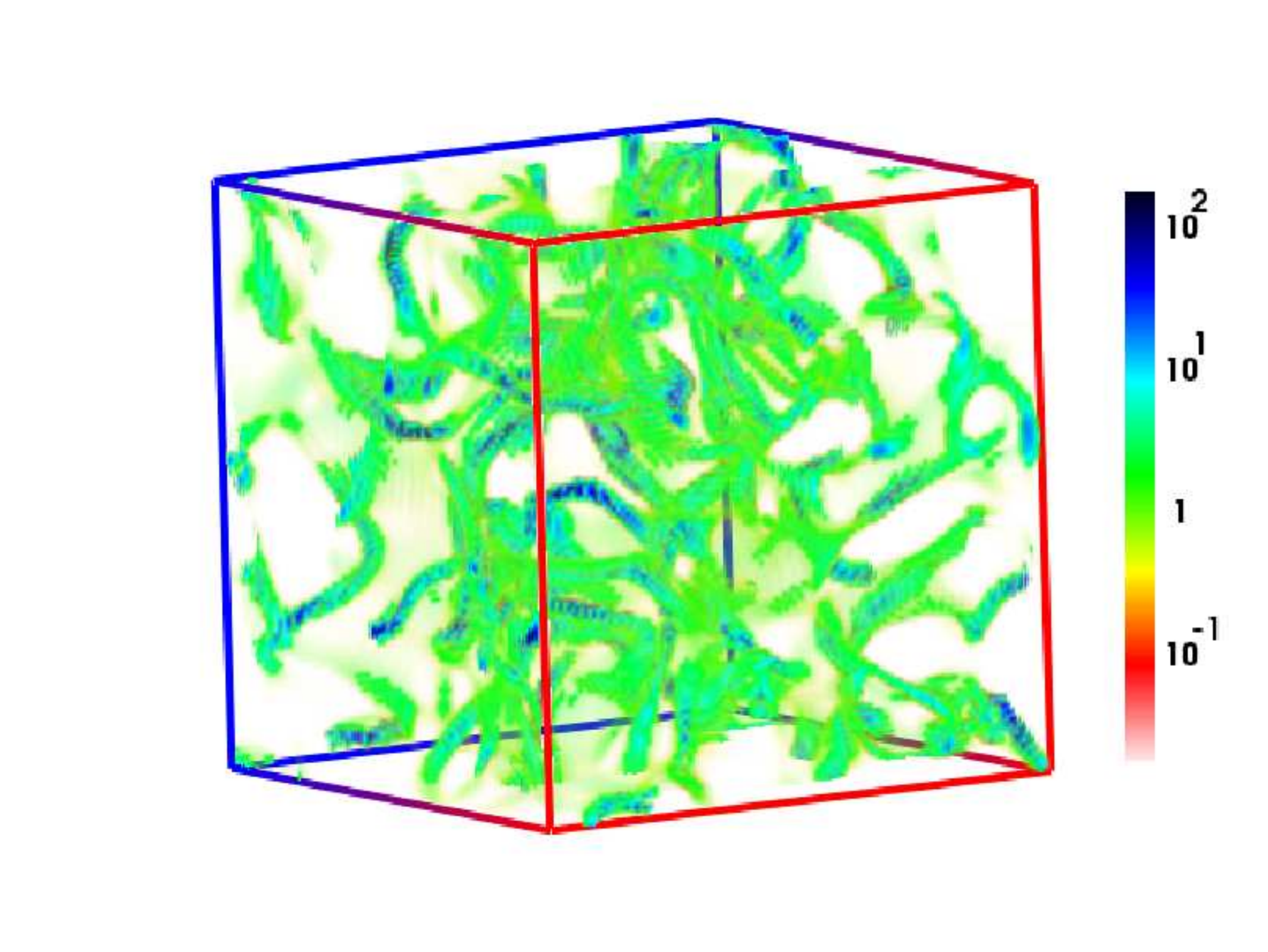,width=9cm}}
\centerline{
\psfig{file=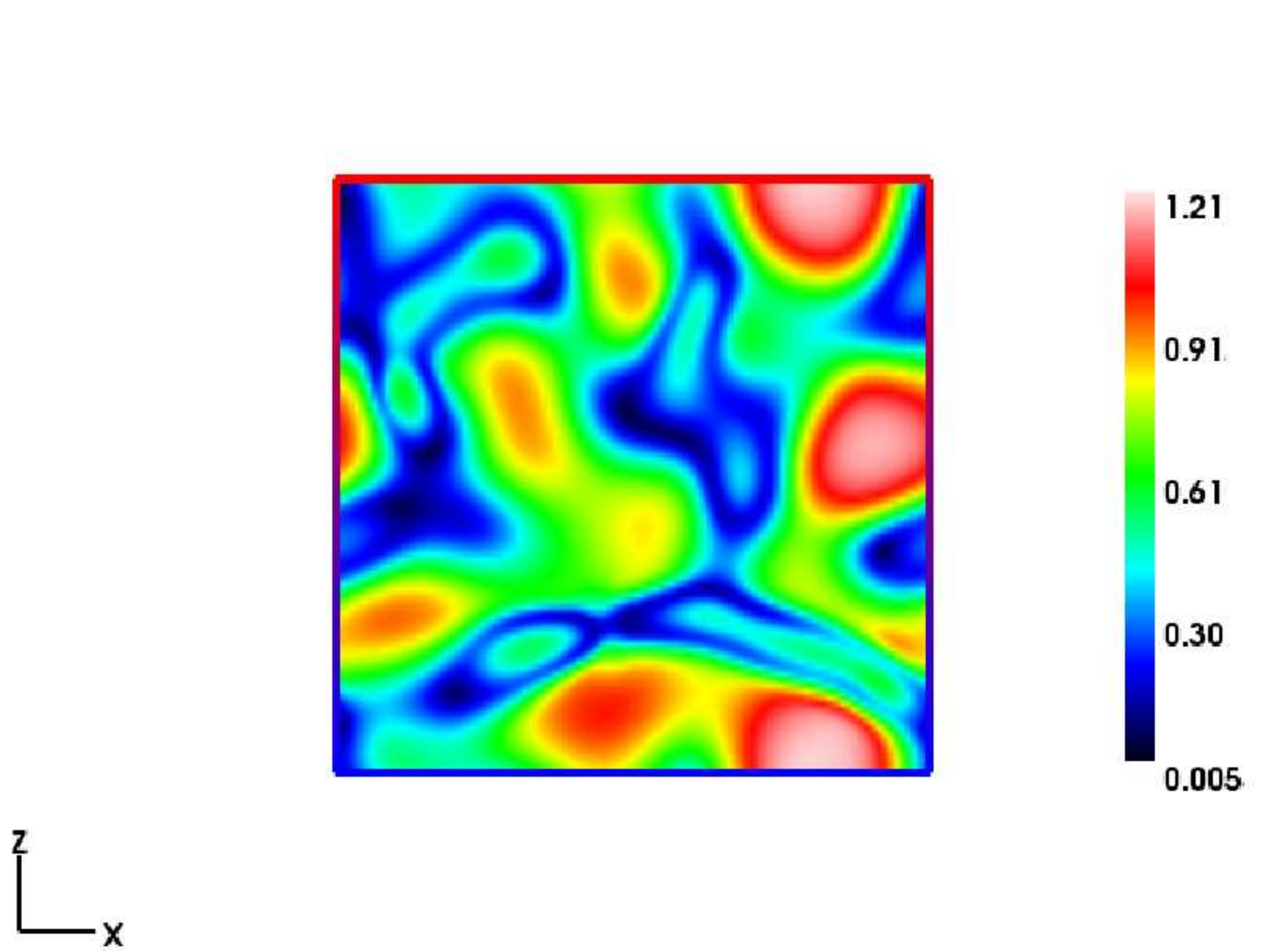,width=9cm}
\psfig{file=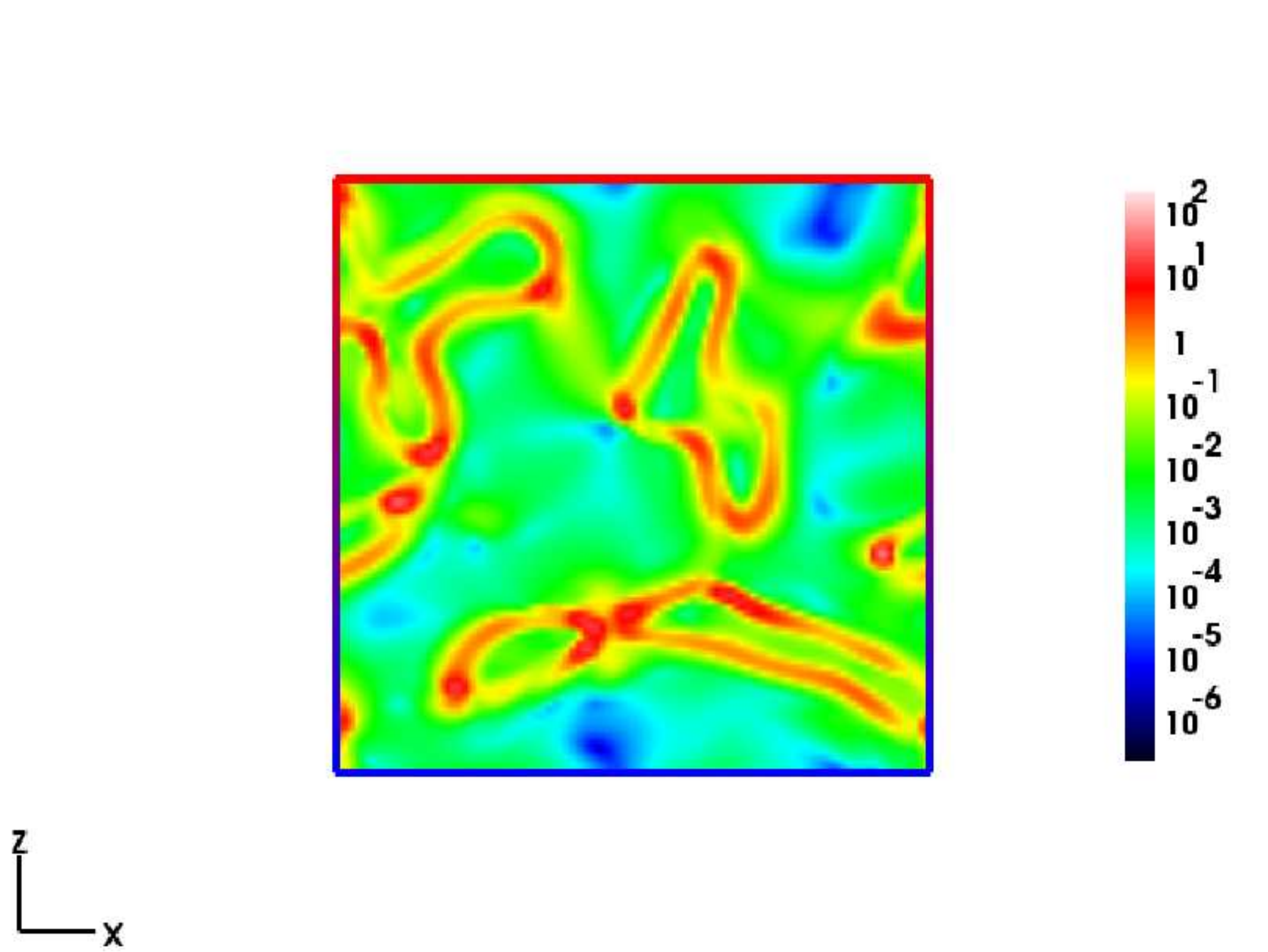,width=9cm}
\caption{Top: (Left) The location of the bubbles in the Higgs field
  norm (in red) with a lower cutoff set at $0.7\, v$ and the locus of
  points with twice the magnetic energy density ($|\vec B(\vec x)|^2$)
  (in blue) higher than $0.01\, m^4$.  (Right) Locus of points where
  the magnetic energy density is above $0.03\, m^4$.  Bottom: (Left)
  Two-dimensional contour plots of the Higgs field norm.  (Right)
  Two-dimensional contour plots of the magnetic energy density.  Data
  correspond to $mt=15$ and $\mh=2\mw$.}}
\label{fig:higgsymag}}

\FIGURE{
\centerline{
\psfig{file=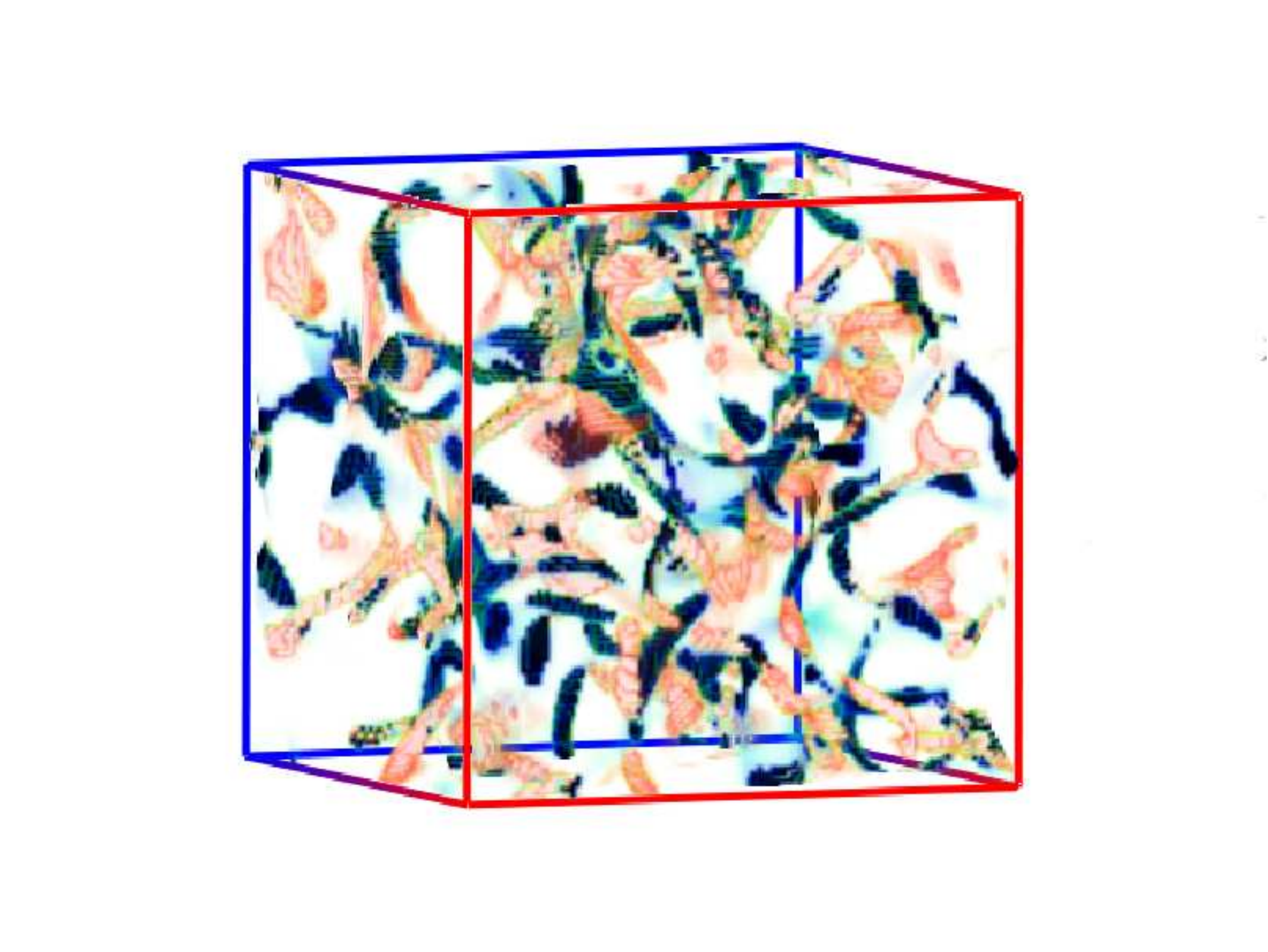,width=9cm}
\psfig{file=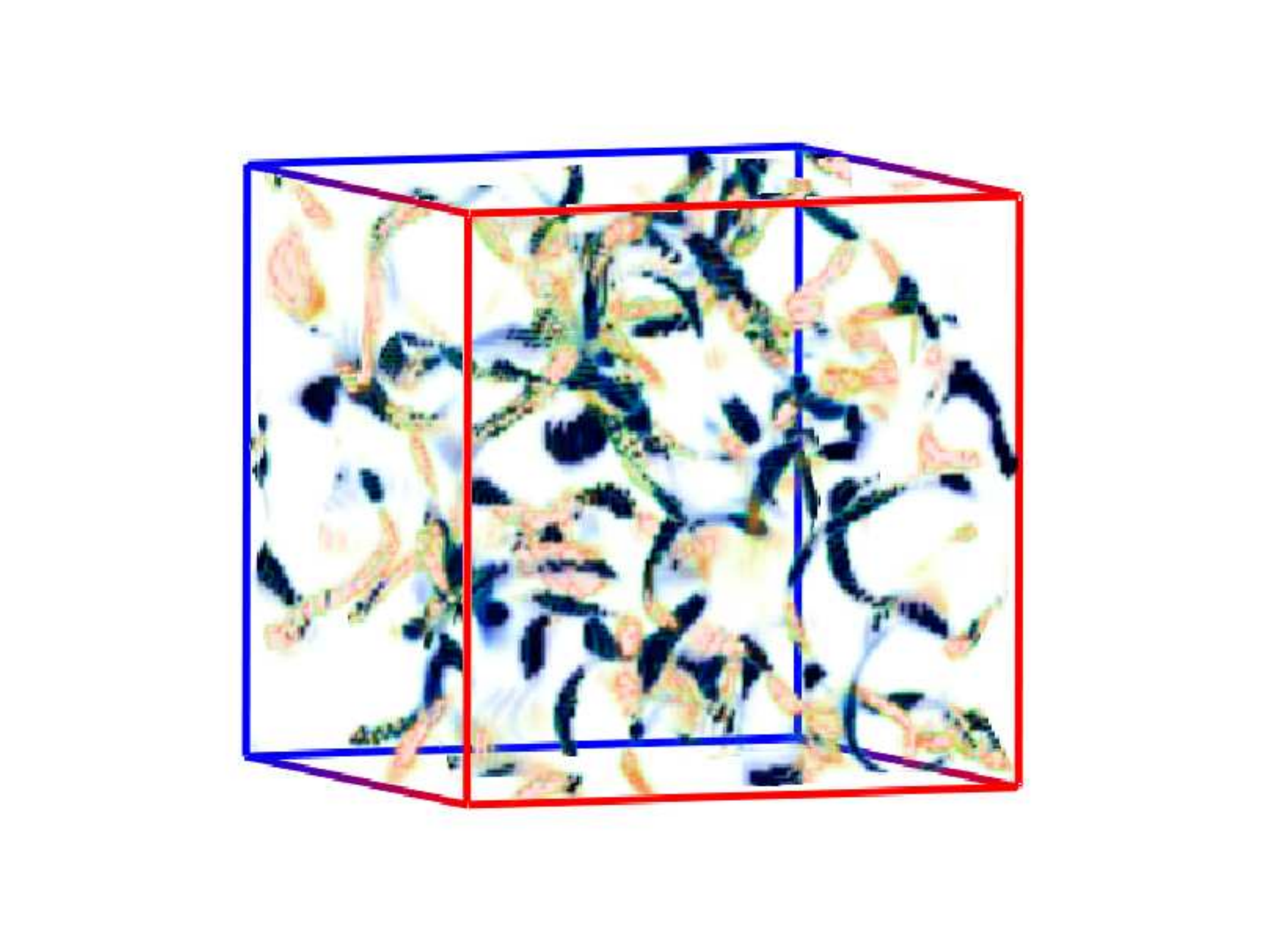,width=9cm}}
\centerline{
\psfig{file=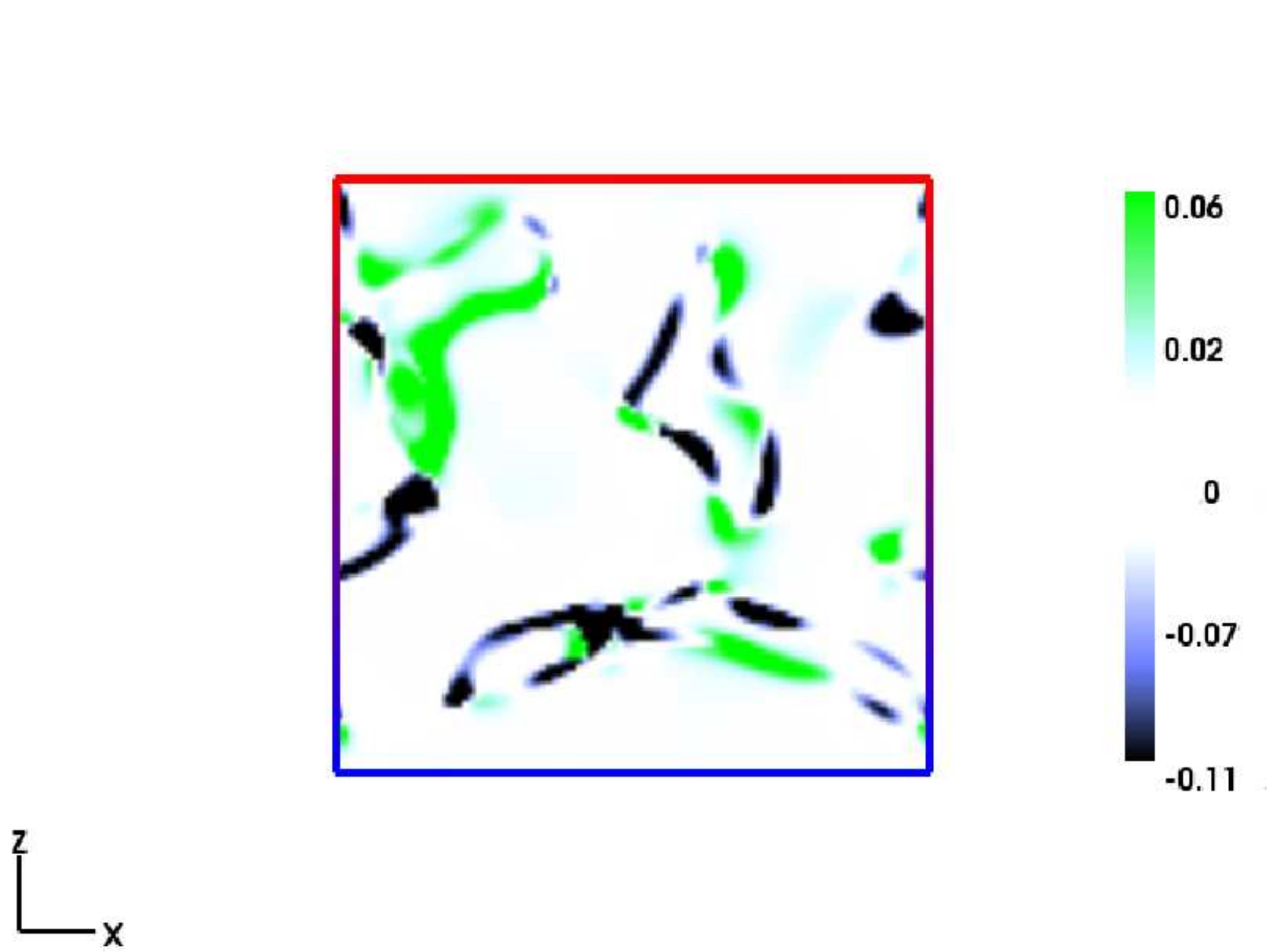,width=9cm}
\psfig{file=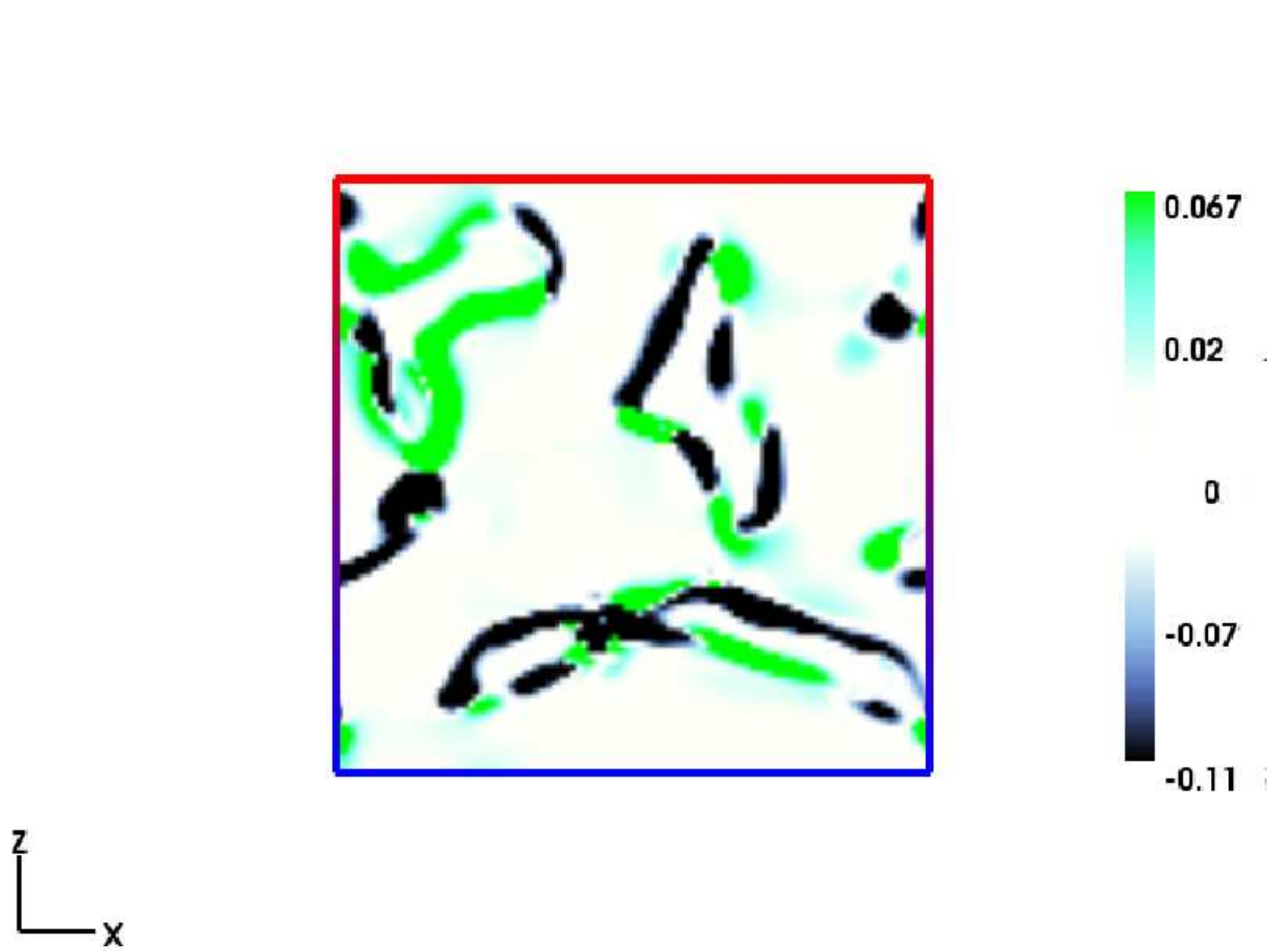,width=9cm}
\caption{Top: (Left) Helicity of the magnetic field.  (Right) Helicity
  of the $Z$-boson field.  Bottom: (Left) Two dimensional contour
  plots of the helicity of the magnetic field.  (Right) Two
  dimensional contour plots of the helicity of the $Z$-boson field.
  Data correspond to $mt=15$, for $\mh=2\mw$.}}
\label{fig:helt15}}

\subsection{Magnetic strings through symmetry breaking}

We will now focus upon the evolution of the system from the initial
Gaussian random field situation until the onset of symmetry breaking.
To have a global picture of the process we show in
Fig. \ref{fig:energy} the time evolution of the expectation value of
the Higgs field from the initial time $mt_i=5$ of our classical
evolution. Notice the strong initial oscillations for times smaller
than $mt=20$, which are then progressively damped at larger times.
The figure also displays the fraction of the total energy density
carried by electromagnetic fields. We split it into its magnetic and
electric components, and for the latter we analyze separately
longitudinal and transverse parts.~\footnote{ The technicalities
involved in the lattice definition of transverse and longitudinal
fields as in the definition of the $W$ bosons charge densities and
currents are discussed in Appendix~\ref{app2}.} We observe that
between $mt=10$ and $mt=15$, there is an explosive growth of the
electromagnetic fields correlated with the first minimum in the
oscillation of the Higgs field expectation value.  The data in the
figure corresponds to $\mh=3\mw$, but similar behaviour is observed
for the other ratios studied.

\FIGURE{
\centerline{
\psfig{file=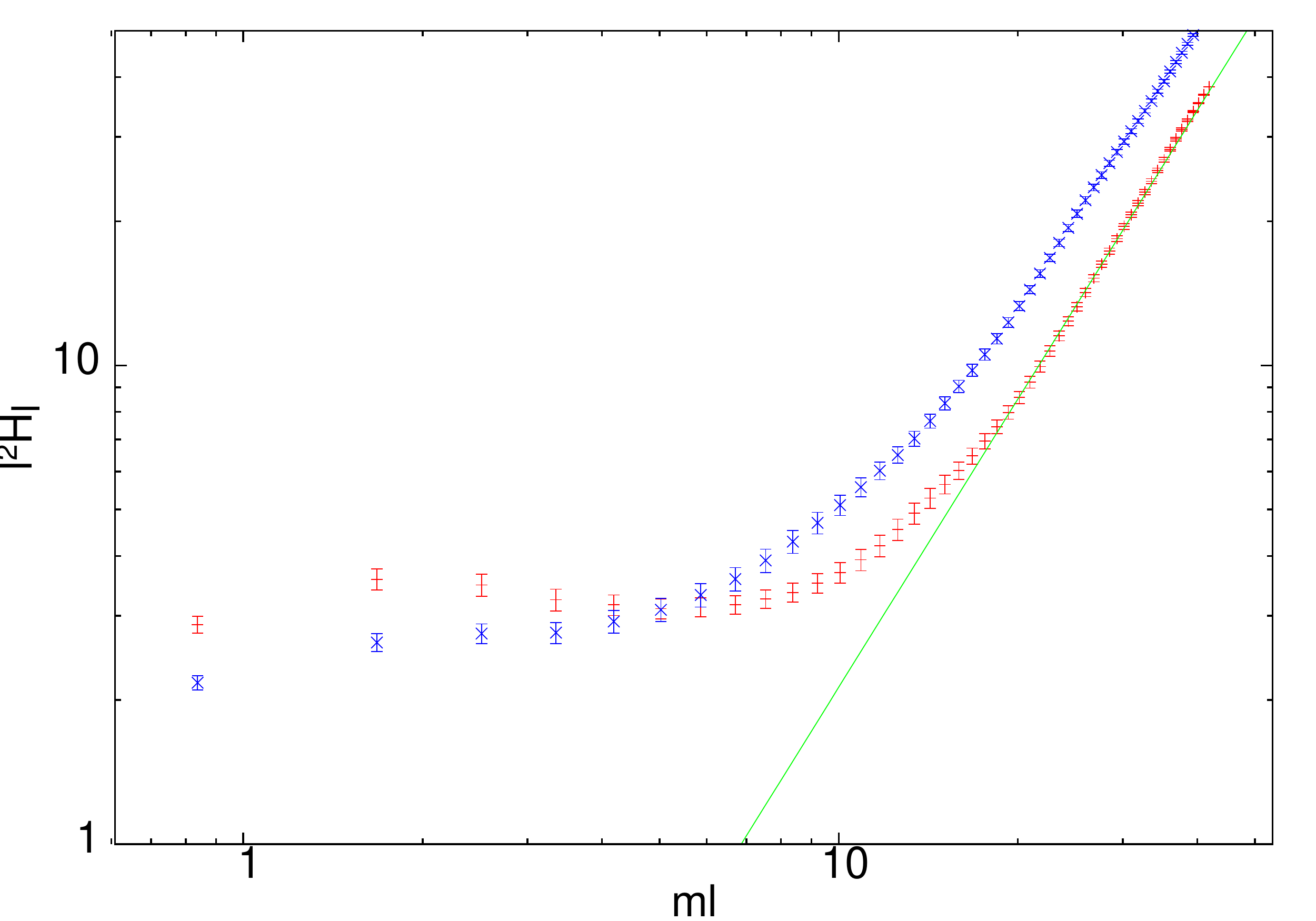,angle=-90,width=9cm}
\caption{ Averaged $l^2 H_l(r_0)$, Eq.~(\ref{HL}), at $mt=15$. We also
show the $l^2$ asymptotic behaviour for the
  $\mh/\mw=4.65$ model. The other data corresponds to $\mh/\mw = 2$.}}
\label{fig:avplato}}

We will now present the spatial structure observed for the magnetic
fields at $mt=15$ after the strong oscillation region. The
corresponding distribution of the Higgs field modulus has been
presented in Ref.~\cite{jmt}.  There we showed that the initial
Gaussian peaks lead to bubbles which expand and collide with
neighbouring ones. This is illustrated in the top left of
Fig.~\ref{fig:higgsymag} where we display a snapshot of the Higgs
field norm at $mt=15$.  At this time bubble shells (in red), that have
grown out of the peaks in the initial Gaussian random field, fill
almost all the volume of the box.  Magnetic fields (shown in blue in
the figure) appear as string-like structures localized in the region
between bubbles, where the Higgs field remains closer to the false
vacuum for a longer period of time.  This linkage between magnetic
strings and Higgs field minima is even more evident in the two
dimensional contour plots presented in the bottom half of
Fig.~\ref{fig:higgsymag}.

The structures observed in the regions of maximal magnetic density are
reproduced when looking at the helical part alone. This is exemplified
by the comparison of Fig. \ref{fig:helt15} with
Fig. \ref{fig:higgsymag}.  The figure also shows how the correlation
between magnetic and $Z$ boson fields, implicit in our initial
conditions, is still preserved once gauge fields and non-linearities
have started to play a role.  An interesting observation can be made
here concerning the connection with baryon number generation.
Analysis of the cold EW transition show that sphaleron-like
configurations, with non-trivial Chern-Simons number, are also located
between bubble shells \cite{jmt2}-\cite{smit0}.  For non-zero Weinberg
angle, sphalerons look like magnetic dipoles~\cite{KM84} and it is
tempting to correlate the observed helical magnetic flux tubes with
the alignment of sphaleron dipoles. Although a detailed investigation
of this correlation is beyond the scope of this paper, our results for
the distribution of magnetic helicity do indeed hint in that
direction.  An evaluation of the net helicity at late times and a
discussion on its persistence will be postponed to section
\ref{latet}.

In the previous figures, the closed string-like structure of the
helicity and magnetic field appears much more clearly that in the
Gaussian random field initial condition at $mt_i=5$. To quantify the
string-like character, we have analyzed the following
quantity:
\begin{equation}
\label{HL}
H_l(r_0)= \frac{1}{l^3}\int_{L(r_0)} dx^3 |\vec h(x)| \,,
\label{eq:bav}
\end{equation}
where $\vec h(x)$ denotes the helicity density and the integration is
on a box of length~$l$, centered at a point $r_0$ at the center of one
of the strings. Figure~\ref{fig:avplato} shows the $l$-dependence of
$l^2 H_l(r_0)$, averaged over several configurations.  The figure is
intended to show the one-dimensional character of the distribution in
accordance to our string picture. In that case, $l^2\,H_l(r_0)$ should be
$l$-independent in contrast with the $l^2$-behaviour characteristic of
an isotropic distribution. Both regimes are clearly observed in the
figure.  The stringy behaviour is displayed up to $ml\sim 10$, beyond
which the plot shows how the data tends nicely to a straight line of
slope equal to~2. This is to be expected once the box is big enough to
contain several strings.  This leads to an estimate of the string
separation of $ \mh l \sim 14$, which is a significant fraction of the
total length of the box.

\subsection{Charge lumps around magnetic field lines}

\FIGURE{
\centerline{
\psfig{file=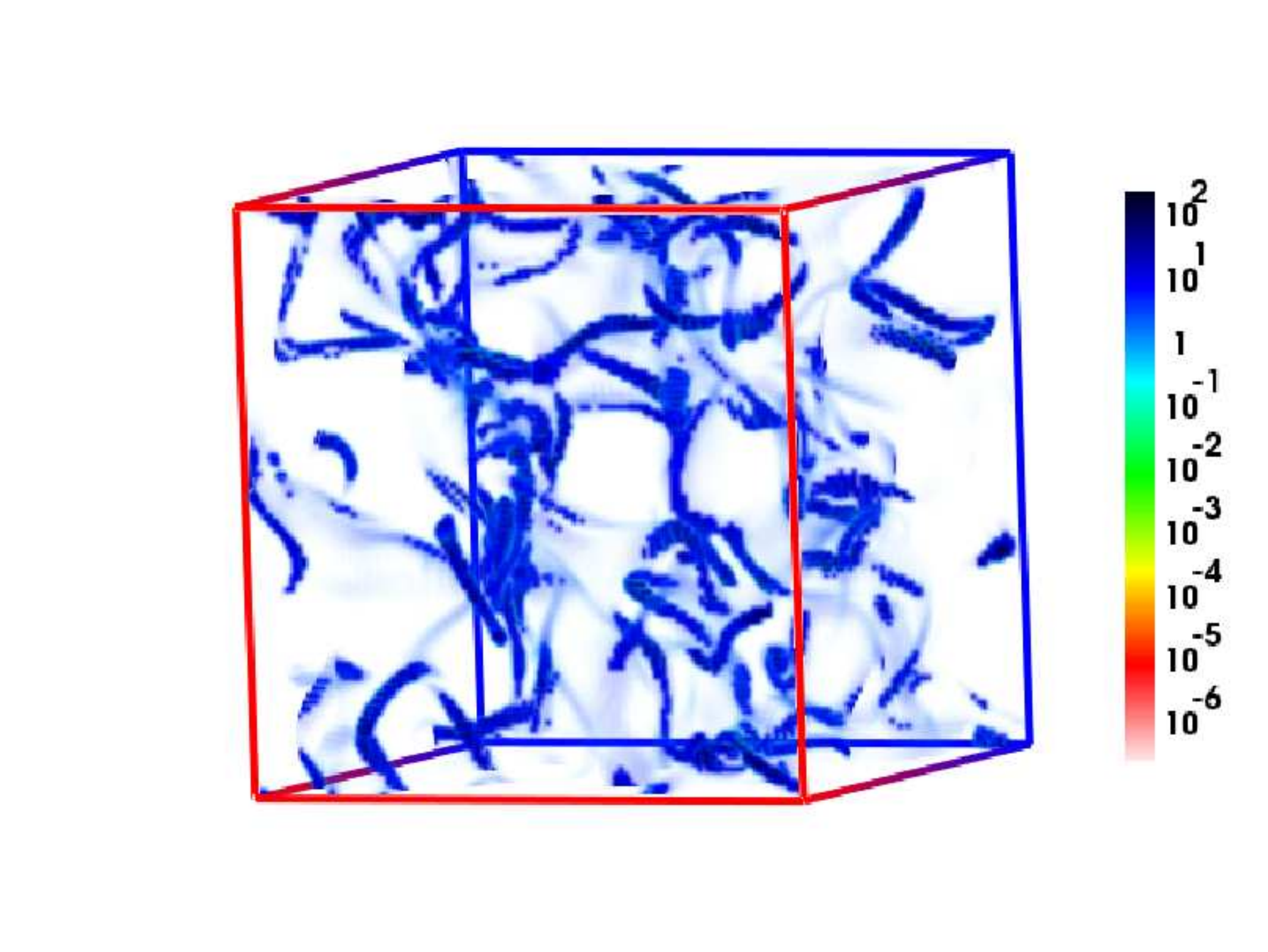,width=9cm}
\psfig{file=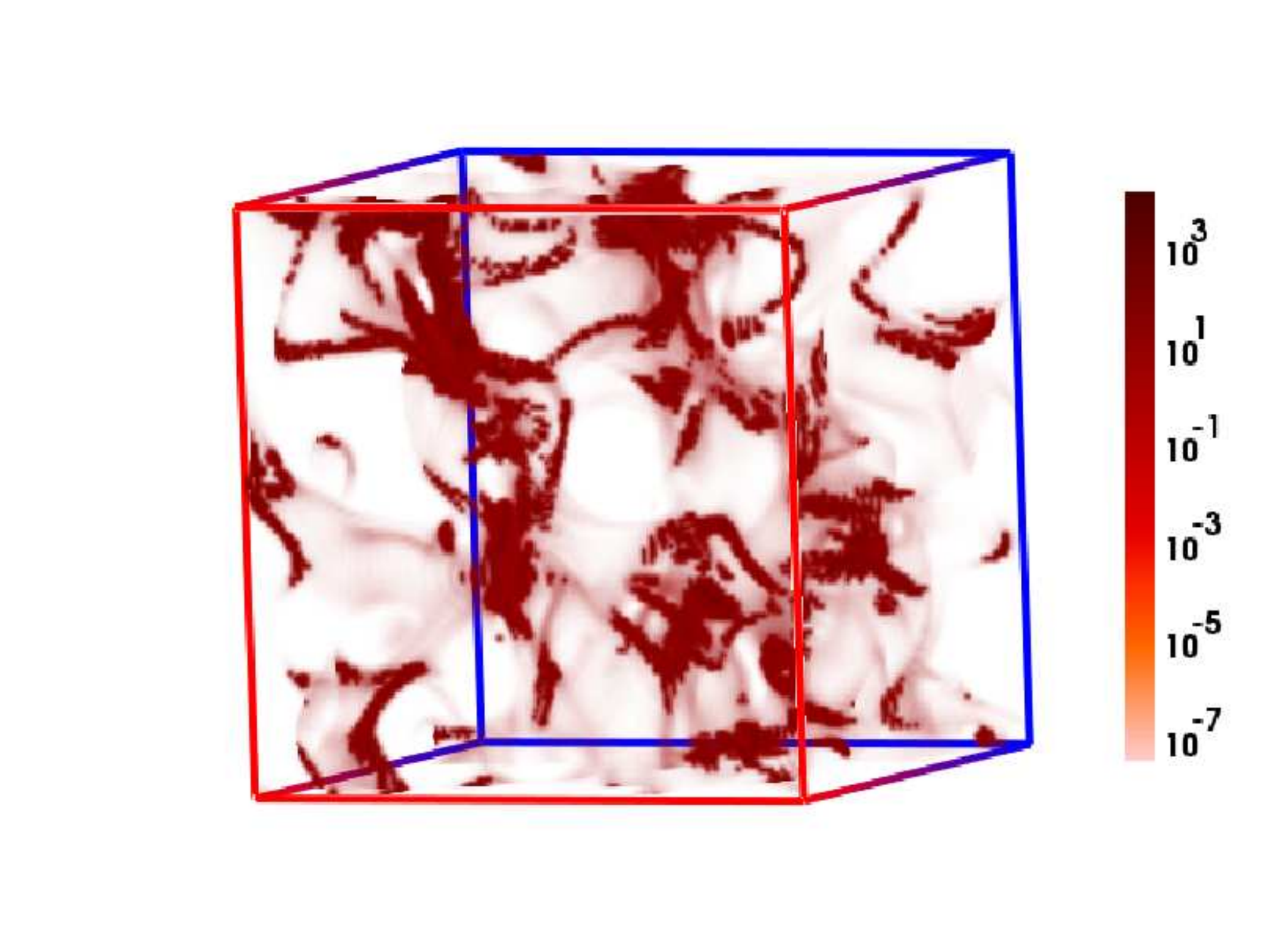,width=9cm}}
\centerline{
\psfig{file=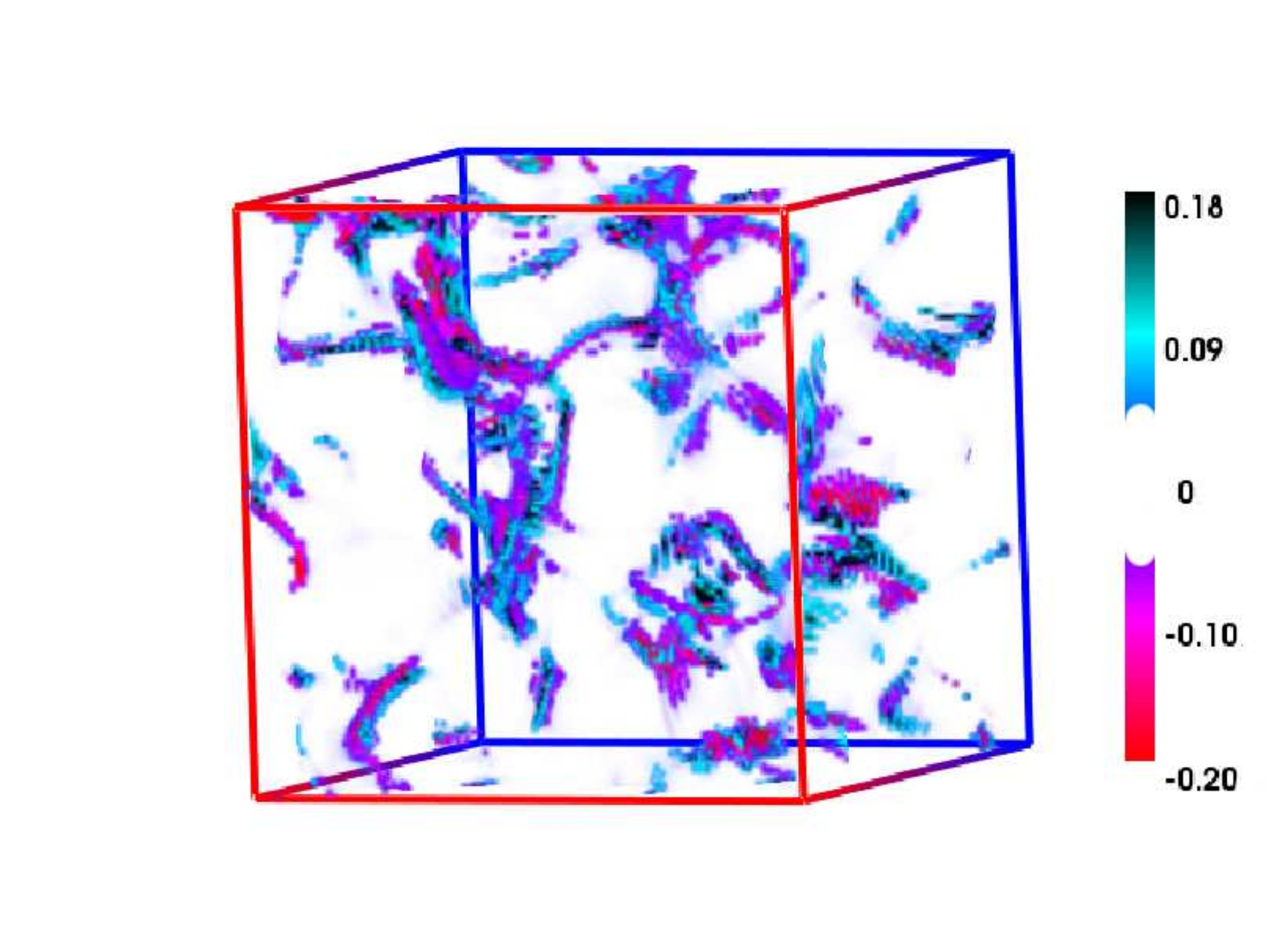,width=9cm}}
\vspace*{-9mm}
\caption{ Top Left: Locus of points with magnetic energy density
$|\vec B(x)|^2$, above $0.01\, m^4$.  Top Right: Locus of points
with electric energy density $|\vec E(x)|^2$ above $0.01\,m^4$.
Bottom: The distribution of $W^\pm$ charge density, tracking the
magnetic field lines. Pink and blue areas represent negative and
positive charge densities respectively. Data correspond to $mt=15$,
 for $\mh=2\mw$.  }
\label{fig:charge1}}

Up to now we have focused on the distribution of magnetic and
$Z$-boson fields, but there is important additional information on the
nature of the primordial plasma during these stages of preheating.
Note that our initial conditions provide a source for charged
$W$-currents and a non-trivial charge density.  It turns out that
there is charge separation at the initial stage. Positive and negative
charges are clustered into separate lumps which track the magnetic
field lines.  Figures~\ref{fig:charge1} and~\ref{fig:charge2} show
this effect at $mt=15$ and $mt=10$, respectively. Note that there 
is a strong correlation between the magnetic field lines and the
distribution of charges of opposite sign around them. The effect is 
seen particularly clear at early times, $mt=10$, where the magnetic 
flux tubes are well defined, and there are fewer of them. The charge
separation is consistent with the effect that would be produced by a
combination of the drift currents induced by gradient and curvature
effects from the magnetic field. The electric field is also strongly
correlated with the location of the charge lumps, as expected. This
charge separation might be responsible for the very slow screening
observed for the longitudinal electric field, which will be discussed
in the next section.

The plasma generated during the first stages of evolution is, as we
have shown, somewhat different from standard MHD plasmas (composed
mainly of protons and electrons, together with photons). Here, long
range string-like structures are observed in the electromagnetic
fields, and opposite $W$-charges cluster in large regions of space
inducing non-trivial electric fields.  It is expected that these
charge lumps will eventually disintegrate when the $W$-fields decay
into light fermions (quarks and leptons), which travel at the speed of
light and diffuse the charge, leading at late times to a standard MHD
plasma.

\FIGURE{
\centerline{
\psfig{file=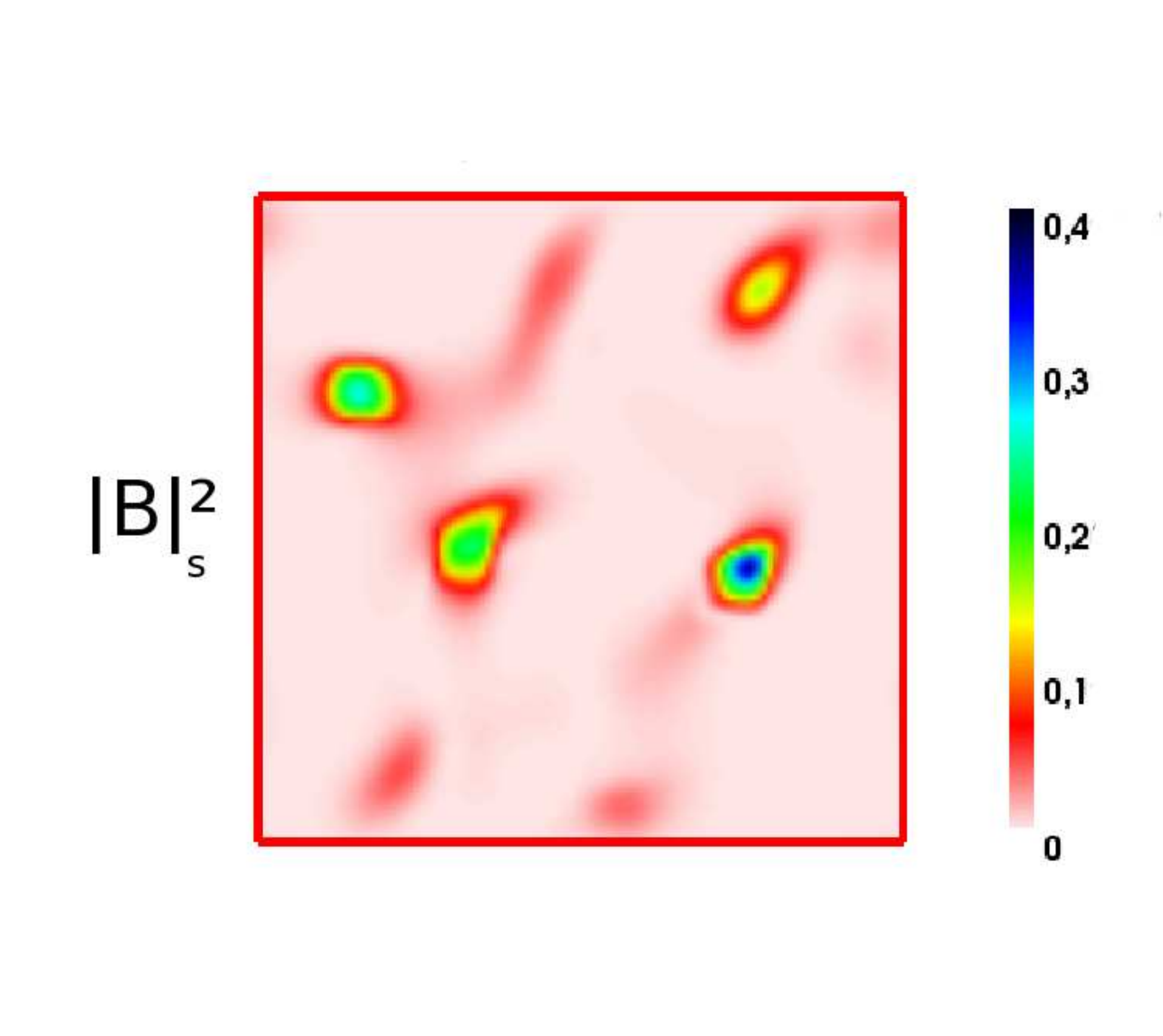,width=8.0cm}
\psfig{file=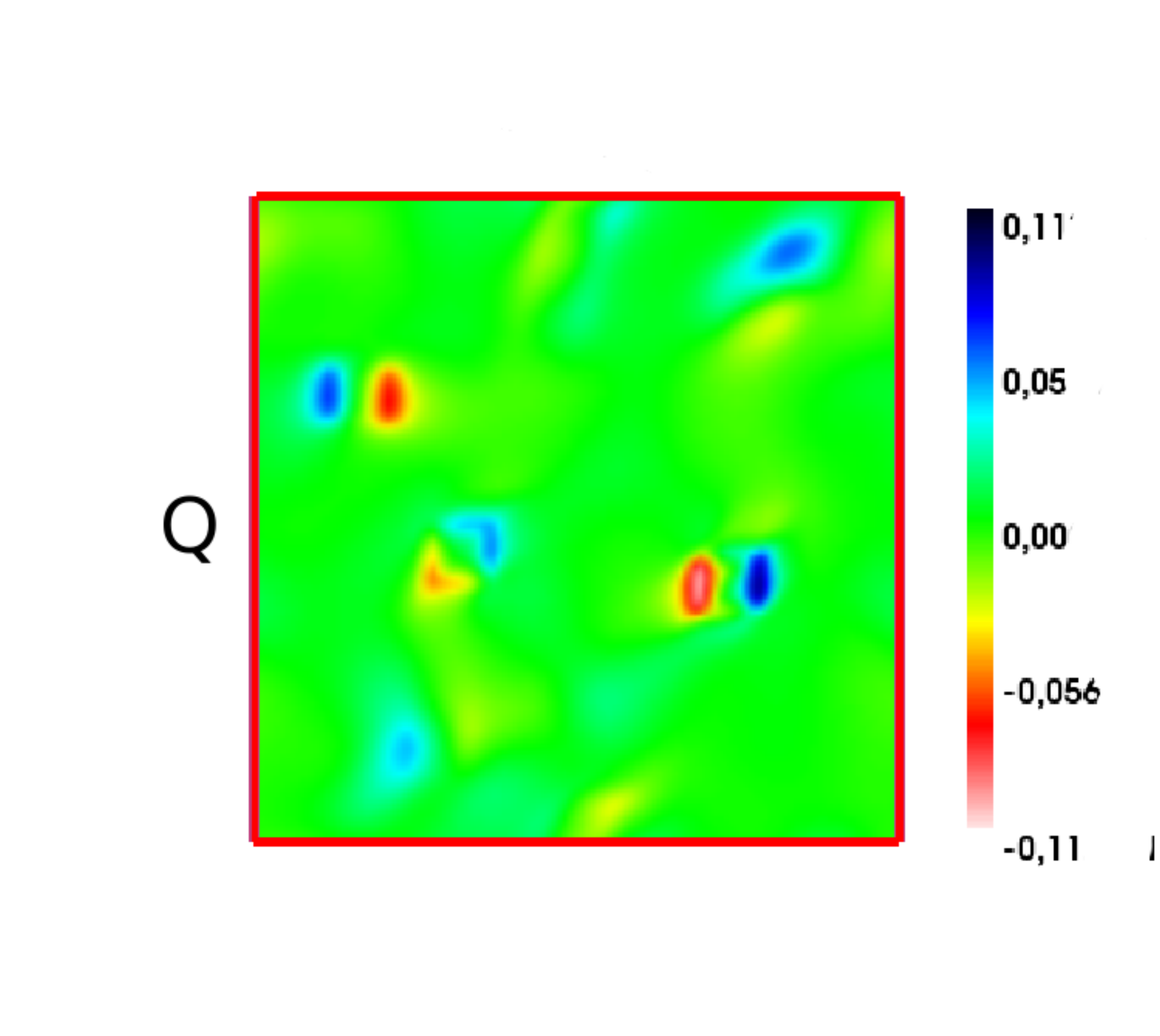,width=8.0cm}}
\vspace*{-0.4cm}
\caption{ Left: Locus of points with magnetic field
density $|\vec B(\vec x)|^2$ above $0.01\,m^4$.
Right: The 2-dimensional $W$-charge distribution
localized in lumps of opposite sign facing each other.
Note that the location of the charge lumps is strongly
correlated with the magnetic field flux tubes. These
figures correspond to early times, $mt=10$, for $\mh=3\mw$. }
\label{fig:charge2}}


\section{Late time evolution}\label{latet}

In order to claim a mechanism for cosmological magnetogenesis, the
essential question is whether the amplitude and correlation length of
the generated fields are enough to seed the large scale magnetic
fields observed today.  In this section we will present evidence that
a significant fraction of long range helical magnetic fields remains
after EW symmetry breaking and is even amplified at later times, a
period in which kinetic turbulence has been observed~\cite{tkachev,ajmt}. 
As we will see below, our estimate for the amplitude of the magnetic
field seed gives a fraction $\sim 10^{-2}$ of the total energy density
at the EW scale. This could be enough to seed the cluster and
supercluster values without the need for a dynamo mechanism.

\FIGURE{
\centerline{
\psfig{file=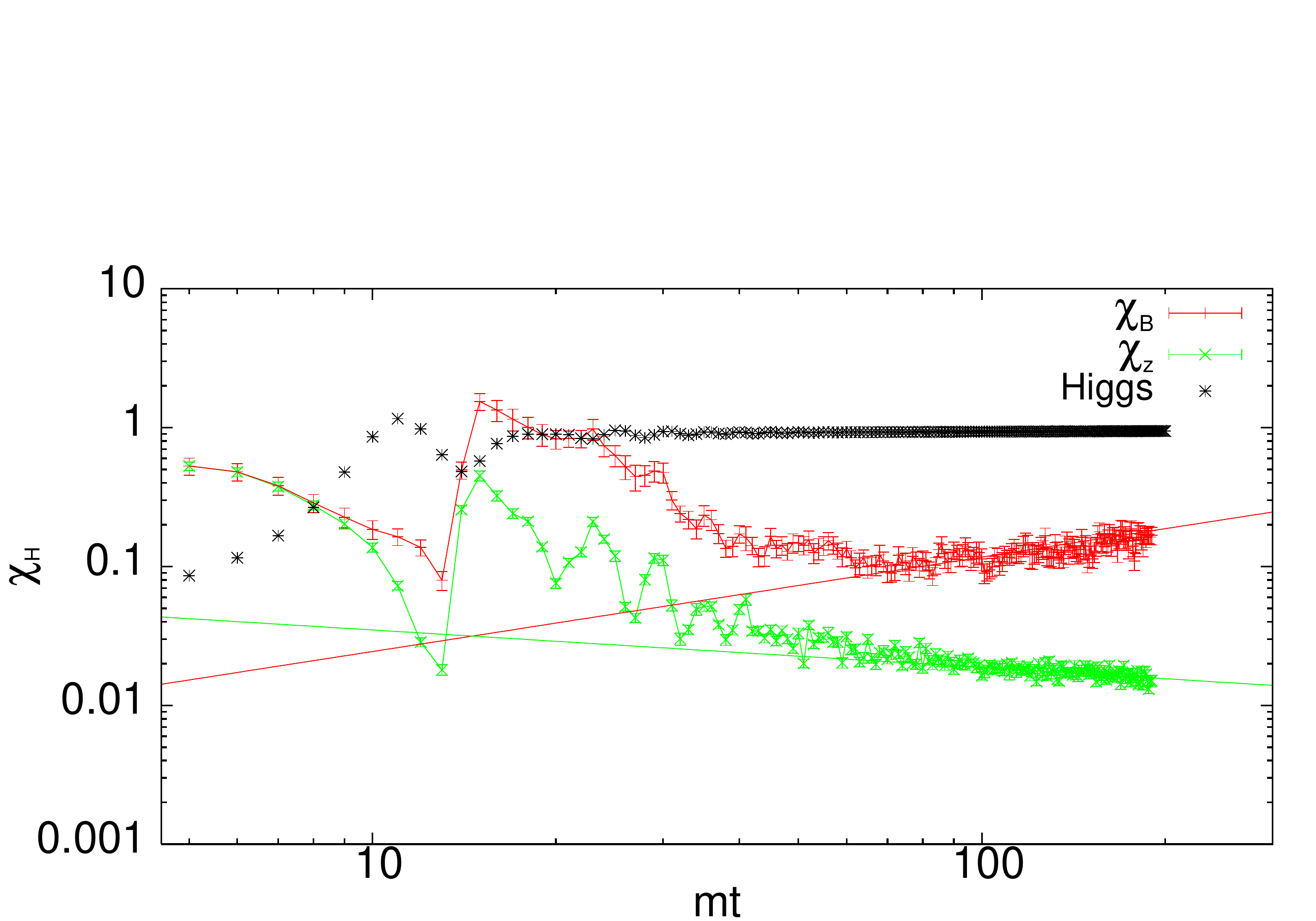,angle=-90,width=7.5cm}
\psfig{file=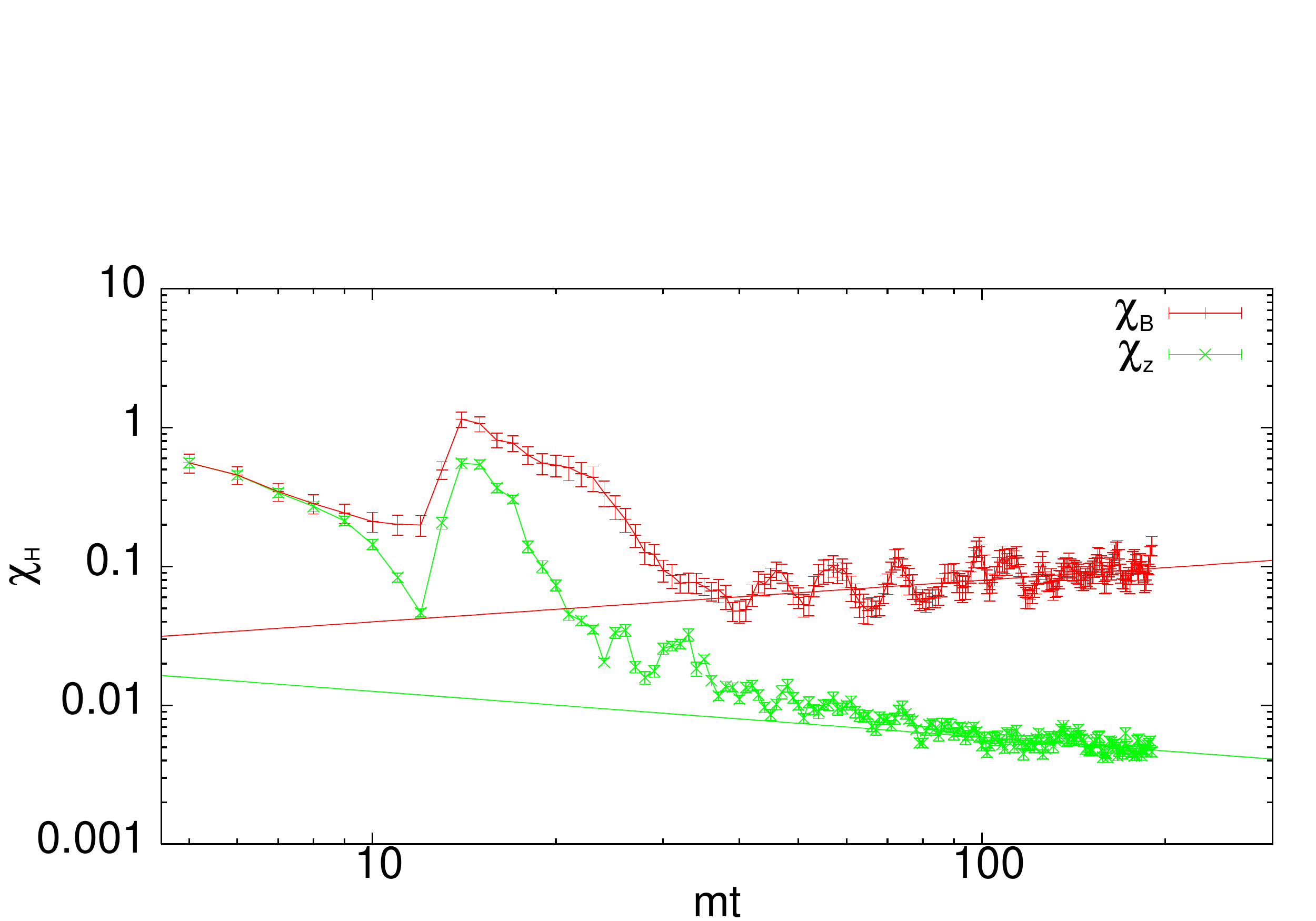,angle=-90,width=7.5cm}
}
\centerline{
\psfig{file=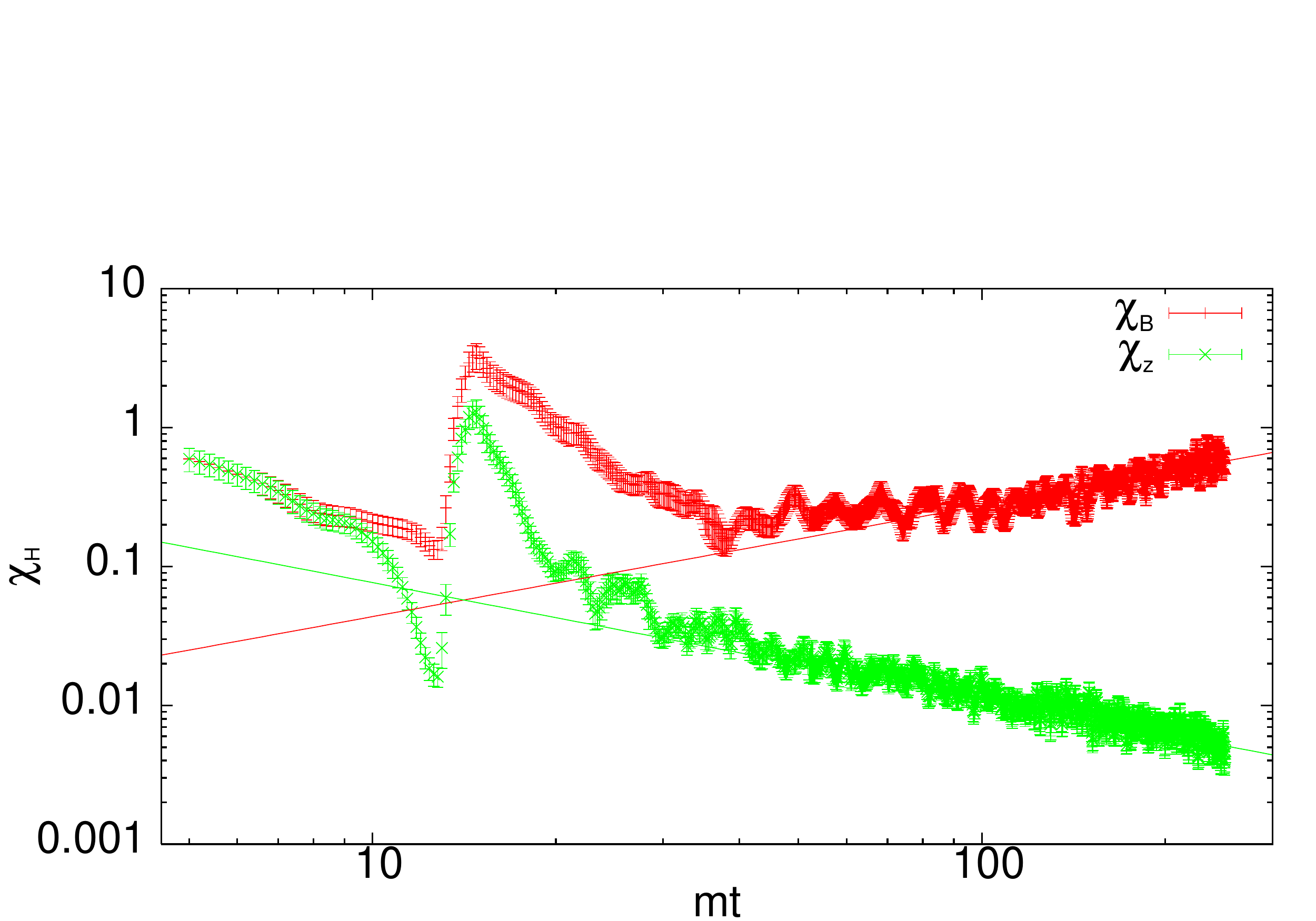,angle=-90,width=8.5cm}
}
\vspace*{-5mm}
\centerline{
\caption{ We display the time evolution of the helical
  susceptibilities for the magnetic field (with fit $t^\alpha$) and
  the $Z$-boson field (with fit $t^\beta$). The latter is rescaled by
  $\tan^4\theta_W$ to match the initial electromagnetic helicity. Top
  left is for $\mh/\mw = 2$, averaged over 80 configurations, with
  $\alpha=0.7(1)$ and $\beta=-0.27(4)$. Top right is for $\mh/\mw =
  4.65$, averaged over 80 configurations, with $\alpha=0.3(1)$ and
  $\beta=-0.33(5)$. Bottom is for $\mh/\mw = 3$, averaged over 200
  configurations, with $\alpha=0.8(1)$ and $\beta=-0.82(4)$. All data
  correspond to $ma=0.42$ and $\pmin=0.15\,m$. The top left figure
  also shows the time evolution of the Higgs mean to illustrate the
  time when SSB takes place. }} \label{fig:lhelicity}}

More difficult is to address the issue of whether the magnetic field
spectrum experiences inverse cascade, i.e.  transference of energy
from high to low momentum
modes \cite{Brandenburg1996}-\cite{Banerjee}. Inverse cascade is
required to make the coherence length of the magnetic field grow
(almost) as fast as the horizon until the time of photon decoupling.
Our approach does not allow to extrapolate the time evolution for
sufficiently long times. Nevertheless, we will provide some evidence
that inverse cascade might be at work. However, additional work is
required to analyze if it can be sustained for a sufficiently long
time.  This might require a full magnetohydrodynamics treatment of the
time evolution for which our set up will provide an initial condition.

\FIGURE{
\centerline{
\psfig{file=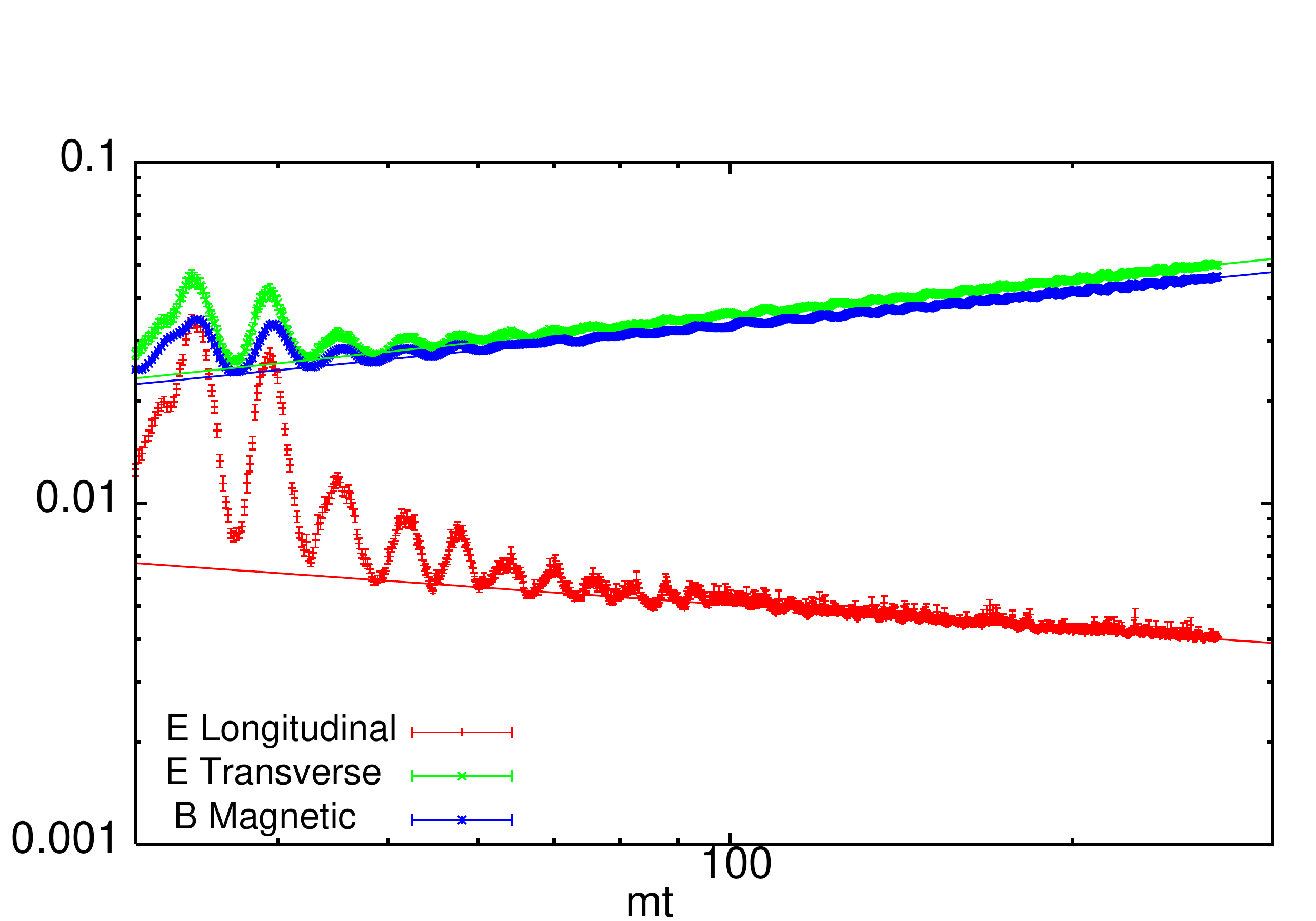,angle=-90,width=11.cm}
\caption{ We display the log-log plot of the time evolution of the
electric (transverse and longitudinal) and magnetic energy densities.
The exponents of the power law fits are:
Transverse electric field: $0.350(1)$; Magnetic field: $0.330(1)$
and Longitudinal electric field: $-0.234(2)$.
For $\mh=3\,\mw$, $ma=0.42$ and $\pmin=0.15\,m$, averaged over 200 
configurations.}}
\label{fig:lenergy}}


\subsection{Magnetic helicity and electromagnetic energy densities}

We will first analyze in detail how electromagnetic fields evolve in
time, paying particular attention to the evolution of the magnetic
field helicity long after SSB.

As mentioned above, the relevant quantity for helicity in the absence
of CP violation is the helical susceptibility $\chi_H$. Its time
evolution, for different values of the $\mh/\mw$ ratio, is displayed
in Fig. \ref{fig:lhelicity}.  At the same time we display the helical
susceptibility of the $Z$ boson magnetic field, rescaled by
$\tan^4\theta_W$ to make it agree with the initial electromagnetic
helicity, see discussion after Eq.~(~\ref{helicity}). The late time
behaviour, after $mt\sim 60$, gives further support to the
Vachaspati-Cornwall's conjecture.  It corroborates that, while the $Z$
boson helicity is damped in time, the magnetic helicity is preserved
and even increases with a power law dependence in time given by
$t^\alpha$ with $\alpha$= $0.7(1), 0.8(1), 0.3(1)$ for
$\mh/\mw=2, 3$ and $4.65$ respectively.  The corresponding helical
susceptibilities at $mt=100$ are $0.11(2), 0.26(1), 0.12(2)\ m^3$. 
Note that the model with $\mh=3\mw$ is more efficient than the
others in generating helicity at late times. This suggests a non
monotonic dependence of the helicity on the Higgs to W mass ratio, a
feature also observed in the generation of Chern-Simons number
\cite{smit0,jmt2}.  In the remaining of this section we will focus on
results for this particular value of the mass ratio. Comments upon the
dependence on $\mh/\mw$ are deferred to section \ref{ratiom}.

The late time evolution of the integrated magnetic, longitudinal and
transverse electric energies, for $\mh=3\mw$ is presented in
Fig. \ref{fig:lenergy}.  A large fraction of the electromagnetic
fields generated after SSB is preserved by the time evolution. From
$mt\sim 60$ onwards, the transverse energy densities increase with
time, again with a power law dependence: $t^{\alpha}$, with $\alpha=
0.350(1)$ and $0.330(1)$ for electric and magnetic energy densities
respectively.  At these late times, transverse electromagnetic fields
are composed of an admixture of radiation and long range seed
fields. In section~\ref{sec:inversec} we will see how to separate
these two components by analyzing the electromagnetic field power
spectra.  Note also that there is a significant fraction of
longitudinal electric fields, even at the later stages of the
evolution. As already mentioned, the slow screening of the
longitudinal component of the electric field is tied to the presence
of large charged lumps around magnetic field lines, see 
Figs.~\ref{fig:charge1} and~\ref{fig:charge2}, which persist even
at late times.

\FIGURE{
\centerline{
\psfig{file=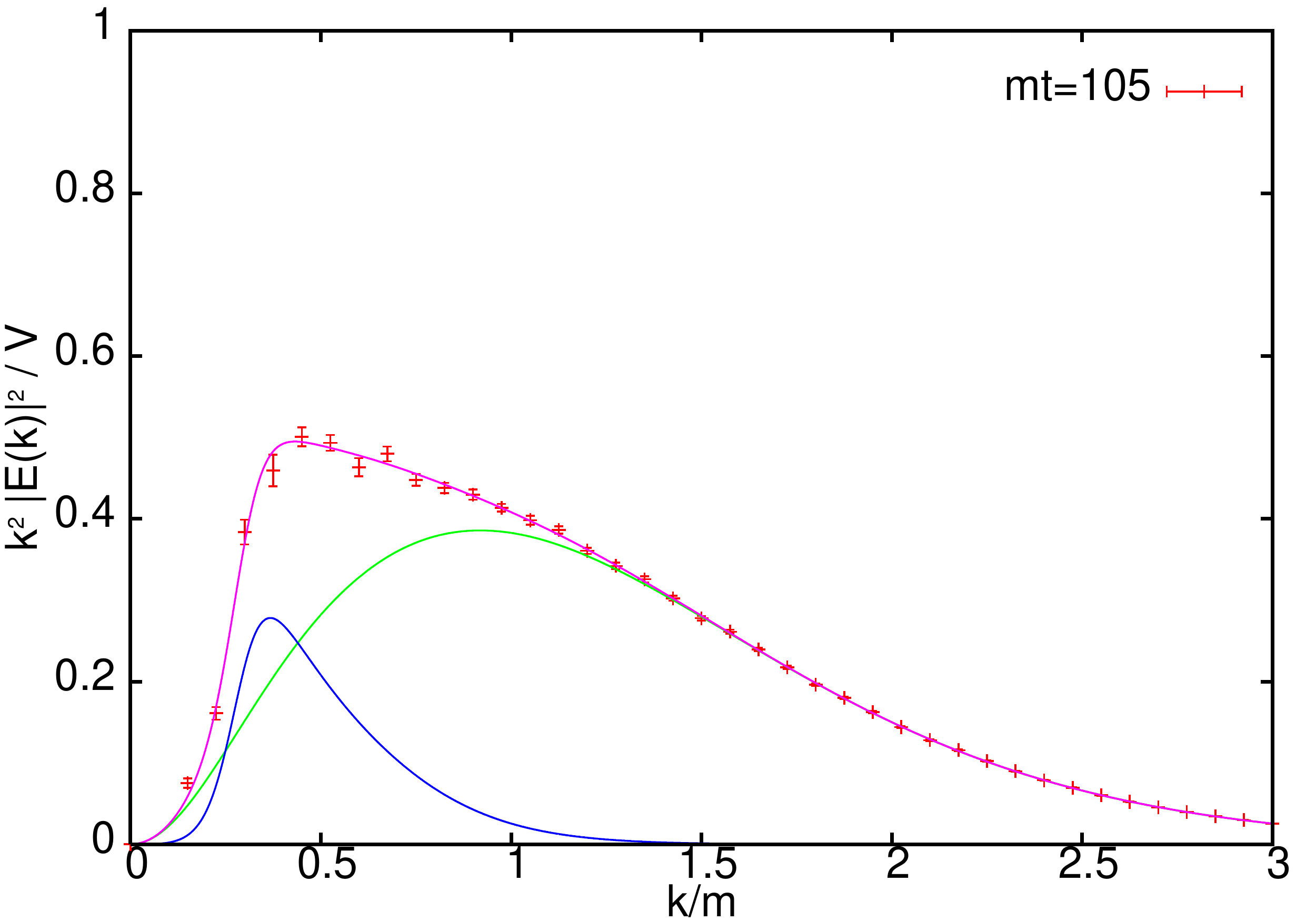,angle=-90,width=8cm}
\psfig{file=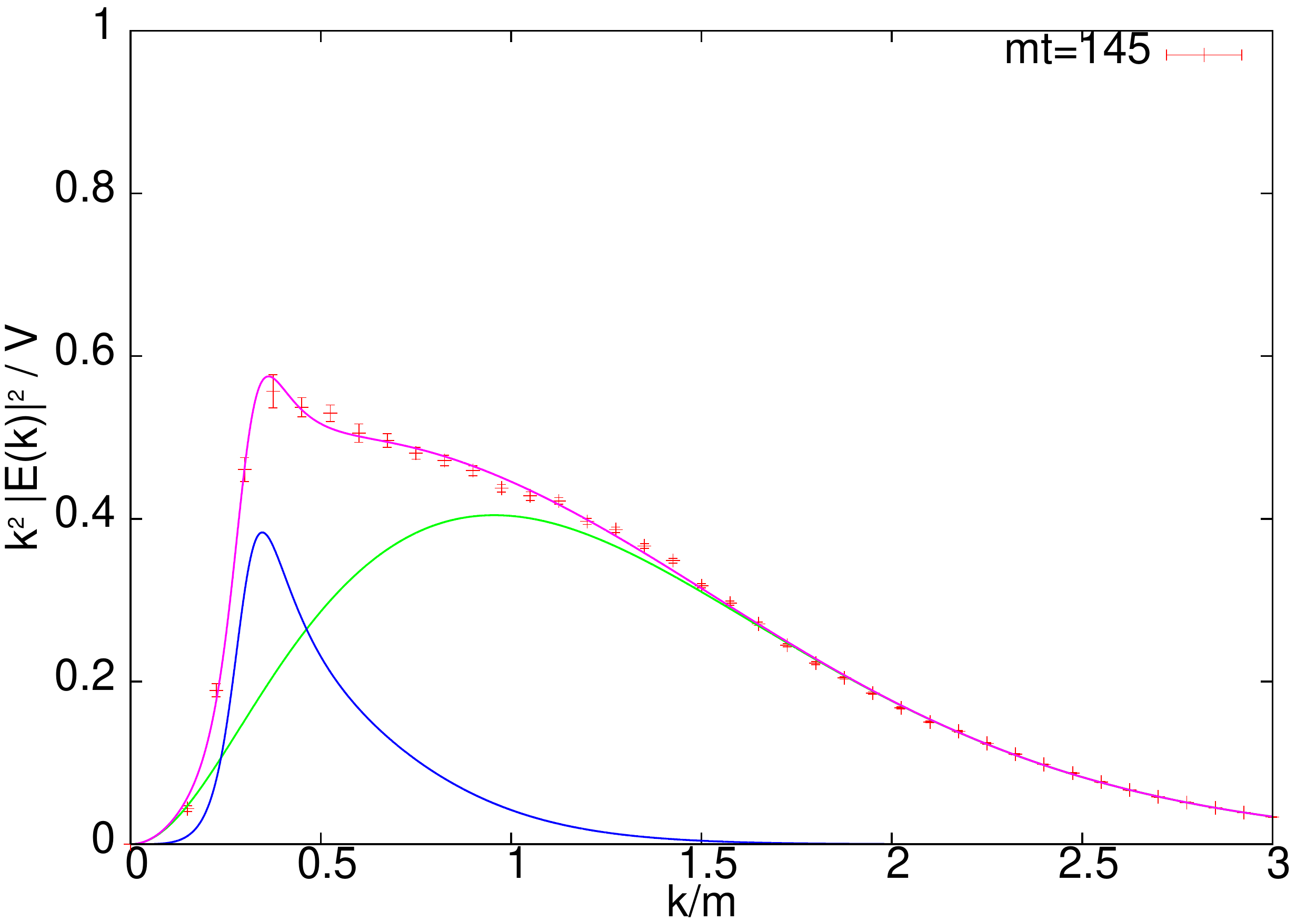,angle=-90,width=8cm}}
\centerline{\psfig{file=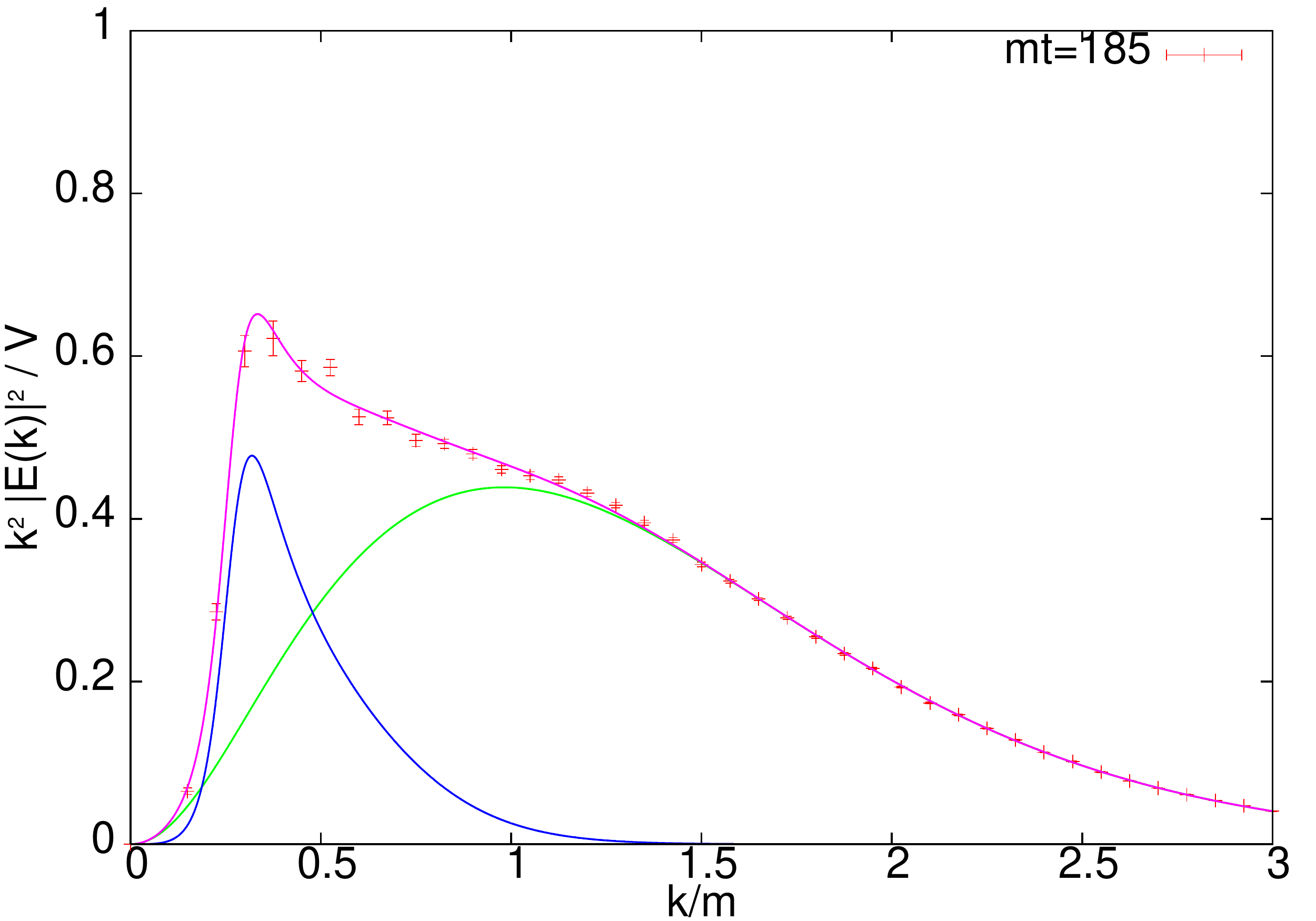,angle=-90,width=8cm}
\psfig{file=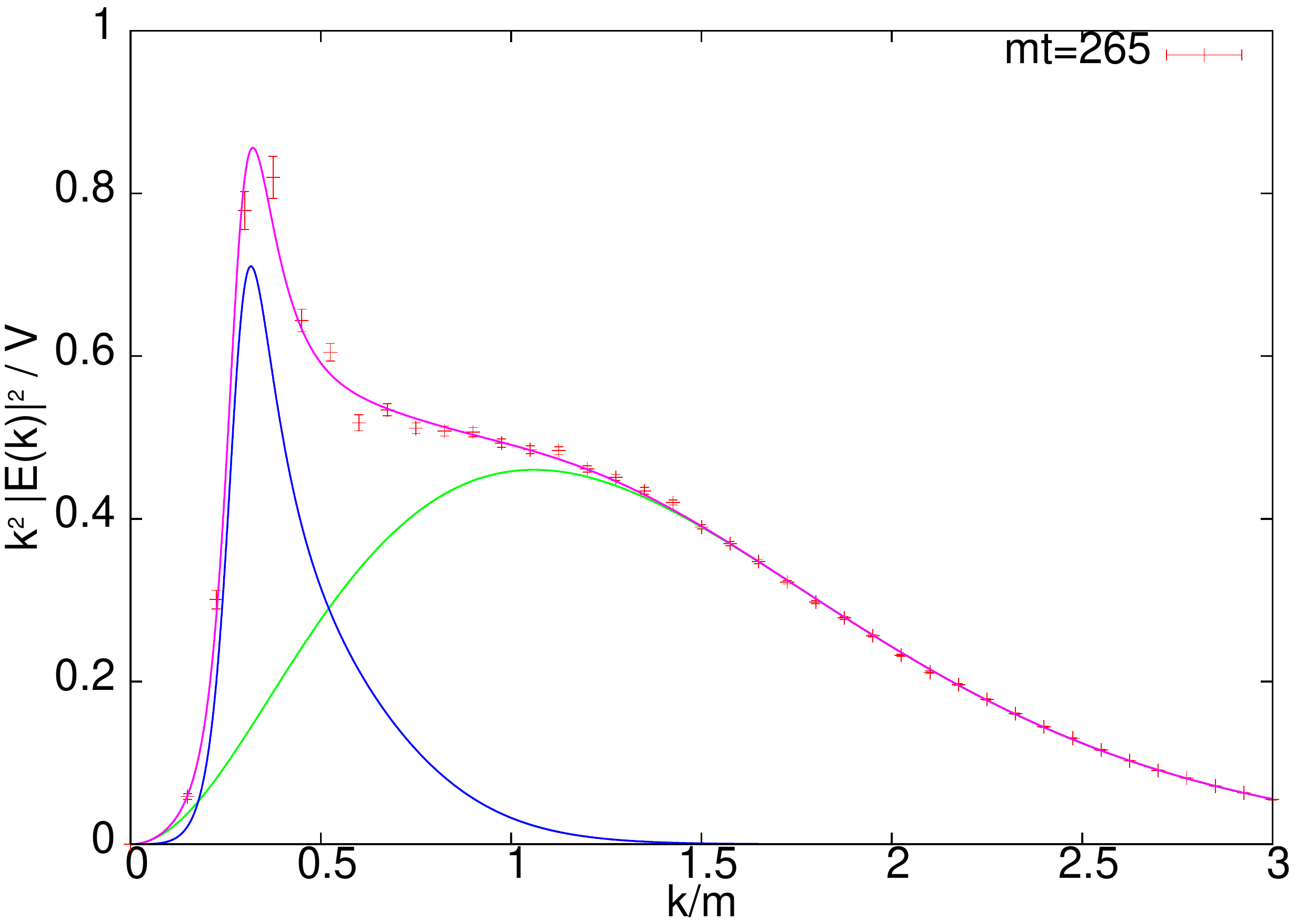,angle=-90,width=8cm}
\caption{ We plot $\langle k^2 |\vec E(k)|^2 \rangle/ \cV $ vs $k$,
averaged over 150 configurations.  The lines represent fits to the
radiation and seed field electromagnetic components according to
Eqs. (\ref{eq:uvphotons}), (\ref{eq:irphotons}) respectively.  Results
are presented at $mt= $ $105$, $145$, $185$ and $265$.  In all cases
$\mh=3\mw$, $ma=0.42$ and $\pmin=0.15\,m$.  }}
\label{fig:spec}}

\FIGURE{
\centerline{
\psfig{file=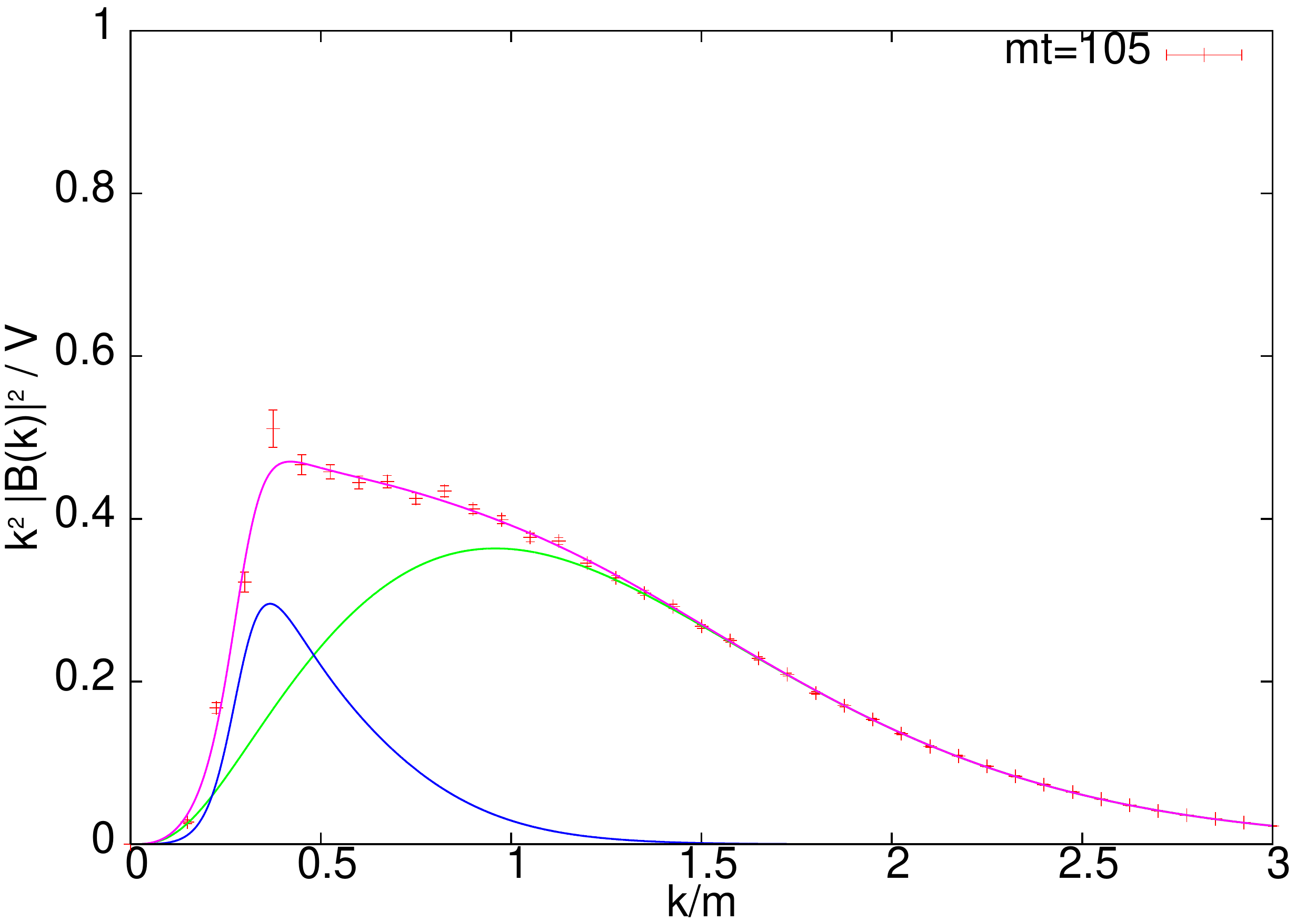,angle=-90,width=8cm}
\psfig{file=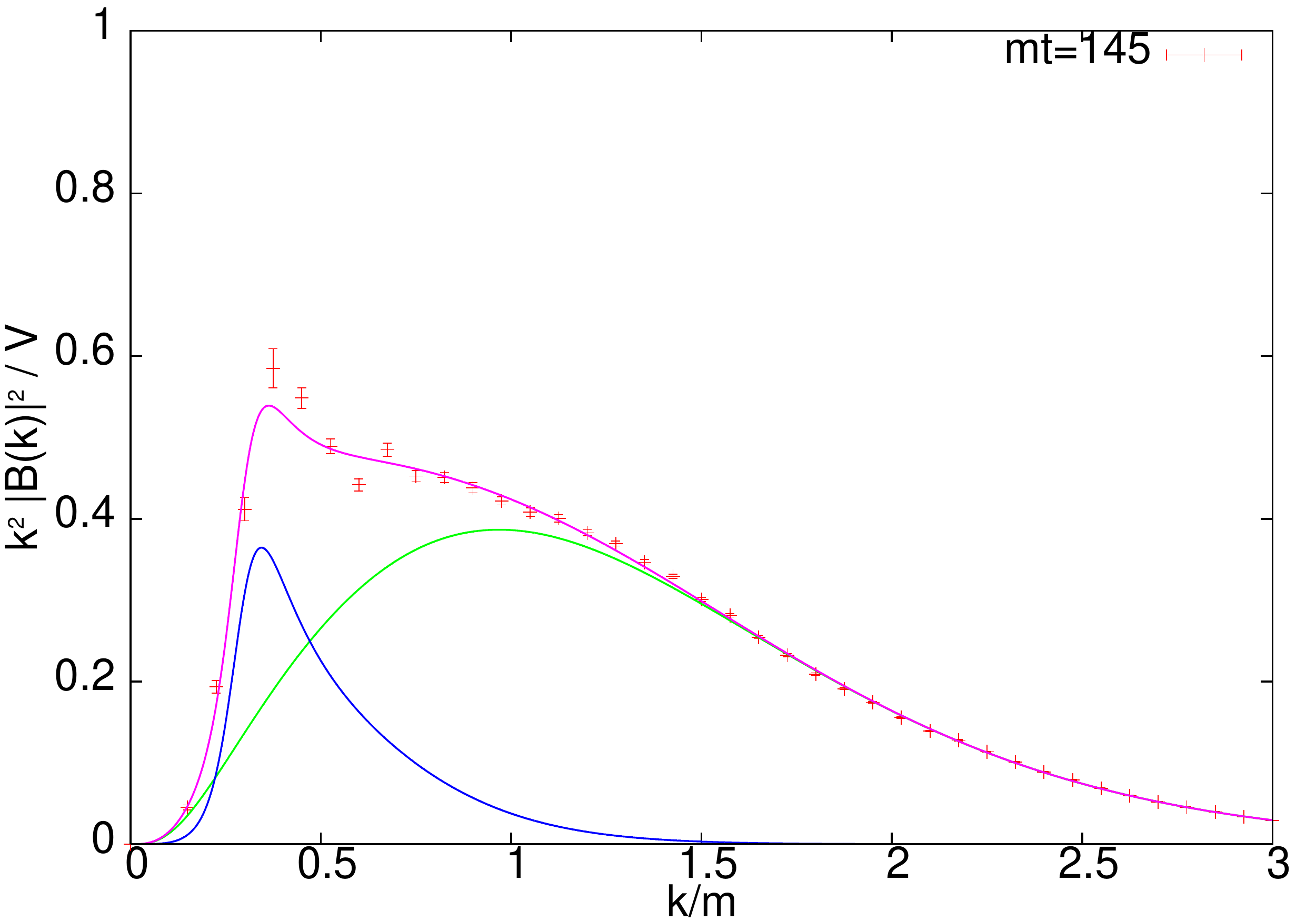,angle=-90,width=8cm}}
\centerline{\psfig{file=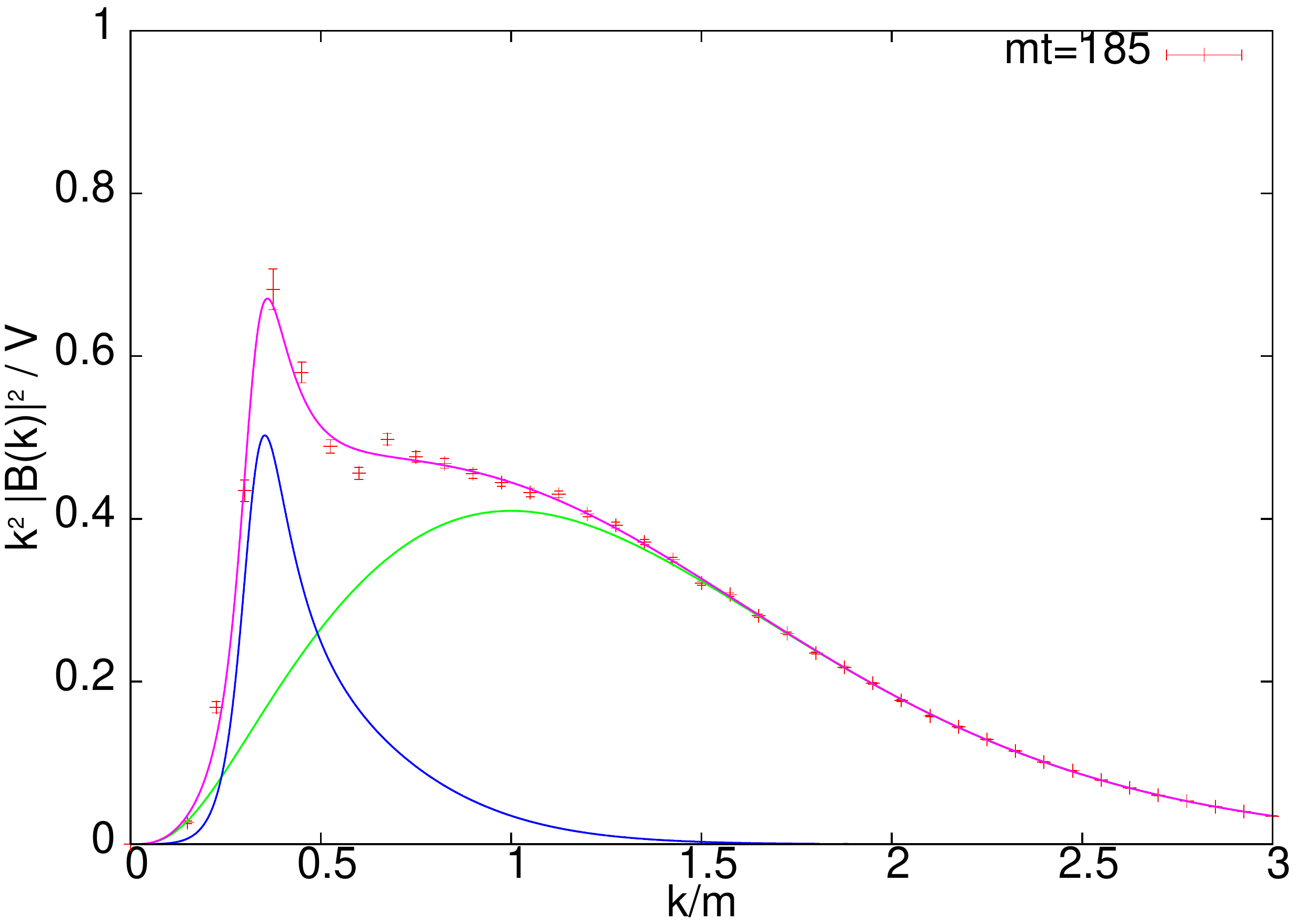,angle=-90,width=8cm}
\psfig{file=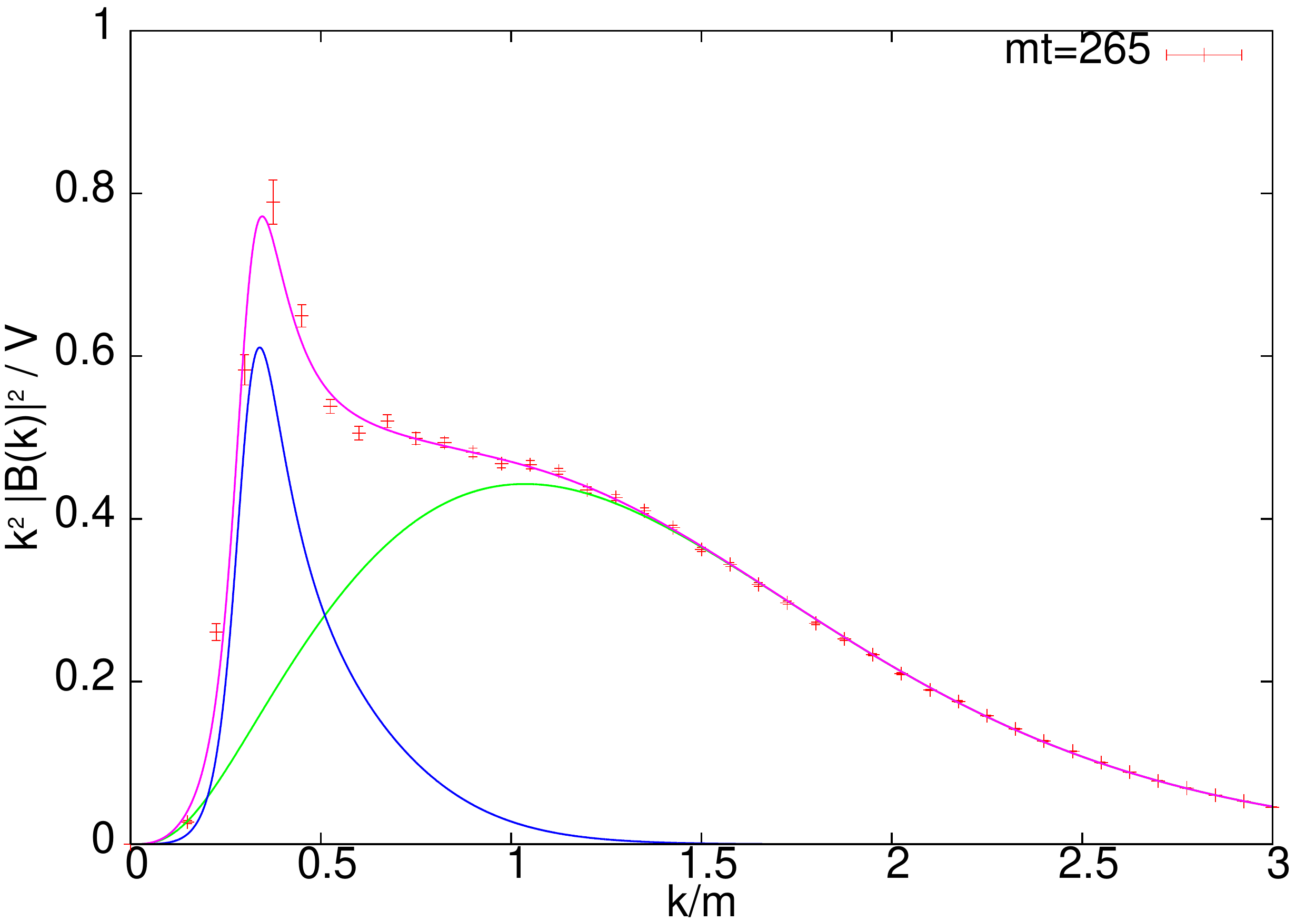,angle=-90,width=8cm}
\caption{ The same as in Fig. \ref{fig:spec} but for the magnetic component: 
$\langle k^2 |\vec B(k)|^2 \rangle / \cV $.
}}
\label{fig:specb}}

\subsection{Electromagnetic field spectrum}
\label{sec:inversec}

To investigate whether inverse cascade is active during the late time evolution
we have analyzed the electromagnetic Fourier spectrum.
Figs. \ref{fig:spec} and \ref{fig:specb} display the time evolution of
$\langle k^2 |\vec E(k)|^2 \rangle/ \cV$ and $\langle k^2 |\vec B(k)|^2\rangle/ \cV $, where $\vec E(k)$ and $\vec B(k)$ are the Fourier components of
the electromagnetic fields and $\cV$ is the physical volume.
The most remarkable feature in the spectrum is the peak at small momenta
that develops with time,  which is distinctly separated from the
high momentum component.  This behaviour suggests that the spectrum contains
two uncorrelated distributions which describe respectively electromagnetic
radiation and the long range electric and magnetic seed fields.
Following this indication,  we have performed fits to the spectrum where
this separation is made explicit:
\be
\vec F (k) = \vec F^{\rm seed}(k) + \vec F^{\rm rad}(k)
\ee
with $\vec F = \vec E $ or $\vec B$.
For the expectation values of the electric and magnetic correlators we obtain
accordingly:
\bea
\langle |\vec E(k)|^2 \rangle &=& \langle |\vec E^{\rm seed}(k)|^2  \rangle + \langle |\vec E^{\rm rad}(k)|^2  \rangle\\
\langle |\vec B(k)|^2 \rangle &=& \langle |\vec B^{\rm seed}(k)|^2 \rangle  + \langle |\vec B^{\rm rad}(k)|^2 \rangle \nonumber
\eea

In the remaining of this section we will describe these two components, 
starting with the electromagnetic radiation and ending with the infrared component
which describes the magnetic field seed.

\subsubsection{Electromagnetic radiation}

The radiation component dominates the electromagnetic energy density,
its contribution being a factor of 5-10 larger than the one coming
from seed fields.  Its profile is very well described by:
\bea
{1 \over \cV }\langle |\vec E^{\rm rad}(k)|^2\rangle &=& {2 w_E  \over e^{\beta (w_E -\mu_E)} -1} \, \, \label{eq:uvphotons}
 \\
{1 \over \cV }\langle |\vec B^{\rm rad}(k)|^2\rangle &=& {2 k  \over  e^{\beta (w_B -\mu_B)} -1} \, , \nonumber
\eea
with $w_{E(B)} = \sqrt{k^2 + m_{E(B)}^2}$ and parameters given in
Table \ref{tab:temp}.  As illustrated in figures \ref{fig:spec} and
\ref{fig:specb}, this distribution fits very well the high momentum
part of the spectrum but fails in reproducing the low momentum peak.
Eq. (\ref{eq:uvphotons}) represents free massive thermal radiation
with non zero chemical potential at temperatures slightly rising with
time, which we interpret as an effect induced by the plasma of the 
$W$-fields.

\TABLE{
\begin{tabular}{||c||c|c|c||c|c|c||}\hline\hline
\hm $mt$ \hm&\hm $T_E/m$ \hm &\hm $m_E/m$ \hm &\hm $\mu_E/m$&\hm $T_B/m$ \hm &\hm $m
_B/m$ \hm &\hm $\mu_B/m$
 \\ \hline \hline
\hm 105\hm &\hm 0.32(1)\hm &\hm 0.77(1)\hm &\hm 0.61(1) \hm
&\hm 0.32(1)\hm &\hm 0.66(1)\hm &\hm 0.60(1)  \hm
\\ \hline
\hm 125\hm &\hm 0.33(1)\hm &\hm 0.74(1)\hm &\hm 0.58(1) \hm
&\hm 0.33(1)\hm &\hm 0.61(2)\hm &\hm 0.57(2) \hm
\\ \hline
\hm 145\hm &\hm 0.34(1)\hm &\hm 0.75(1)\hm &\hm 0.58(1)  \hm
&\hm 0.33(1)\hm &\hm 0.60(2)\hm &\hm 0.56(2) \hm
\\ \hline
\hm 165\hm &\hm 0.34(1)\hm &\hm 0.76(2)\hm &\hm 0.59(1)  \hm
&\hm 0.34(1)\hm &\hm 0.61(2)\hm &\hm 0.57(2) \hm
\\ \hline
\hm 185\hm &\hm 0.34(1)\hm &\hm 0.82(1)\hm &\hm 0.63(1) \hm
&\hm 0.34(1)\hm &\hm 0.65(2)\hm &\hm 0.60(2) \hm
\\ \hline
\hm 205\hm &\hm 0.35(1)\hm &\hm 0.84(1)\hm &\hm 0.64(1) \hm
&\hm 0.34(1)\hm &\hm 0.64(2)\hm &\hm 0.59(2)\hm
\\ \hline
\hm 245\hm &\hm 0.35(1)\hm &\hm 0.93(1)\hm &\hm 0.68(1) \hm
&\hm 0.35(1)\hm &\hm 0.64(1)\hm &\hm 0.59(2) \hm
\\ \hline
\hm 265\hm &\hm 0.36(1)\hm &\hm 0.93(1)\hm &\hm 0.67(1) \hm
&\hm 0.35(1)\hm &\hm 0.65(2)\hm &\hm 0.59(2) \hm
\\ \hline \hline
\end{tabular}
\caption{Parameters of the fit to the high momentum part of
the transverse electric and magnetic spectra in 
Eq.~(\ref{eq:uvphotons}), for $\mh=3\mw$, $ma=0.42$ and 
$\pmin=0.15\,m$. Errors in parenthesis combine both systematic 
and statistical effects.}
\label{tab:temp}
}

Similar information can be extracted from the distribution of local
values of the norm of the transverse electric and magnetic fields. For
free photons this should follow a Maxwellian distribution (see
Appendix~\ref{app3}):
\be
P(B)= \sqrt{\frac{2}{\pi}}\bigg( \frac{3}{\langle B^2\rangle}\bigg)^{3/2} B^2 \ 
e^{- {3 B^2 \over 2 \langle B^2 \rangle}},  \hspace*{1cm}
\label{eq:maxwell}  \,.
\ee
where $B=|\vec B(\vec x)|$. Our data does indeed reproduce this
behaviour at late times.  In Fig.~\ref{fig:maxw} we display the time
evolution of the distribution of magnetic field norms, starting from
$mt=5$.  Although initially the distribution
differs significantly from the Maxwellian one, it is approached as time
evolves and photons thermalise.  There is, however, a systematic
mismatch when we fit the tail of the Maxwellian distribution, even at
large values of $mt$.  This signals again a deviation from free
radiation, like the one observed in the low momentum part of the
magnetic and electric spectra. It is in this deviation where the
contribution of the seed magnetic fields reside.

\FIGURE{
\centerline{
\psfig{file=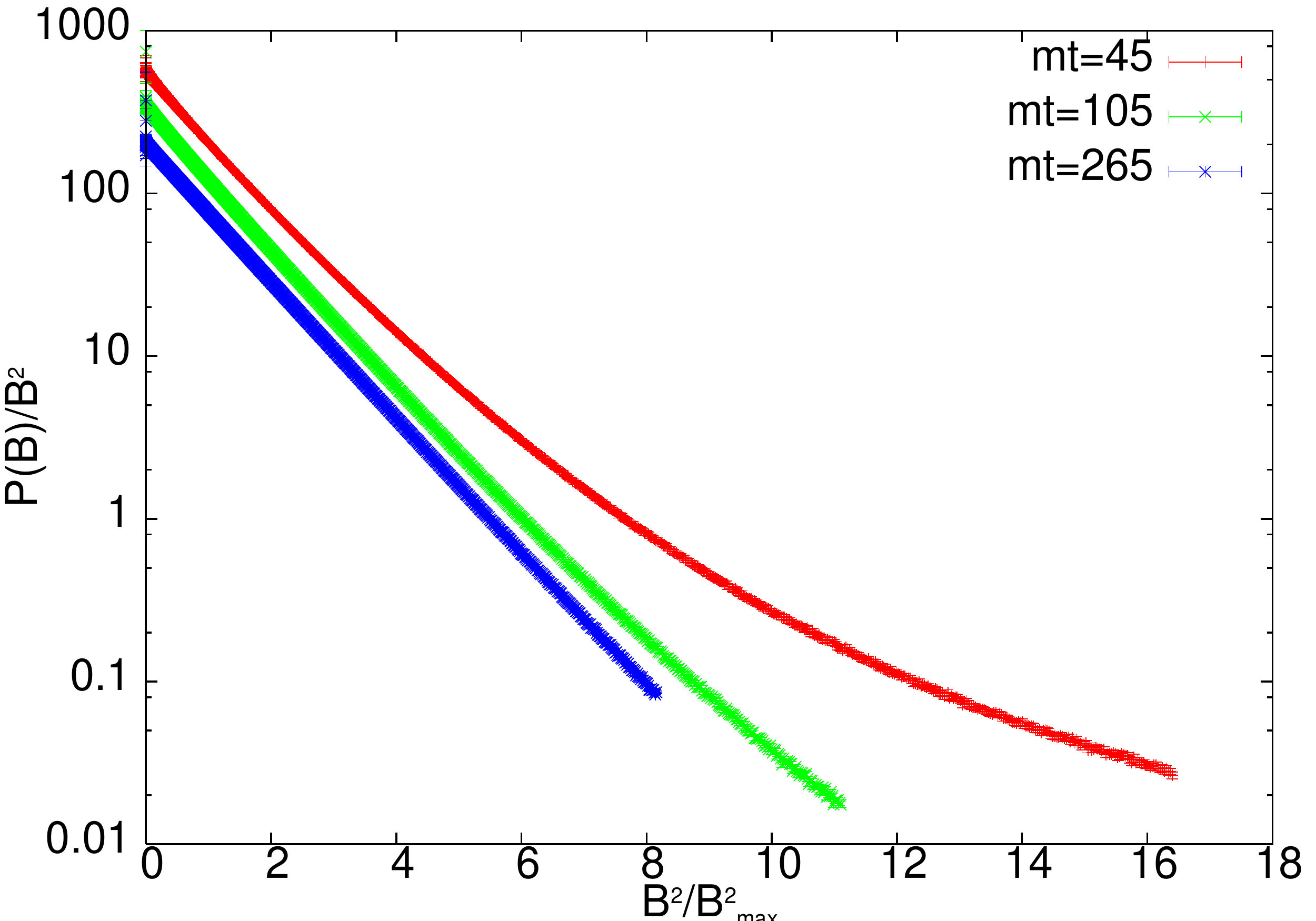,angle=-90,width=8cm}
\psfig{file=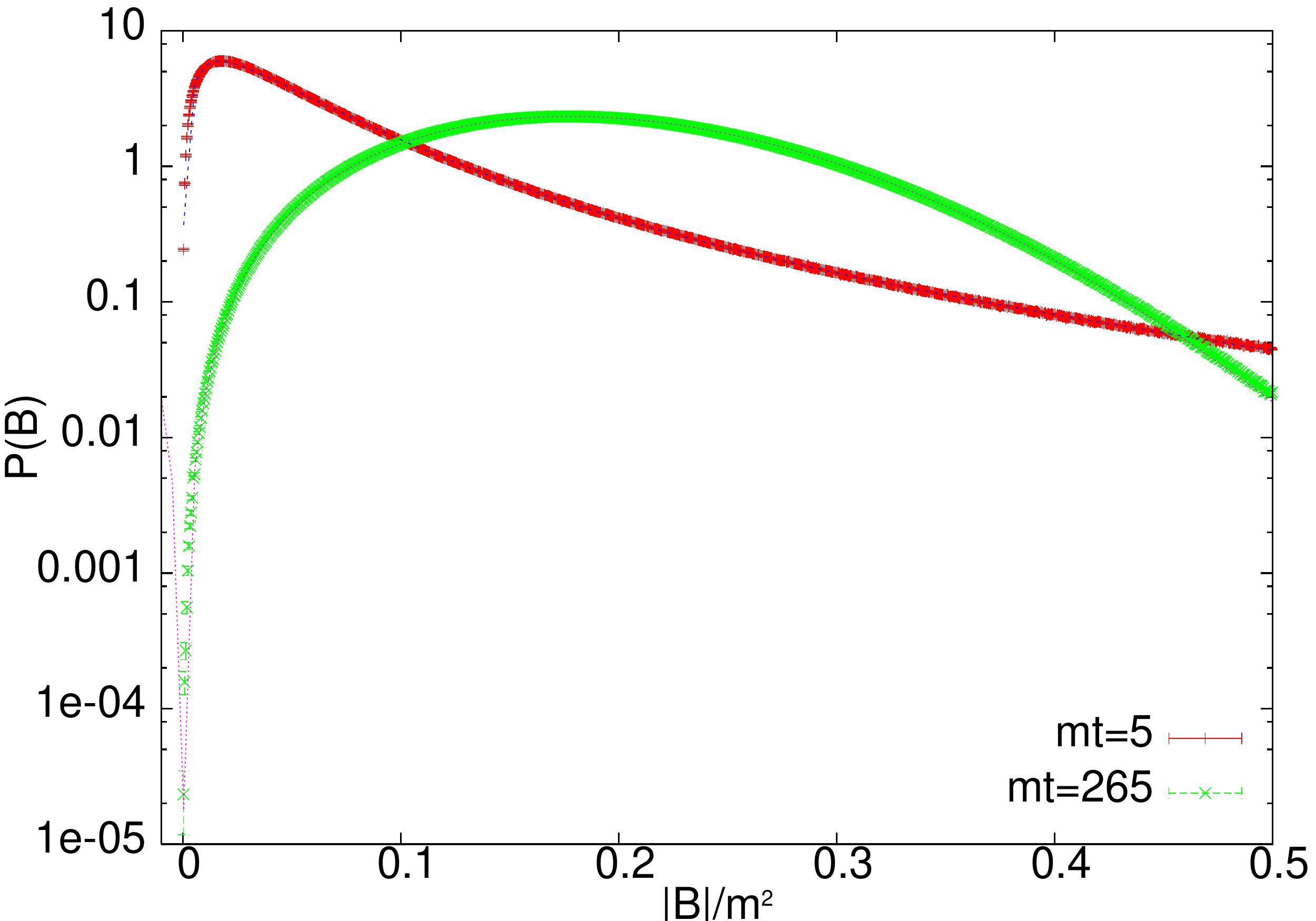,angle=-90,width=8cm}
\caption{ We show the time evolution of the distribution of magnetic
  field norms.  Left: For $\mh = 3 \mw$ we display the log of
  $P(B)/B^2$ vs $B^2/B_{\rm max}^2$ (i.e. normalized to the value at
  the peak of the distribution) .  Right: For $\mh = 3 \mw$ we compare
  the initial distribution of the local magnitude of the magnetic
  field at $m
t=5$ with the one obtained at $mt=265$, the latter fitted
  to a Maxwellian distribution.  The fit to the $mt=5$ data is
  described in Appendix~\ref{app4}.
}}
\label{fig:maxw}}

\subsubsection{Electric and magnetic seeds}

We turn now to the analysis of the infrared part of the spectrum, which is the 
relevant one for the generation of the LSMF seed field.
This low momentum part has been fitted to: 
\bea
{1 \over \cV}\langle |\vec E^{\rm seed}(k)|^2\rangle = {2 k  \over e^{\hat \beta_{E} (\hat w_{E} -\hat \mu_{E})} -1} \,  ,
\label{eq:irphotons} \\
{1 \over \cV}\langle |\vec B^{\rm seed}(k)|^2\rangle = {2 k  \over e^{\hat \beta_{B} (\hat w_{B} -\hat \mu_{B})} -1} \,  ,
\nonumber
\eea
with $\hat w_{E(B)} = \sqrt{(k-k^0_{E(B)})^2 + \hat m_{E(B)}^2}$ and
parameters given in Tables \ref{tab:temp2}, \ref{tab:temp2b}. This could
represent again massive radiation at non-zero chemical potential if it were 
not for the peculiar shift $k_0$ in the frequency $\hat w$. We interpret the
value of $k_0 \sim 0.3 m$ as a characteristic momentum scale of the long range
electromagnetic fields.

\TABLE{
\begin{tabular}{||c|c|c|c|c||}\hline\hline
\hm $mt$ \hm &$\hat T_E/m$ \hm &\hm $\hat m_E/m$ &\hm $\hat \mu_E/m$&\hm $\hat k^0_E/m$ \hm
 \\ \hline \hline
\hm 105  \hm &\hm 0.11(1)  \hm &\hm 0.33(5)  \hm & \hm 0.30 (4)  \hm & \hm 0.29(1)\hm 
\\ \hline
\hm 125  \hm &\hm 0.13(1)  \hm &\hm 0.24(4)  \hm &\hm 0.22(3)  \hm & \hm 0.29(1) \hm
\\ \hline
\hm 145  \hm &\hm 0.14(1)  \hm &\hm 0.21(5)  \hm &\hm 0.18(3)  \hm & \hm 0.30(1) \hm
\\ \hline
\hm 165  \hm &\hm 0.13(1)  \hm &\hm 0.25(5)  \hm &\hm 0.23(3)  \hm & \hm 0.29(1) \hm
\\ \hline
\hm 185  \hm &\hm 0.09(2)  \hm &\hm 0.49(8)  \hm &\hm 0.48(6)  \hm & \hm 0.27(1) \hm
\\ \hline
\hm 205  \hm &\hm 0.11(1)  \hm &\hm 0.36(6)  \hm &\hm 0.35(3)  \hm & \hm 0.29(1) \hm
\\ \hline
\hm 225  \hm &\hm 0.10(2)  \hm &\hm 0.39(10)  \hm &\hm0.38(3)   \hm & \hm 0.28(1) \hm
\\ \hline
\hm 245  \hm &\hm 0.11(1)  \hm &\hm 0.37(7)  \hm &\hm 0.35(3)  \hm & \hm 0.30(1) \hm
\\ \hline
\hm 265  \hm &\hm 0.10(1)  \hm &\hm 0.45(7)  \hm &\hm 0.44(4)  \hm & \hm 0.28(1) \hm
\\ \hline \hline
\end{tabular}
\caption{
Parameters of the fit to the low momentum part of
the transverse electric spectrum in Eq. (\ref{eq:irphotons}),
for $\mh=3\mw$, $ma=0.42$ and $\pmin=0.15\,m$.
}
\label{tab:temp2}
}

\TABLE{
\begin{tabular}{||c|c|c|c|c||}\hline\hline
\hm $mt$ \hm&$\hat T_B/m$ \hm &\hm $\hat m_B/m$&\hm $\hat \mu_B/m$&\hm $\hat k^0_B/m$ \hm
 \\ \hline \hline
\hm 105  \hm &\hm 0.11(1)  \hm &\hm 0.32(7)  \hm & \hm 0.30(3)  \hm & \hm 0.29(1)  \hm      
\\ \hline
\hm 125  \hm &\hm 0.13(1)  \hm &\hm 0.24(7)  \hm &\hm 0.21(4)  \hm & \hm 0.31(1) \hm
\\ \hline
\hm 145  \hm &\hm 0.13(1)  \hm &\hm 0.24(6)  \hm &\hm 0.22(3)  \hm & \hm 0.29(1) \hm
\\ \hline
\hm 165  \hm &\hm 0.13(1)  \hm &\hm 0.27(6)  \hm &\hm 0.23(4)  \hm & \hm 0.29(1) \hm
\\ \hline
\hm 185  \hm &\hm 0.13(2)  \hm &\hm 0.18(10)  \hm &\hm 0.16(8)  \hm & \hm 0.32(3) \hm
\\ \hline
\hm 205  \hm &\hm 0.11(1)  \hm &\hm 0.31(7)  \hm &\hm 0.29(4)  \hm & \hm 0.30(1) \hm
\\ \hline
\hm 225  \hm &\hm 0.11(1)  \hm &\hm 0.26(5)  \hm &\hm 0.25(4)  \hm & \hm 0.31(1) \hm
\\ \hline
\hm 245  \hm &\hm 0.10(1)  \hm &\hm 0.37(9)  \hm &\hm 0.36(2)   \hm & \hm 0.29(1) \hm
\\ \hline
\hm 265  \hm &\hm 0.11(2)  \hm &\hm 0.33(9)  \hm &\hm 0.32(3)  \hm & \hm 0.30(1) \hm
\\ \hline \hline
\end{tabular}

\caption{
Parameters of the fit to the low momentum part of
the magnetic  spectrum in Eq. (\ref{eq:irphotons}).
For $\mh=3\mw$, $ma=0.42$ and $\pmin=0.15\,m$.
}
\label{tab:temp2b}
}

A quantitative estimate of the energy density and correlation length
of the seed electromagnetic fields can be obtained from our fits to
the low momentum part of the spectrum. The mean energy density is 
computed from the integral of the seed field spectrum as
\be 
\langle \rho_{\rm seed}^F \rangle  = {1 \over 2 \cV} \sum_{\vec k} {|\vec F^{\rm seed}(k)|^2 \over  \cV}\, ,
\label{eq:seedampl}
\ee
with $F=E (B)$. The correlation length, $\xi_{E(B)}$, is extracted from
\be
\xi= {2 \pi \over \bar k}\, \, , \, \, {\rm with } \,\,  \bar k^2= 
{\sum_{\vec k} k^2 \ |\vec F^{\rm seed}(k)|^2 \over \sum_{\vec k} |\vec F^{\rm seed}(k)|^2}\, .
\label{eq:seedcl}
\ee
Table~\ref{tab:seeds} and Fig.~\ref{fig:seeds} summarise our
results. We have tested finite volume independence by comparing two
different physical volumes: $\pmin=0.125\,m$ and $\pmin=0.15\,m$. The
numbers in Table~\ref{tab:seeds} come from an average of the results
obtained at these two physical volumes, with errors given by the
dispersion between them.

We obtain a magnetic seed whose mean energy density increases linearly
with time. Within the time ranges we have analysed, its fraction to
the total comes out to be of order $\sim 10^{-2}$.  Assuming the
magnetic field expands as radiation, this would give magnetic fields
today of order $0.5\, \mu G$, which are in the range of the observed
ones in galaxies, and even in clusters of galaxies, where no-extra
amplification through a dynamo mechanisms is expected.

\vspace*{2mm}

\TABLE{
\begin{tabular}{||c||c|c||c|c||}\hline\hline
\hm $mt$ \hm& $ \langle \rho_{\rm seed}^E \rangle (
\times 10^2)$ &$ m \xi_E $&$ \langle \rho_{\rm seed}^B\rangle (\times 10^2)$&$m \xi_B$
\\ \hline \hline
105 &  $0.62(5)$ & $25.3(1)$ & $0.58(3)$ & $25.7(6)$
\\ \hline \hline
125 &  $0.73(2)$ & $25.2(1)$ & $0.61(1)$ & $24.5(9)$
\\ \hline \hline
145 &  $0.76(4)$ & $24.8(9)$ & $0.72(2)$ & $24.8(3)$
\\ \hline \hline
165 &  $0.76(4)$ & $26.0(10)$ & $0.77(1)$ & $25.4(6)$
\\ \hline \hline
185 &  $0.83(1)$ & $27.6(1)$ & $0.79(2)$ & $26.0(10)$
\\ \hline \hline
205 &  $0.89(2)$ & $27.7(2)$ & $0.79(6)$ & $27.2(5)$
\\ \hline \hline
225 & $0.91(5)$ & $27.9(5)$ & $0.87(1)$ & $28.0(5)$
\\ \hline \hline
245 & $1.06(9)$ & $27.6(4)$ & $0.88(1)$ & $28.1(2)$
\\ \hline \hline
265 & $1.12(7)$ & $27.9(2)$ & $0.92(2)$ & $28.4(7)$
\\ \hline \hline
\end{tabular}
\caption{Fraction of total energy and correlation length of the seed
  electromagnetic fields.  They are both derived from the infrared
  spectrum as described in Eqs. (\ref{eq:seedampl}) and
  (\ref{eq:seedcl}).  The results are obtained by averaging (over 150
  configurations) the values obtained for $\pmin=0.15\,m$ and
  $\pmin=0.125\,m$, with errors reflecting the dispersion
  between them. Data correspond to $\mh=3\mw,\ ma=0.42$.  }
\label{tab:seeds}
}

\FIGURE{
\centerline{
\psfig{file=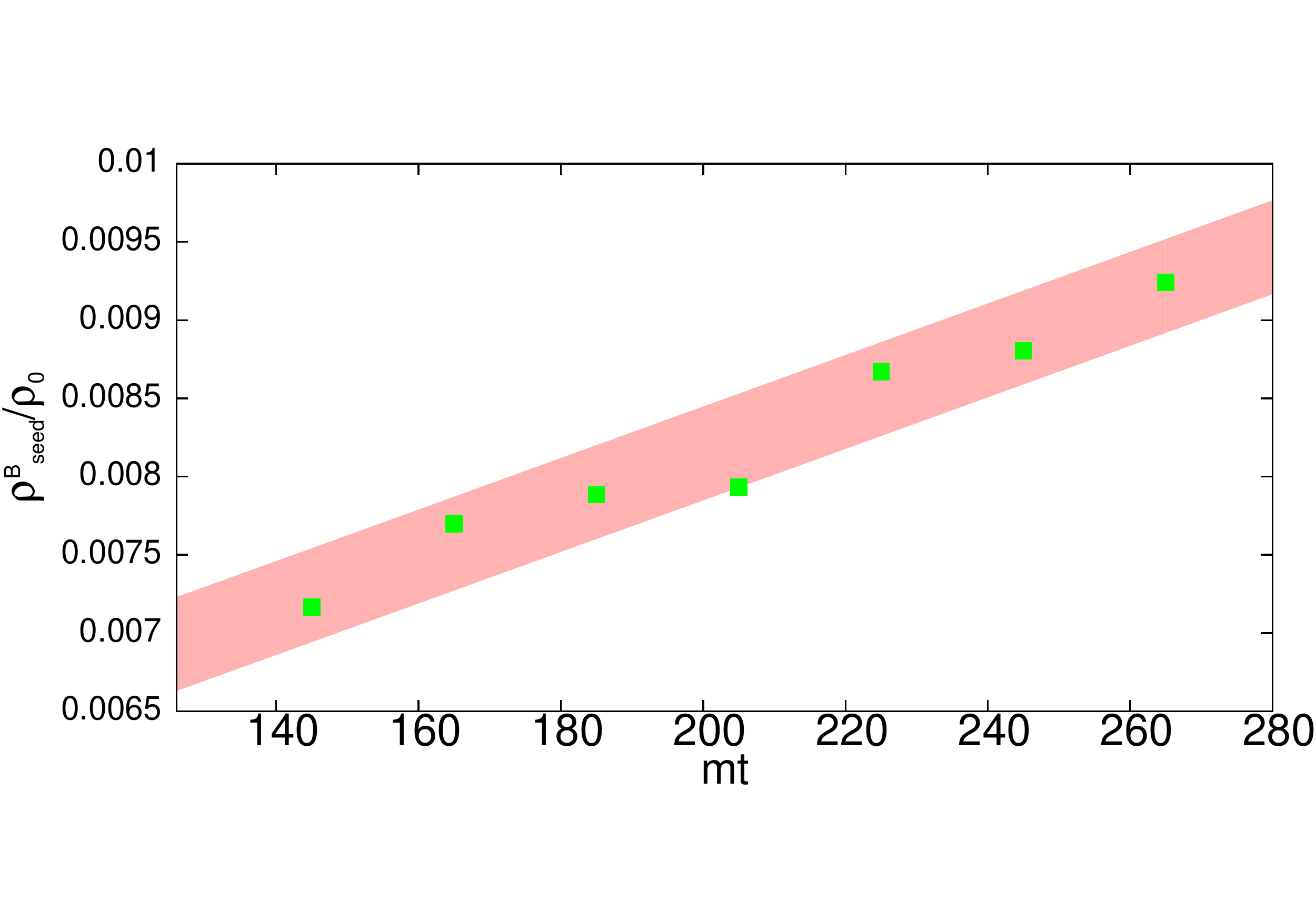,angle=-90,width=8cm}
\psfig{file=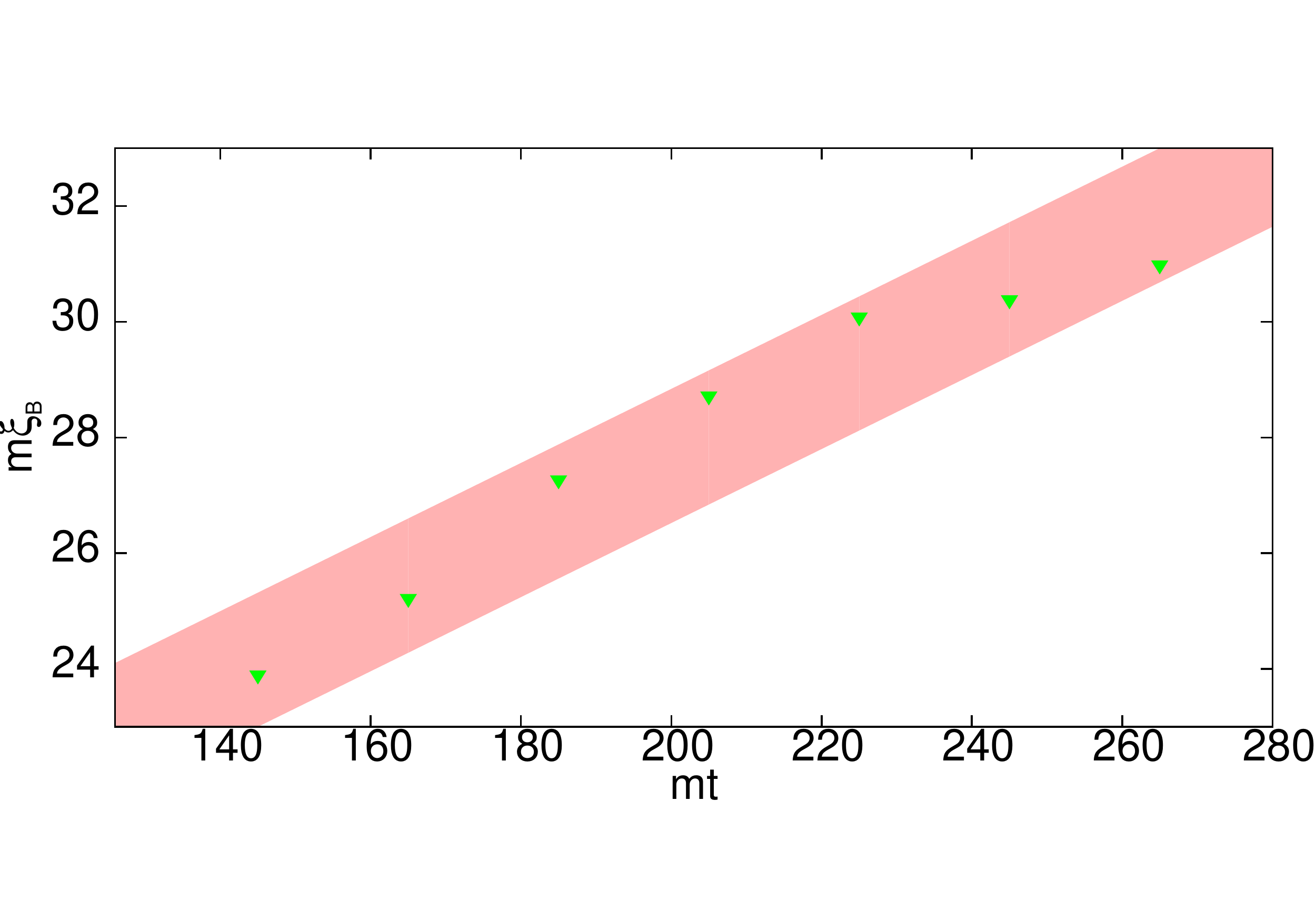,angle=-90,width=8cm}
\caption{ We show the time evolution of $\rho_{\rm seed}^B$ (left) and
  $m\xi_B$ (right), for $\mh=3\mw,\;ma=0.42$. The results are obtained
  by averaging the values obtained for $\pmin=0.15\,m$ and
  $\pmin=0.125\,m$, with bands representing the dispersion in the
  errors.  The fits are $ \rho_{\rm seed}^B /\rho_0= 0.0035(5)+
  2.3(3)\times 10^{-5} mt $ and $m \xi_B= 20.1(4) + 0.033(2) mt $
  respectively.  }}
\label{fig:seeds}}

\vspace*{-0.5cm}

Concerning the correlation length, it is difficult to make a
definitive statement about the presence of inverse cascade, given
the small time scales we can explore with our numerical simulation.
Nevertheless, within the time span we have
analyzed, our results clearly show a linear increase of the magnetic
correlation length with time (see Fig.~\ref{fig:seeds}). This result
is robust under changes of $\pmin$ and lattice spacing. The
observed growth is described by $m\xi_B(t)= 20.1(4)+ 0.033(2) mt$,
giving at $mt=265$ a characteristic length scale for seed magnetic
fields of order $m\xi_B (mt=265) \sim 30(1) $.  This is much larger
than the thermal correlation length, $m\xi_{\rm thermal} \sim 10 $,
and represents a significant fraction of the physical volume. It also
implies a considerable increase from the initial value at $mt=5$,
obtained from the initial spectrum to be $m\xi_B(mt=5)\sim 17$.  From
these results we can safely conclude that the time evolution has
succeeded in amplifying the correlation length of the magnetic seed
generated at SSB. Nevertheless, a more detailed study, including
plasma effects, would be required to determine whether $\xi$ will be
further amplified at late times.

\FIGURE{
\centerline{
\psfig{file=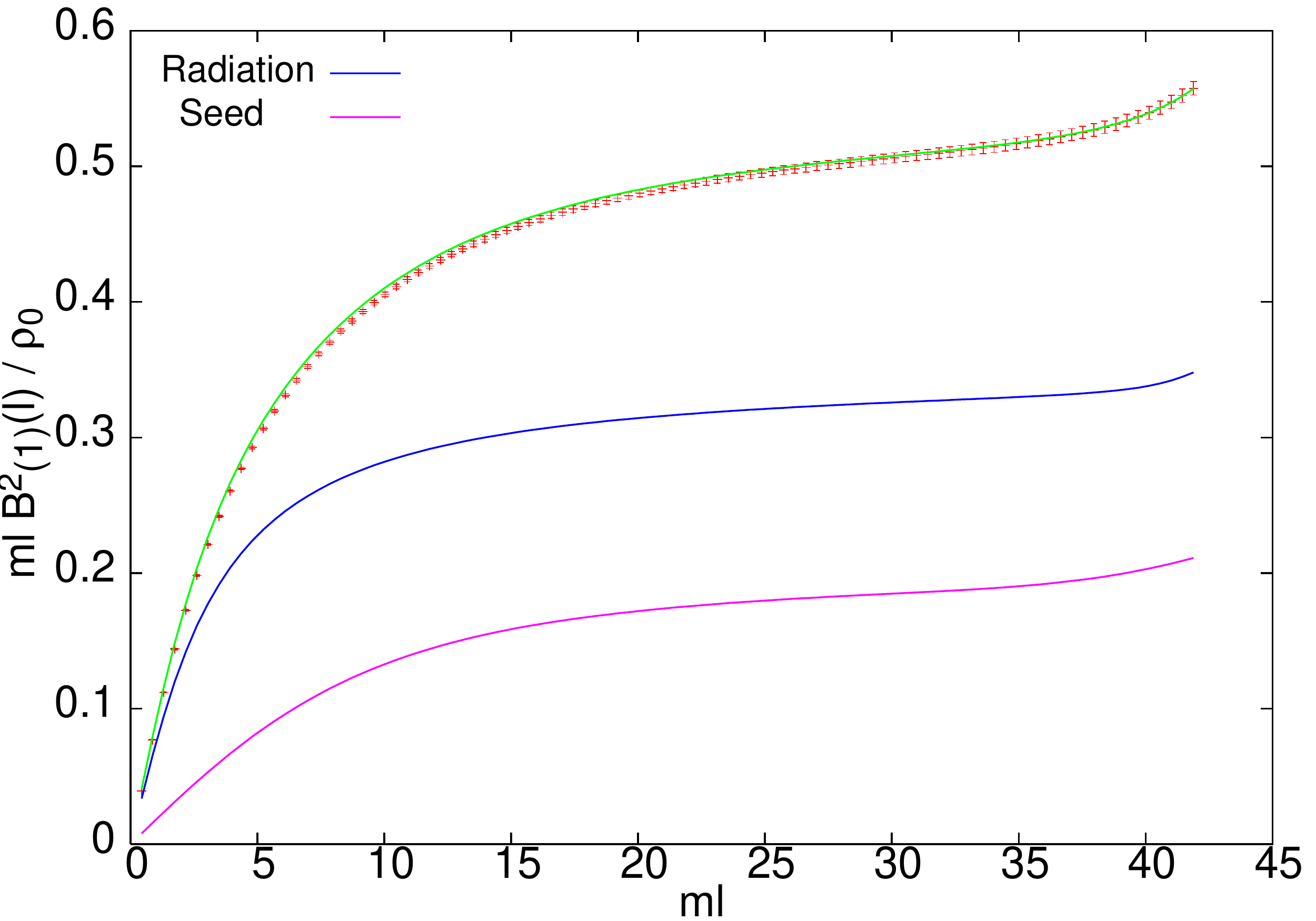,angle=-90,width=6cm}
\psfig{file=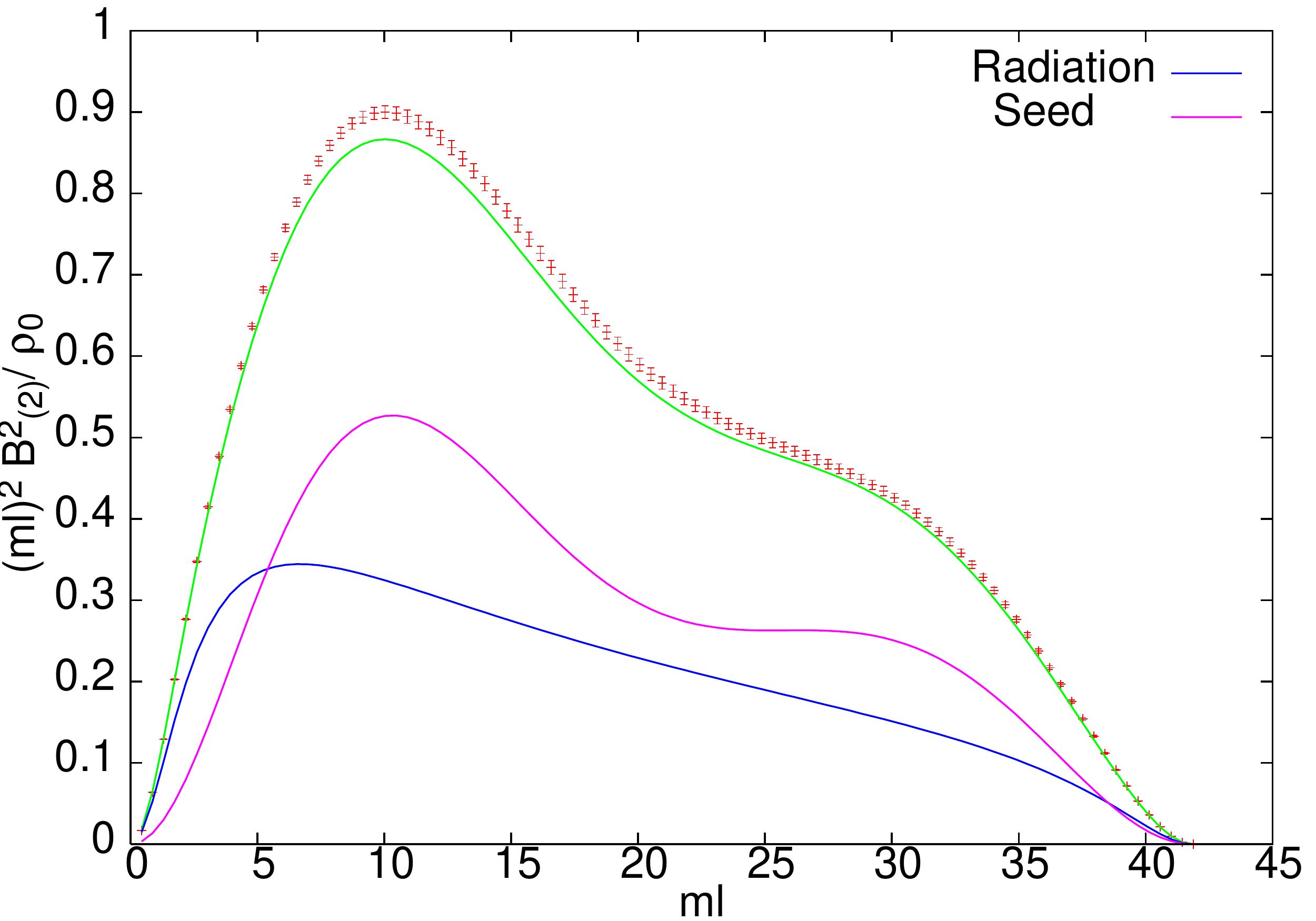,angle=-90,width=6cm}
\psfig{file=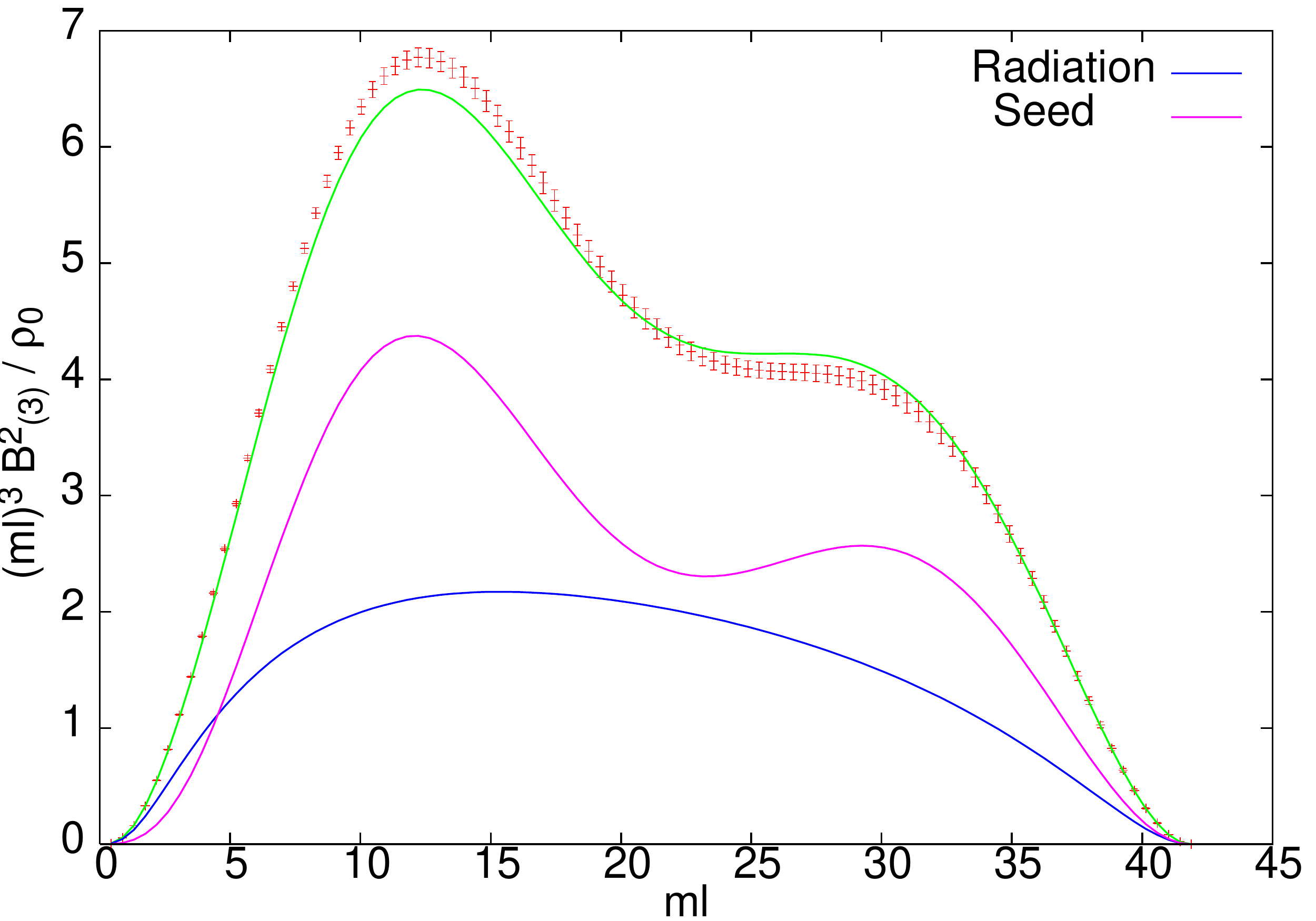,angle=-90,width=6cm}
\caption{ We show the dependence with $ml$ of the three spatial
averages (\ref{eq:lineav})$-$(\ref{eq:volav}), for $mt=245$.
The lines are extracted from our fits to the
infrared and radiation parts of the spectrum. Note that the
fall-off at large distances is just a volume effect. }}
\label{fig:lineav}}

In addition to the direct analysis of the  spectrum we have also followed an 
alternative strategy to separate both the magnitude and the scale of the magnetic
remnant from the radiation bath. A common way to do this, which has been 
extensively used in the literature, is through the computation of several
spatial averages of the electromagnetic fields. Following Ref. 
\cite{HindmarshEverett1998}, we introduce the following averages:
\begin{itemize}
\item
A line average:
\be
B_{(1)} (l) =  {1 \over l } \int_C \vec B \cdot d\vec x \,,
\label{eq:lineav}
\ee
with $C$ a straight line of length $l$.    
\item
The average magnetic
flux over a surface of area $l^2$:
\be
B_{(2)} (l)=  {1 \over l^2 } \int_S \vec B \cdot d\vec S \,,
\label{eq:surfav}
\ee
\item
A volume average:
\be
\vec B_{(3)}(l) =  {1 \over l^3 } \int_S \vec B  d^3x \,.
\label{eq:volav}
\ee
\end{itemize}
As discussed in Ref.~\cite{HindmarshEverett1998}, 
the spatial and statistical averages
$  \langle B_{(i)}^2(l) \rangle$ can be easily computed
in  terms of the spectra of the fields.
For instance, the line average for a volume $\cV$ is given by:
\be
\langle B_{(1)}^2 (l)\rangle = {1 \over \cV} \sum_{\vec k} 
{|B_k|^2 \over \cV} W^2(k_1,l)
\ee
with
\be
W(k_i,l) =  {2 \sin(k_i l/2) \over k_i l}.
\ee
Analogous expressions can be found for the other two quantities. The
advantage of these averages is that they filter out the high momentum
part of the spectrum and allow to recover, at large $l$, information
about the low momentum modes. We have checked that our fits to the
spectrum correctly reproduce the spatial averages. This is illustrated
in Fig. \ref{fig:lineav}, where we present results for the three
averages at $mt=245$ compared with the predictions obtained from our
fits to the spectrum. The quality of the agreement can be considered
very good given that the continuum lines are directly obtained from
the fits to the spectrum (Eqs. (\ref{eq:uvphotons}),
(\ref{eq:irphotons}) and Tables \ref{tab:temp} - \ref{tab:temp2b}),
and not as a result of a fit to the spatial averages.

\FIGURE{
\centerline{
\psfig{file=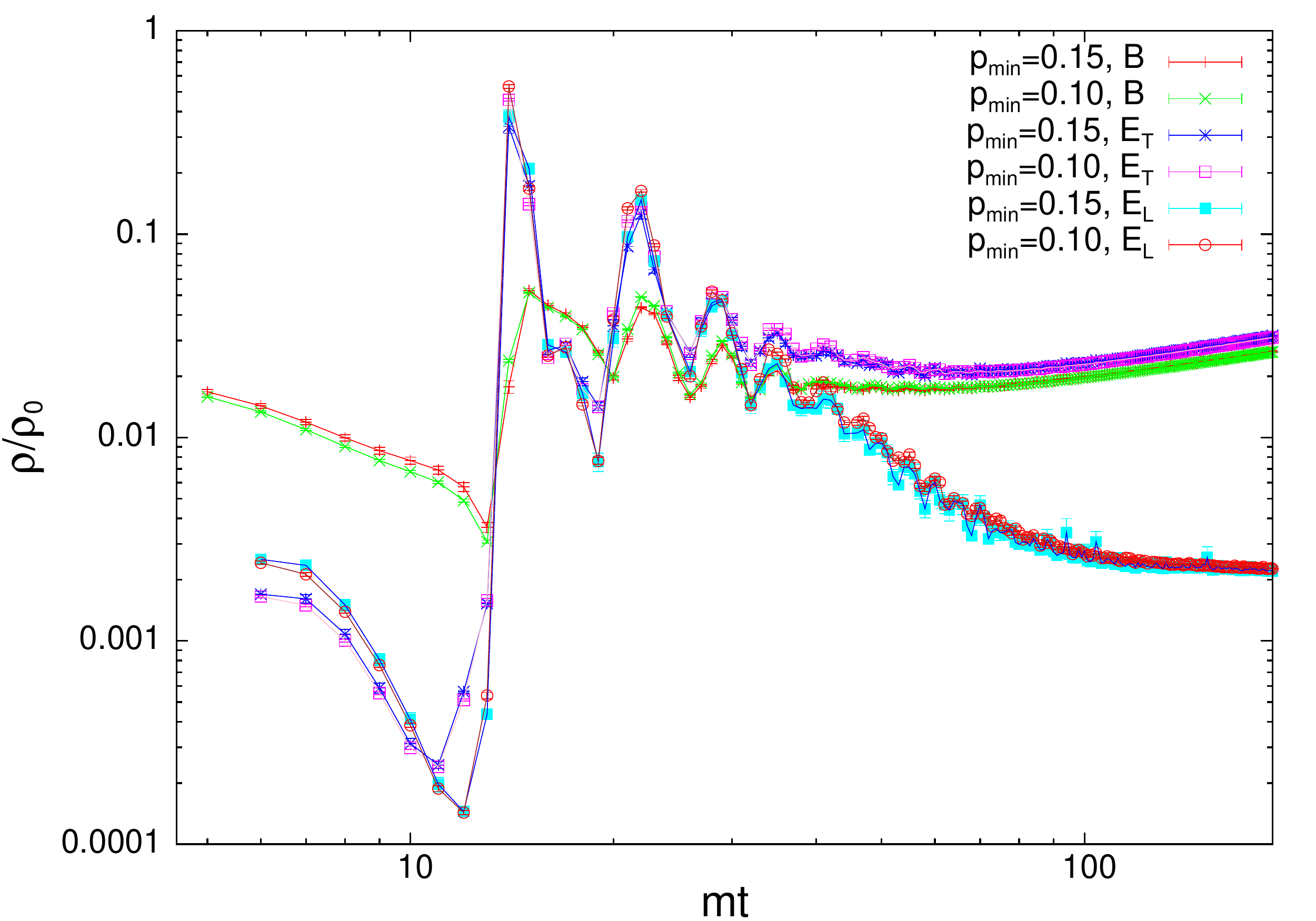,angle=-90,width=8cm}
\psfig{file=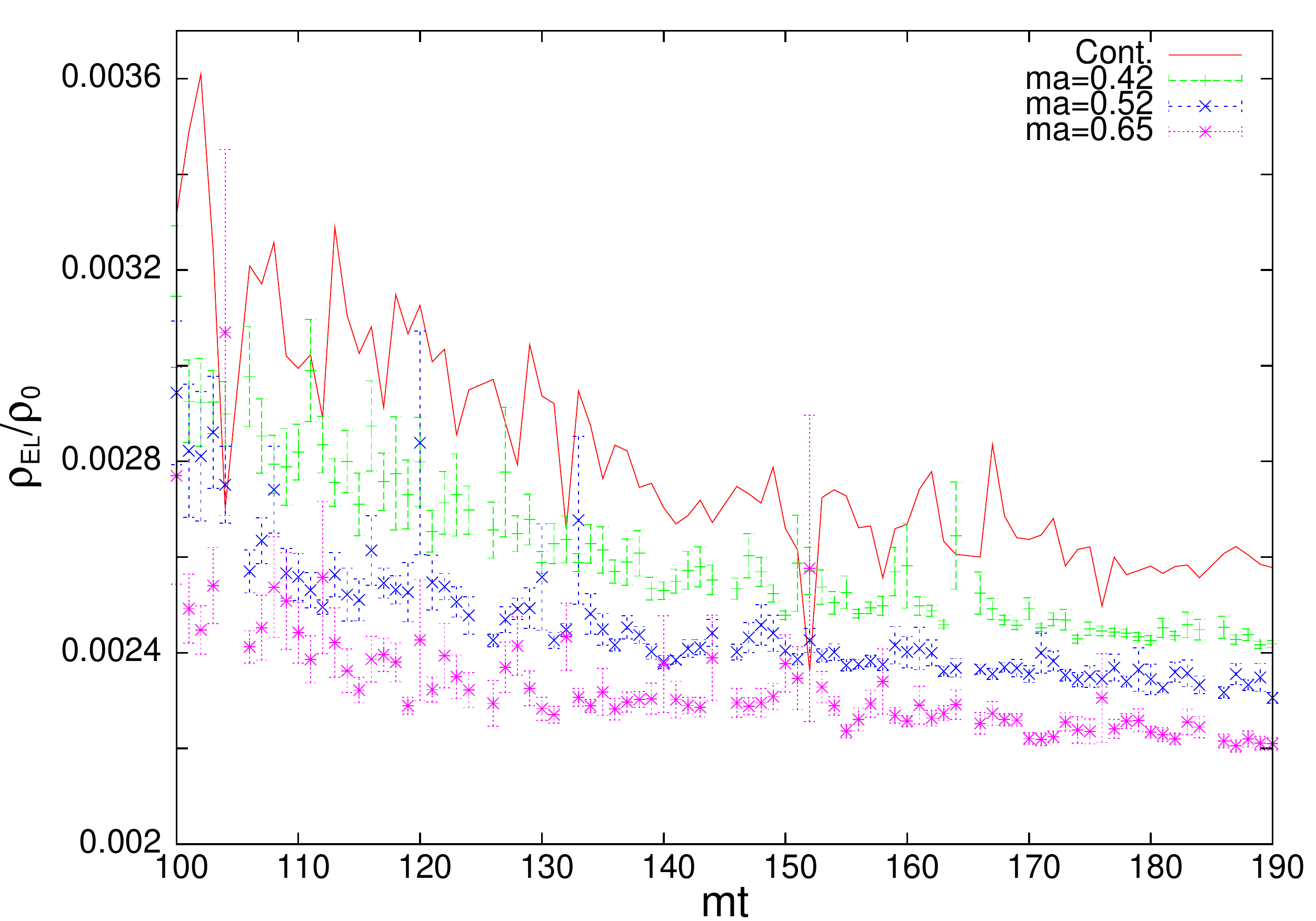,angle=-90,width=8cm}
}
\centerline{
\psfig{file=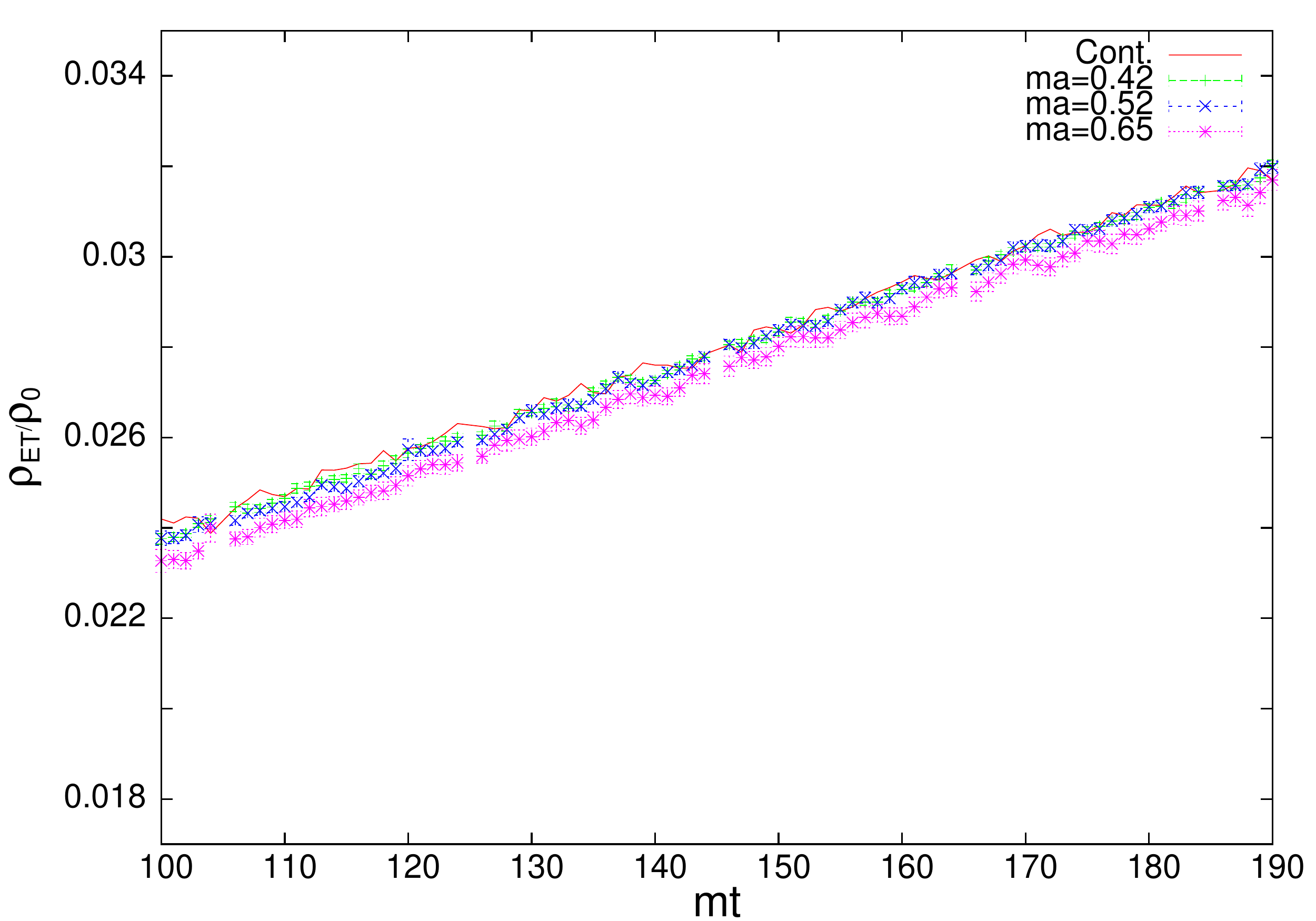,angle=-90,width=8cm}
\psfig{file=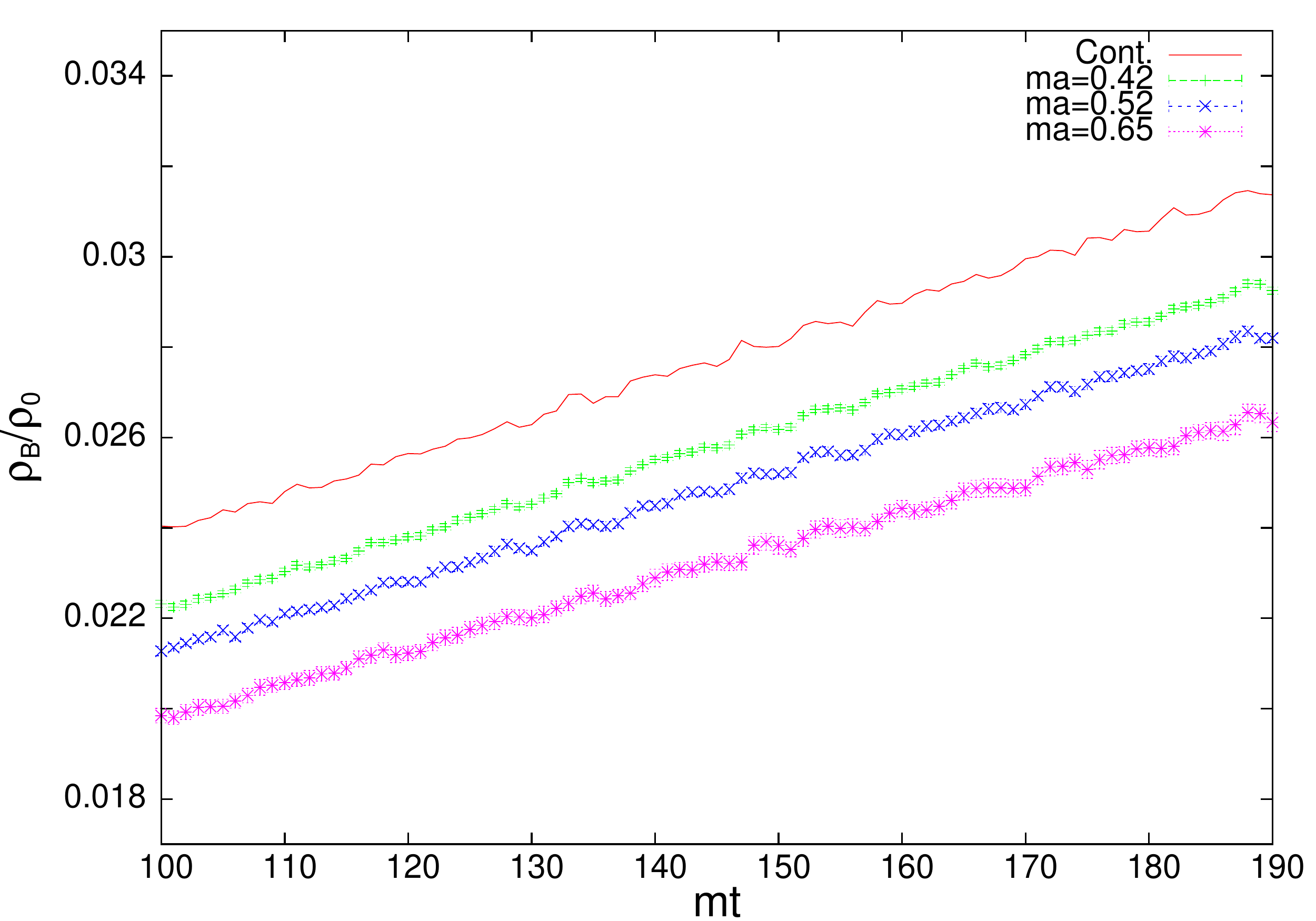,angle=-90,width=8cm}
\caption{Comparison of the fraction of total energy carried by
  electric (transverse and longitudinal) and magnetic fields. Top
  left: for two different values of the minimum momentum: $\pmin=0.1$
  and 0.15 for fixed $ma=0.65$. Top Right and down: 3 different
  lattice spacings $ma$= 0.65, 0.52,0.42, for the longitudinal,
  transverse and magnetic components of the energy. The lines are the
  extrapolation of the results to the continuum $a\rightarrow 0$
  limit.  For $\mh = 2 \mw$ which, from the point of view of lattice
  artefacts, is the worst case situation.  }}
\label{fig:ener_art}}

To summarize, we have found evidence of the presence of a long range
helical magnetic field, whose amplitude and correlation length are
linearly increasing with time. This is accompanied by the growth of a
similar long range electric field. The fate of these electromagnetic
field depends on the subsequent evolution of the plasma which is not
addressable within our classical approximation and would require a
magnetohydrodynamics treatment including the effects of fermion
fields. Our results for the power spectrum of the seed fields can be
used as initial conditions for a MHD treatment as the one developed in
Ref.~\cite{ChristenssonHindmarsh2005}.


\section{Dependence on methodological and model parameters}\label{arte}

\FIGURE{
\centerline{
\psfig{file=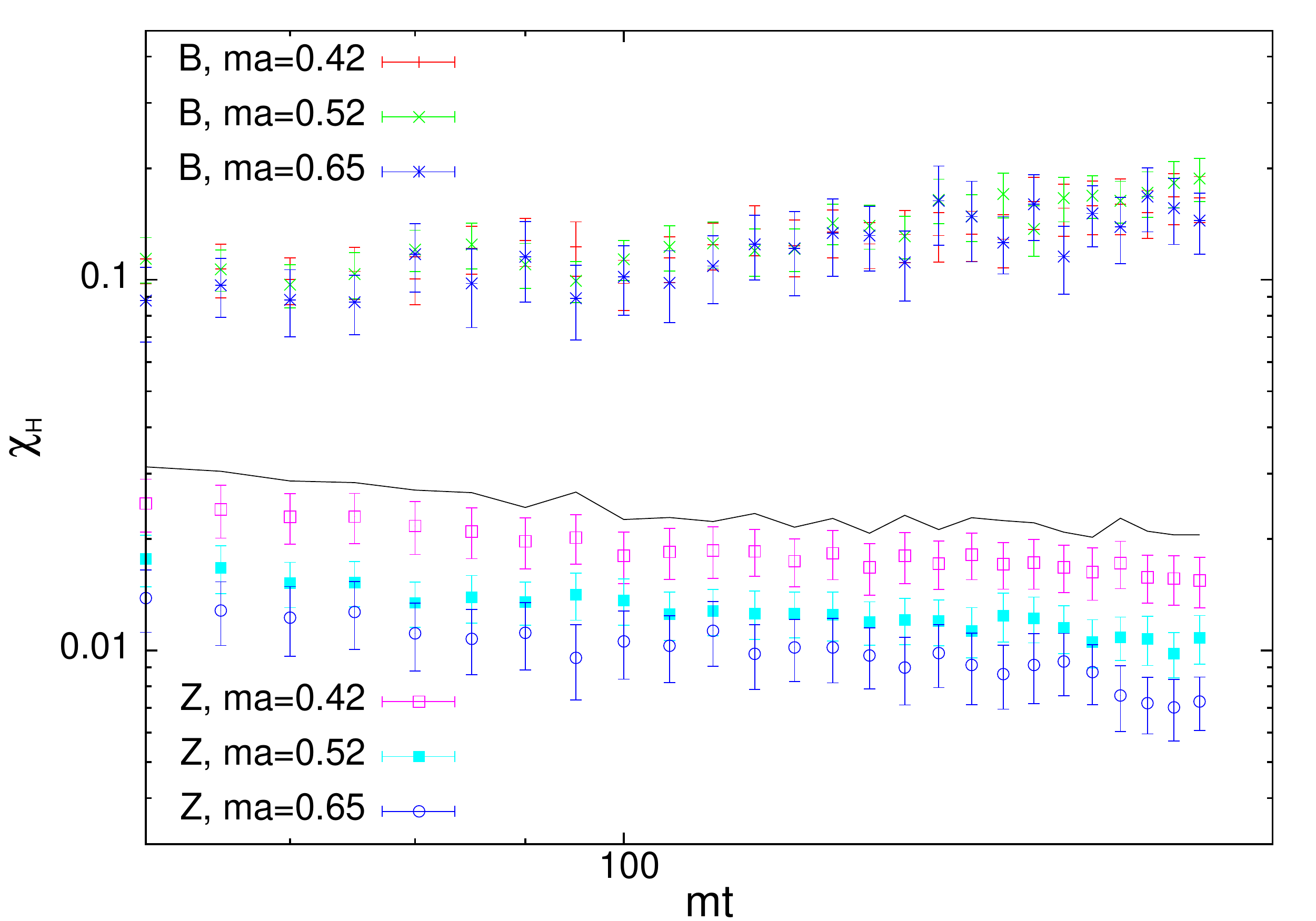,angle=-90,width=8cm}
\psfig{file=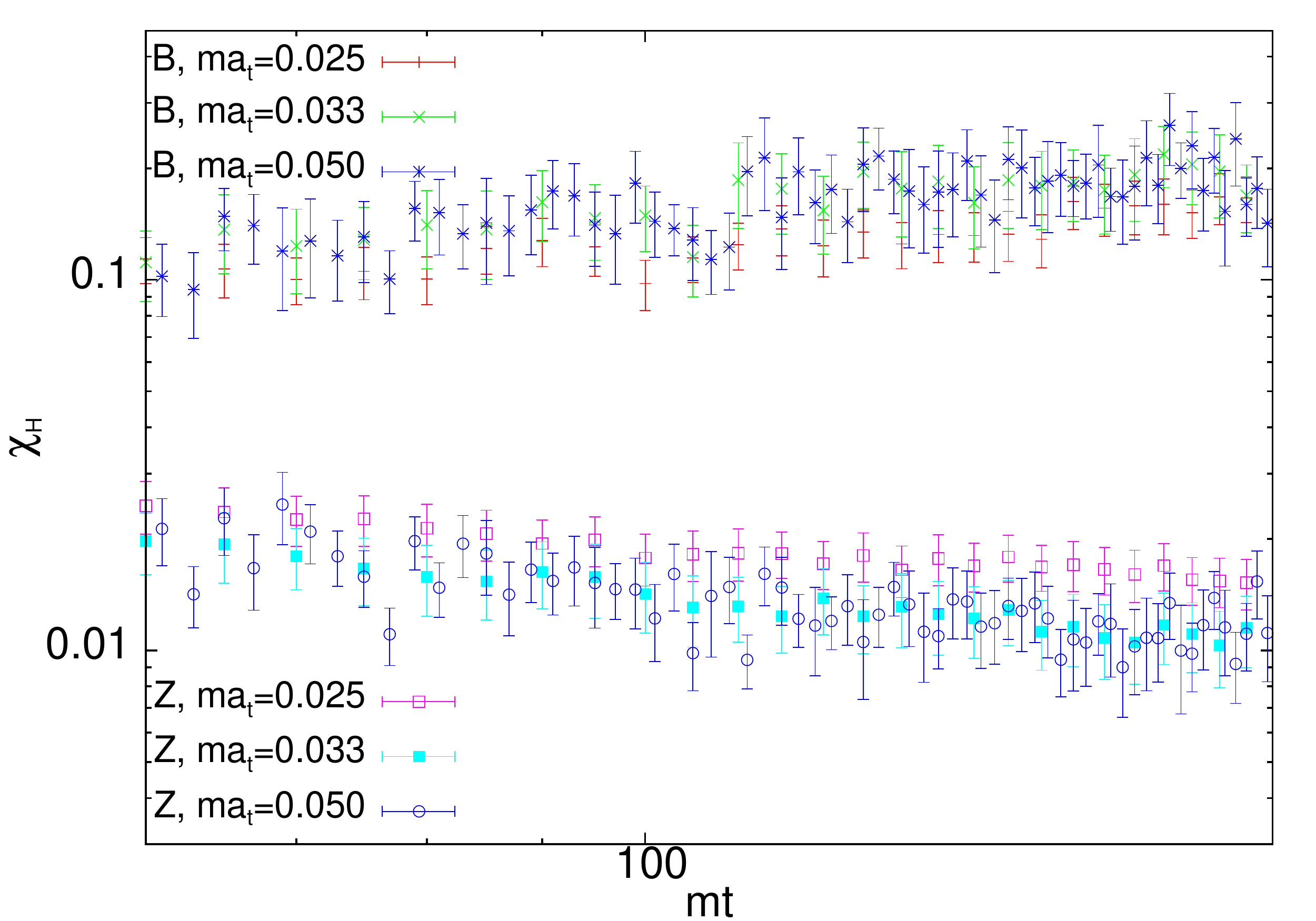,angle=-90,width=8cm}
\caption{Left: Lattice spacing dependence of the magnetic
  susceptibility for $m_H=2m_W, ma= 0.65,\, 0,52,\, 0.42$ and $N =
  64,\, 80,\, 100$.  Right: Temporal lattice spacing dependence of the
  magnetic susceptibility for $ma_t=0.05, 0.025.$ }}
\label{fig:bhel_art}}

In this section we study the (in-)sensitivity of our results to the
lattice and finite volume artefacts. We conclude that all our
qualitative results are unaffected by both types of approximations. 
Furthermore, we estimate the size of the systematic
errors induced by these cut-offs. The lattice artefacts, though
sizable, follow the expected ${\cal O}(a^2)$ dependence allowing an
extrapolation of the most relevant quantities to the continuum limit.

We also analyze the dependence of our magnetic field production
mechanism on the Higgs to $W$-boson mass ratio $\mh/\mw$. It follows
from our scenario that, initially, the helical susceptibility $\chi_H$
is independent of the Higgs self-coupling. At later times however, we
observe a non-monotonic dependence upon the mass ratio, which is
maximal at our intermediate value $\mh/\mw =3$.

\subsection{Lattice and finite volume artefacts}\label{artefacts}

\FIGURE{
\centerline{
\psfig{file=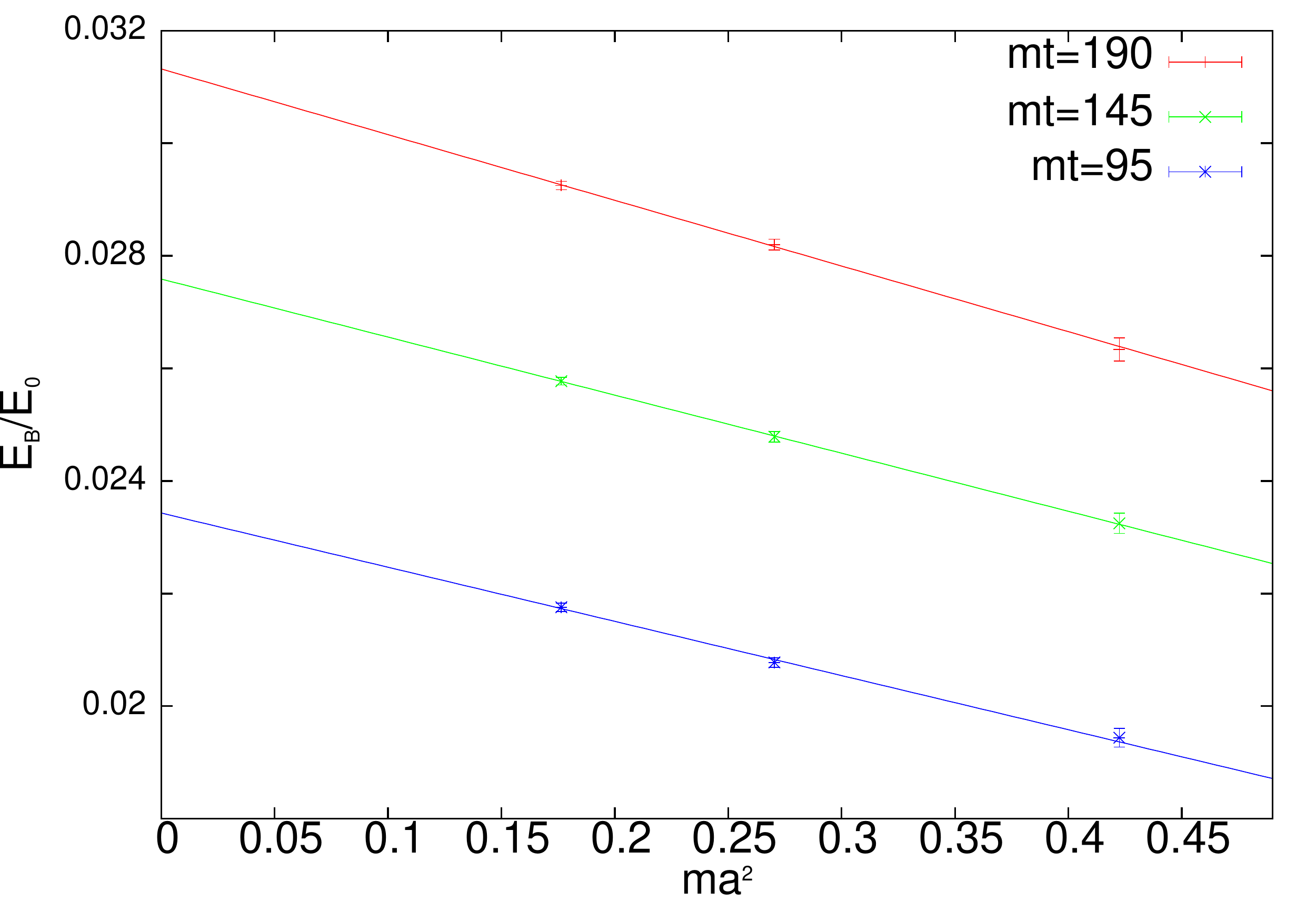,angle=-90,width=8cm}
\psfig{file=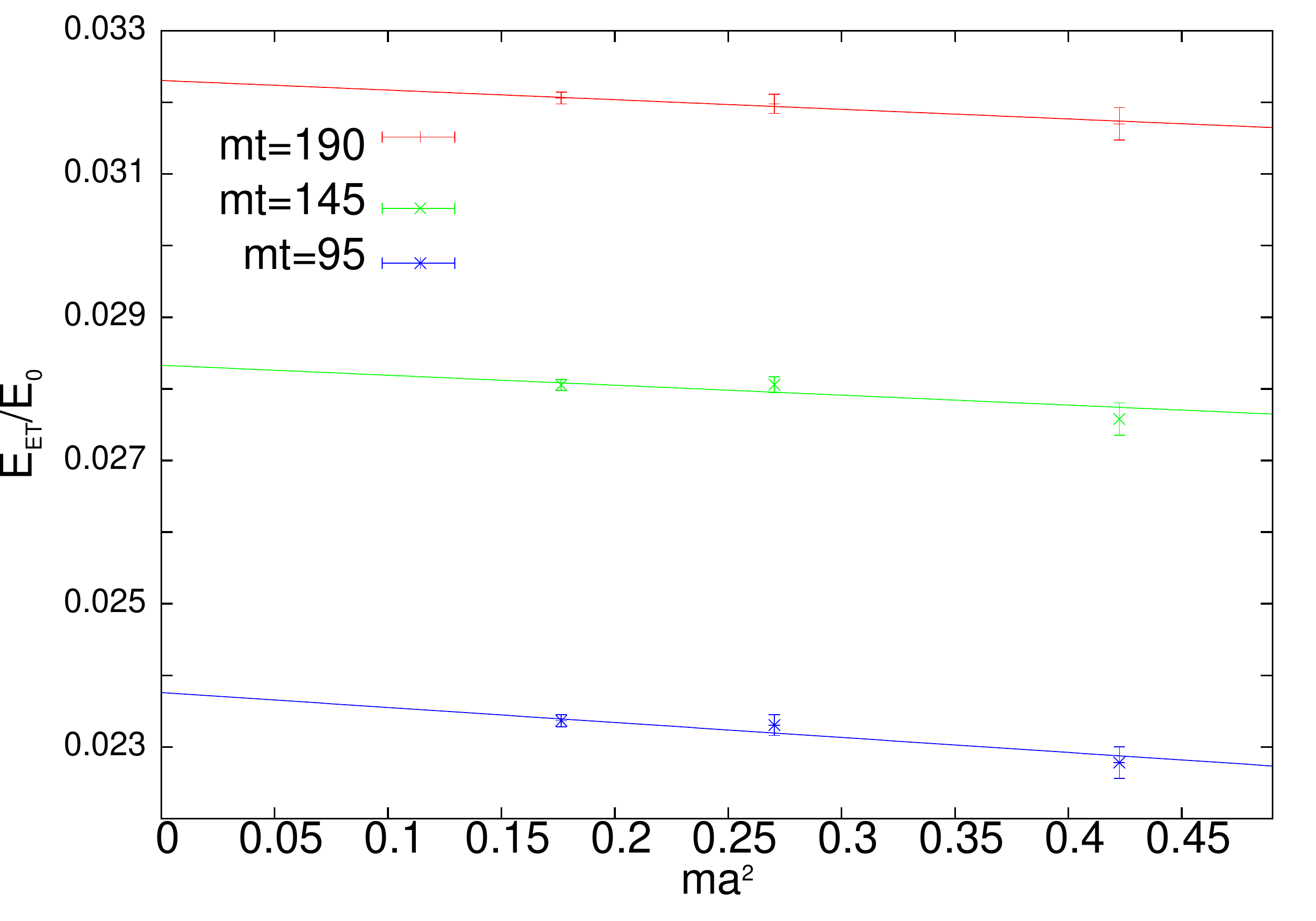,angle=-90,width=8cm}}
\centerline{
\psfig{file=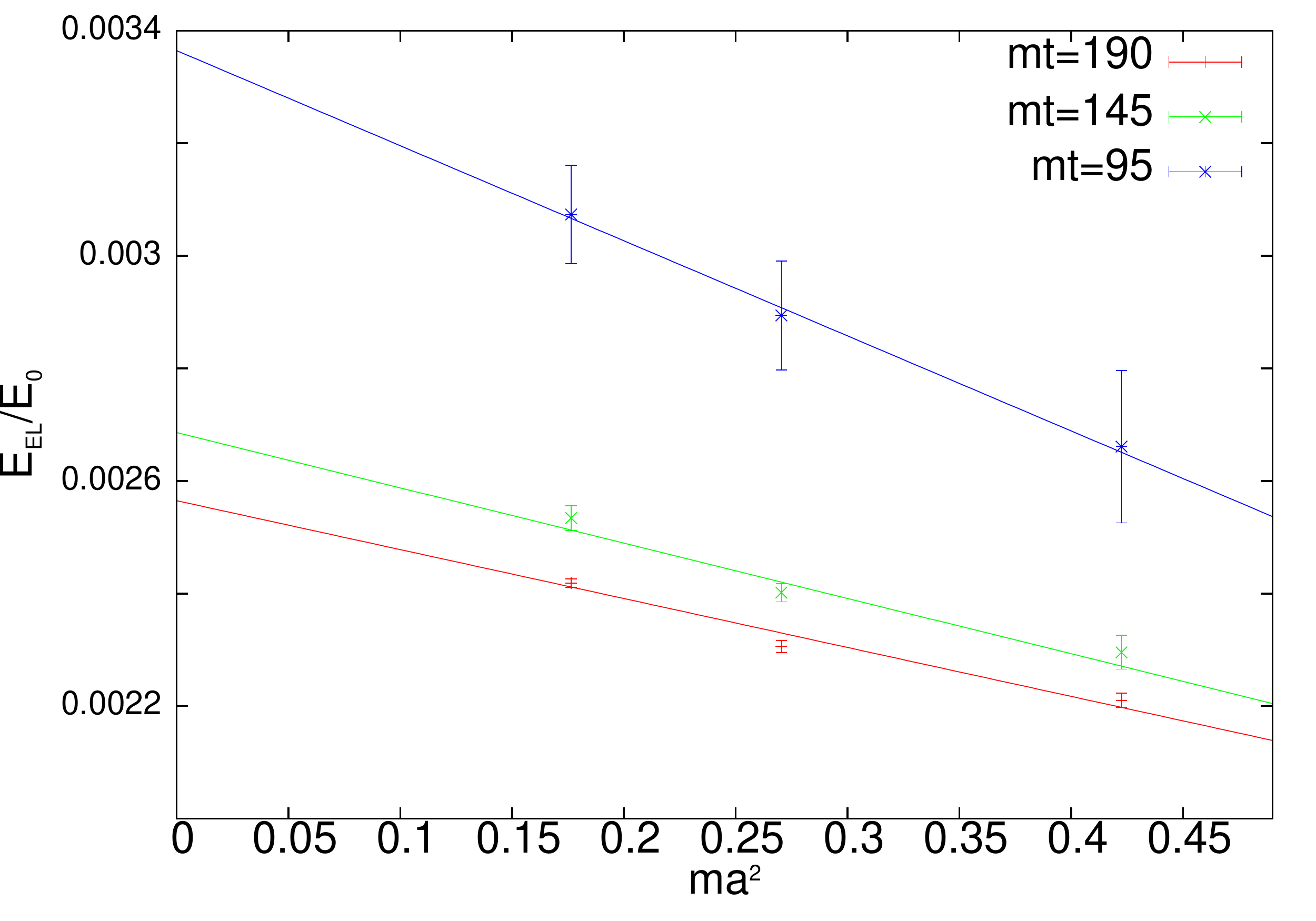,angle=-90,width=8cm}
\psfig{file=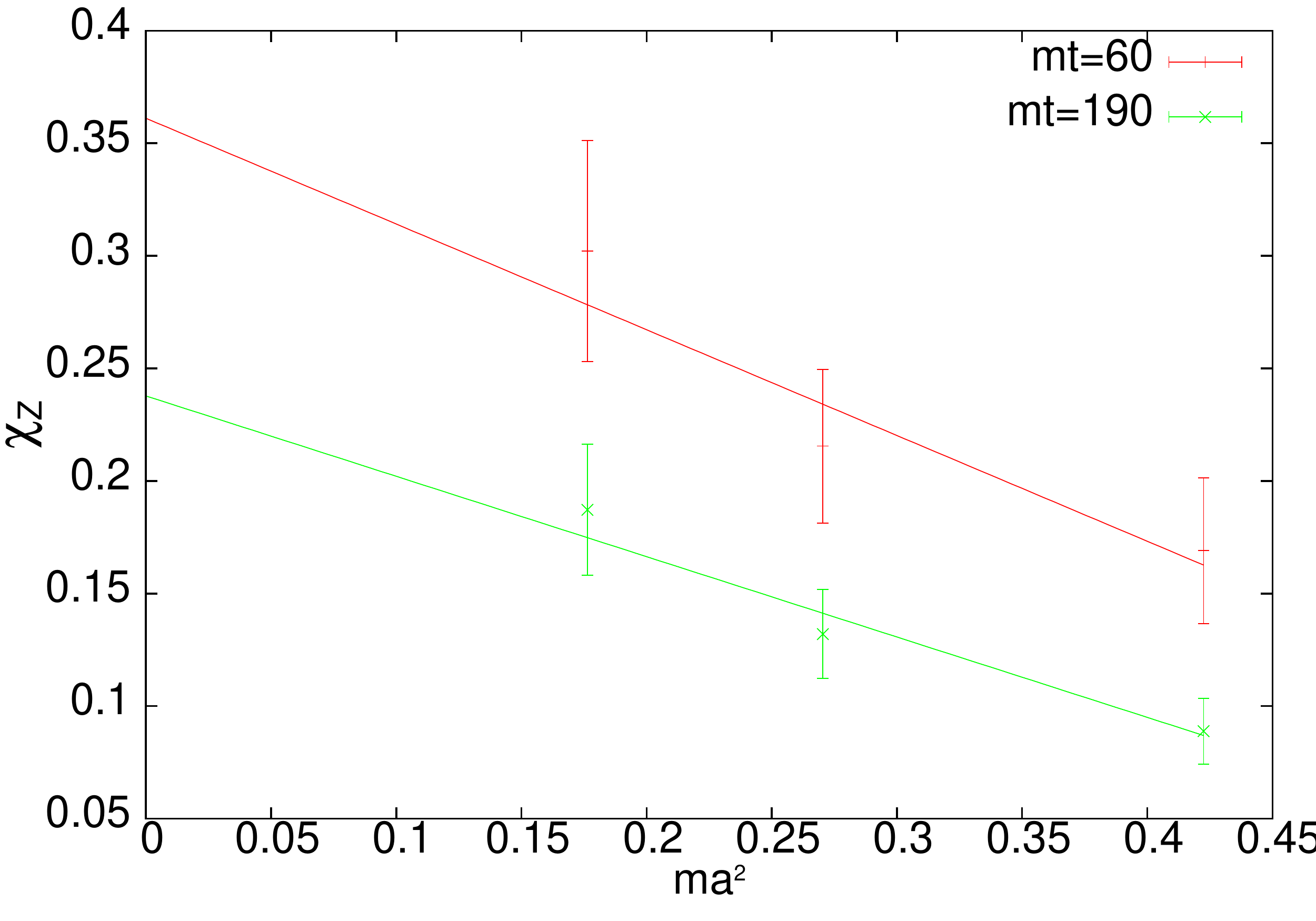,angle=-90,width=8cm}
\caption{Continuum extrapolation of the magnetic, transverse electric,
  longitudinal electric and $Z$-boson susceptibility.  For $\mh= 2\mw$
  and $mt= 95,\,145,\,190$, and $mt=60,\,190$ for the susceptibility.}}
\label{fig:ener_art2}}

In order to determine the size of the errors introduced by our
numerical approach, we have performed simulations at different values
of the physical volume and of the spatial and temporal
lattice spacings. The list of simulation parameters is given in
Table~\ref{lparameters}. The selection of values implies a delicate
compromise among different factors. As shown in Ref.~\cite{jmt}, to
avoid important finite volume effects, we need lattices with momentum
discretization $\pmin = 2\pi /L \leq 0.15\, m$. On the other hand,
concerning lattice artefacts, we have seen in Ref.~\cite{jmt2} that
cut-off independence of certain particular quantities (as the
Chern-Simons number) requires $\mw a \leq 0.3$.  Most of our
lattices satisfy both requirements.

In Figs.~\ref{fig:ener_art} and~\ref{fig:bhel_art} we present results
exhibiting the lattice and finite volume dependence of the
electromagnetic energy densities and of the magnetic helicity. They
correspond to the most disfavourable case of $\mh = 2 \mw$. No
noticeable dependence on the volume is appreciated.  Lattice spacing
artefacts are somewhat stronger but do not change the general pattern
of behaviour.  To analyse this effect in more detail, we display in
Fig.~\ref{fig:ener_art2} the $a^2$ dependence of the electromagnetic
field energy densities and $Z$-boson susceptibility at various times.
In all cases the results are consistent with the expected quadratic
dependence. This allows the extrapolation of the results to the
continuum limit, displayed as a continuous line in
Figs.~\ref{fig:ener_art} and~\ref{fig:bhel_art}. The right-hand side
of the last figure shows that for the case of the magnetic
susceptibility the values obtained for the different lattice spacings
are compatible within statistical errors. Nonetheless, assuming that
the lattice spacing dependence depends smoothly on time, we can obtain
an extrapolation to the continuum limit lying approximately $5\%$
above the values obtained for the smaller spacing.

With respect to finite size effects, long range quantities are
expected to be the most affected.  Thus, it is essential to test that
the low momentum part of the magnetic power spectrum is not biased by
finite volume artefacts.  In Fig.~\ref{fig:spec_vdep} we present
results for $\pmin= 0.125\,m$ and $0.15\,m$. The agreement is very good
for the ratio $\mh/\mw =3$ and preserves the same quality for the
other 2 values of the $\mh$ to $\mw$ mass ratios that we have studied.

\FIGURE{
\centerline{
\psfig{file=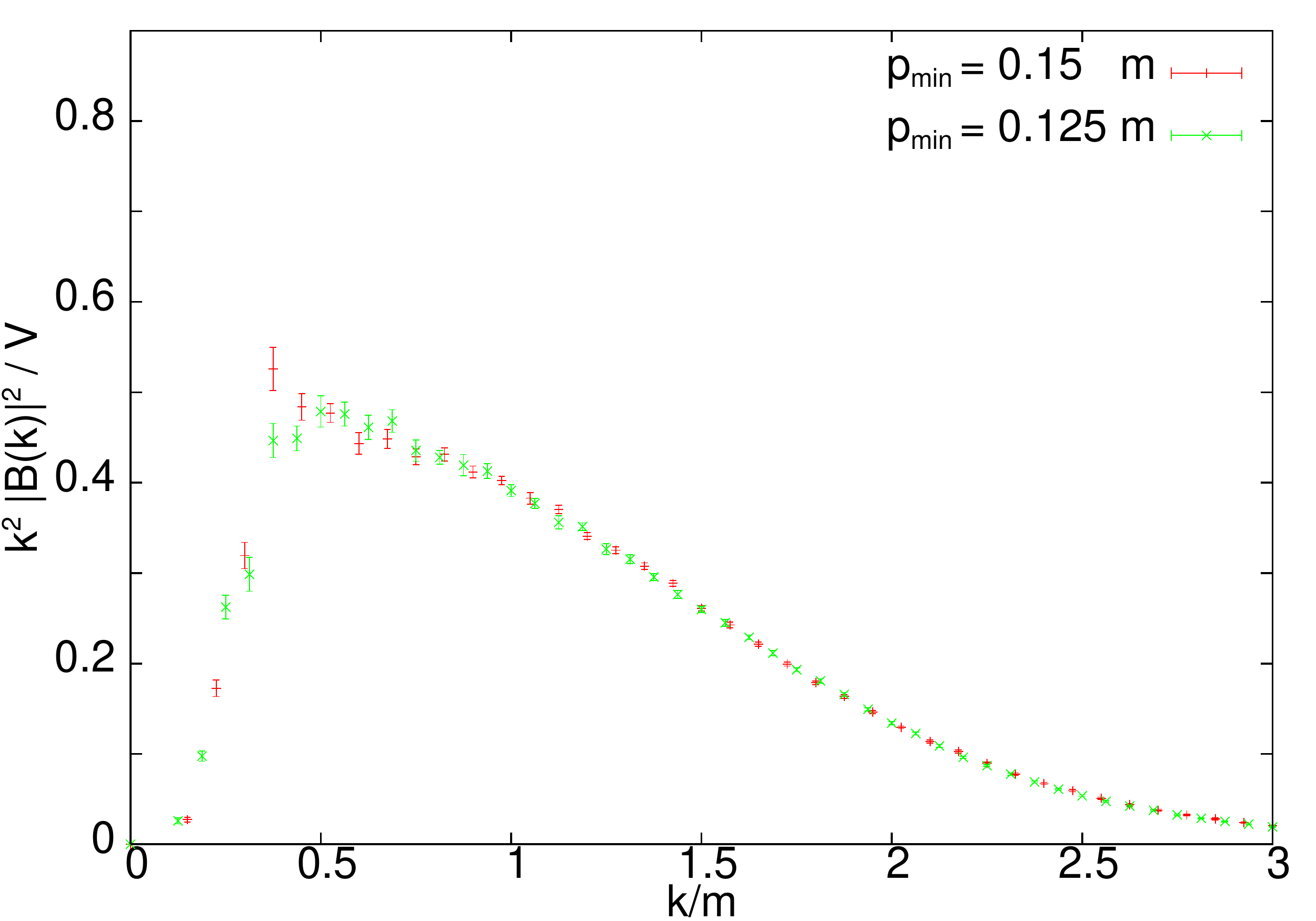,angle=-90,width=8cm}
\psfig{file=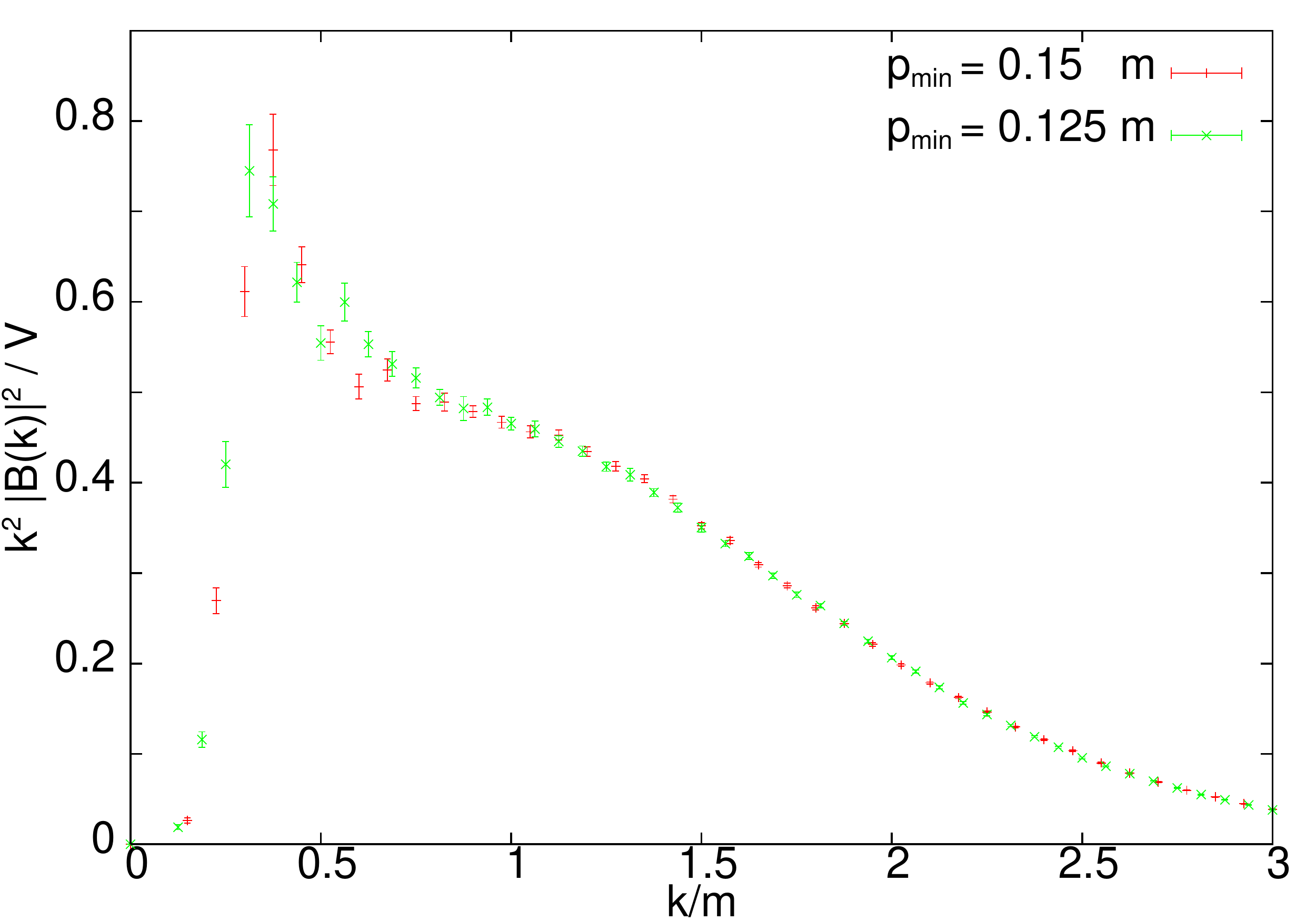,angle=-90,width=8cm}
\caption{ We plot $\langle k^2|B(k)|^2\rangle/\cV$ vs $k$ for the
magnetic component of the electromagnetic energy.  A comparison is
made between results at $\pmin=0.125\,m$ and $\pmin=0.15\,m$.  Results
are presented at $mt=$ 105 (Left) and 265 (Right).  For $\mh = 3 \mw$
and $ma=0.52$.  }}
\label{fig:spec_vdep}}

\subsection{The Higgs to W boson mass ratio}\label{ratiom}

Most of the results presented in the previous sections correspond to a
Higgs to $W$-mass ratio of 3.  Qualitatively the picture remains the
same for the other two ratios analyzed: $\mh=2\,\mw$ and
$\mh=4.65\,\mw$. In Fig.~\ref{fig:mdep} we compare the electromagnetic
energy densities and helical susceptibility as a function of time for
different values of the ratio $\mh/\mw$.  We have chosen here not to
normalize the energy densities to the total one, in order to exhibit the
independence of the initial magnitude of the electromagnetic fields
and helical susceptibility on the value of Higgs self-coupling
$\lambda$, which also determines the mass ratio. Other features of the
initial configuration such as string lengths and widths are also
$\lambda$-independent, and depend only on the mass parameter $M$ that fixes
the Higgs Gaussian random field (see Appendix~\ref{app4}).  This
$\lambda$-independence is preserved in the first Higgs oscillation but
lost afterwards, once non-linearities and the presence of the gauge
fields modifies the dynamics. At late times equipartition would
indicate that the total fraction of energy density carried by the
electromagnetic field would again become $\lambda$-independent.  Since
$\rho_0 = m^4/4\lambda$, the fraction of energy densities in units
of $m^4$ should tend to behave as $1/\lambda$ at late times. This is
indeed the tendency observed in the data.

\FIGURE{
\centerline{
\psfig{file=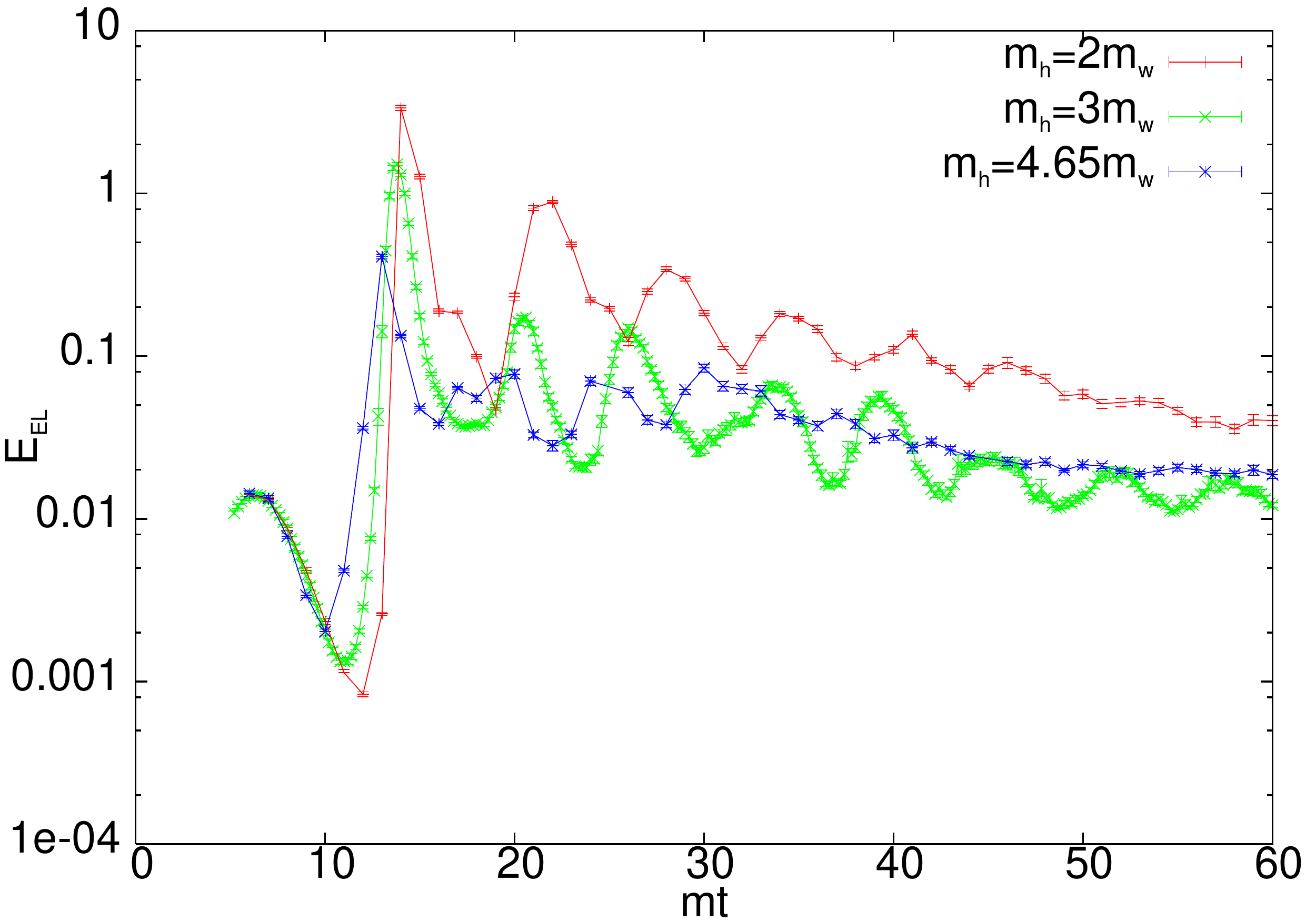,angle=-90,width=7cm}
\psfig{file=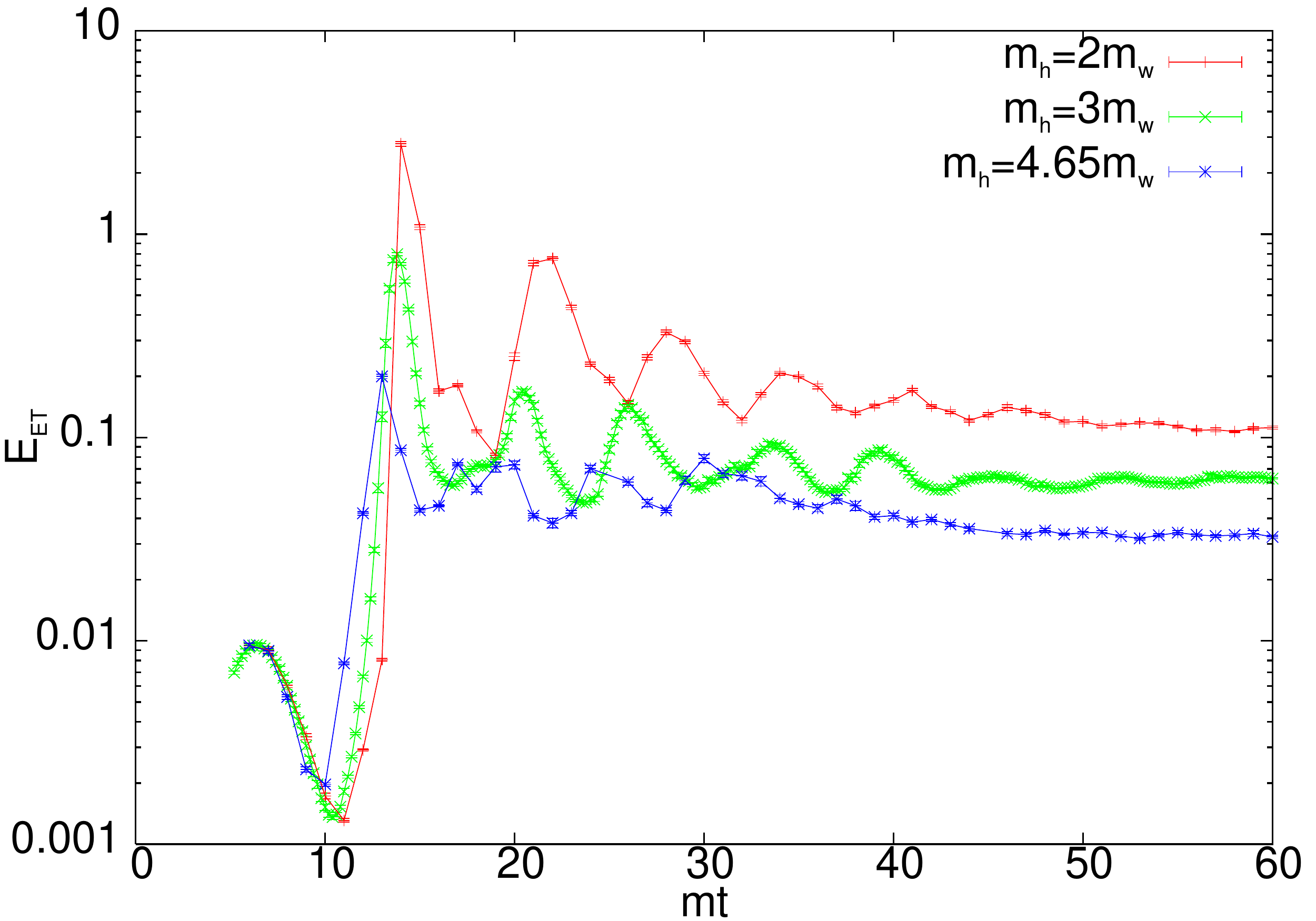,angle=-90,width=7cm}}
\centerline{
\psfig{file=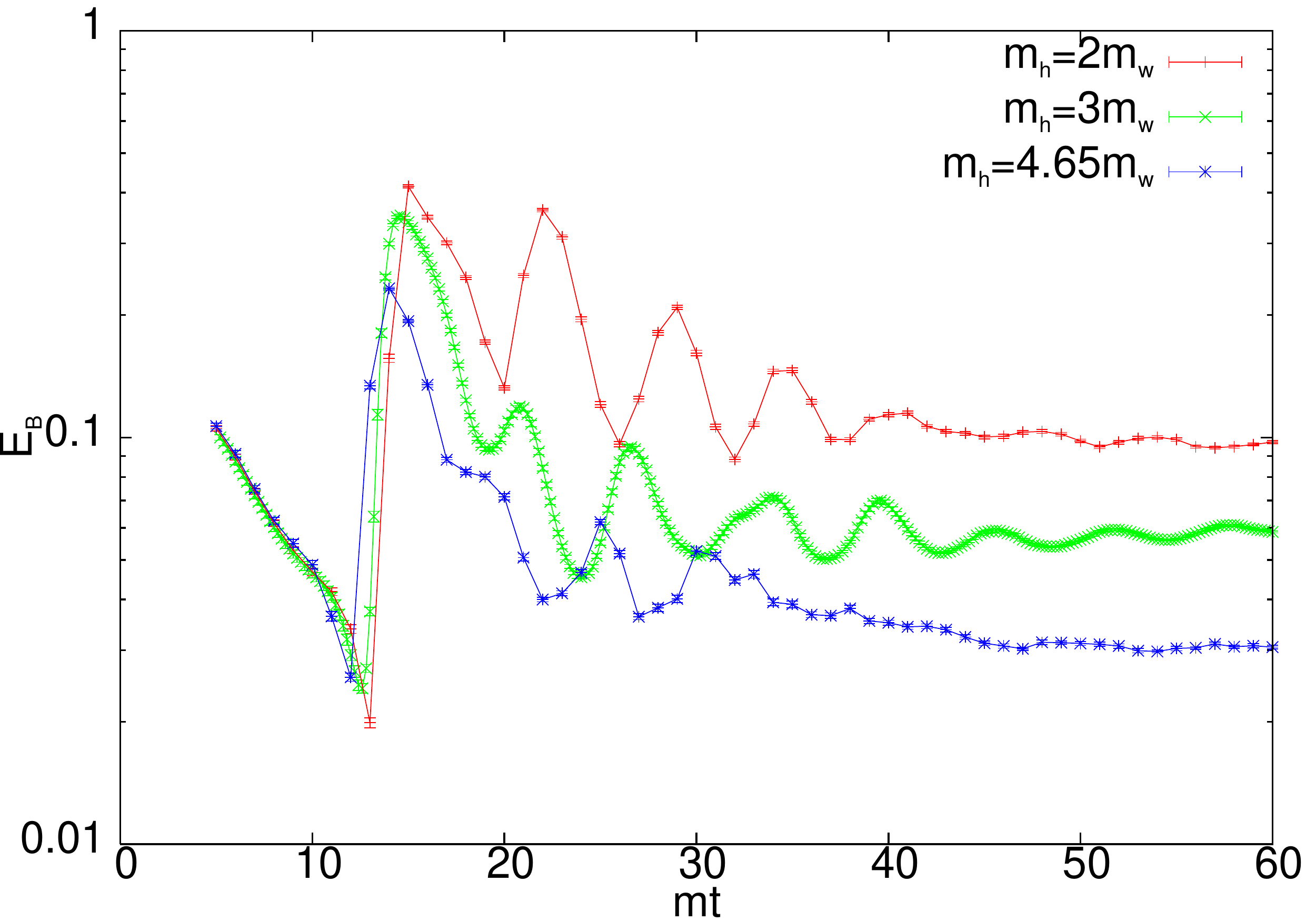,angle=-90,width=7cm}
\psfig{file=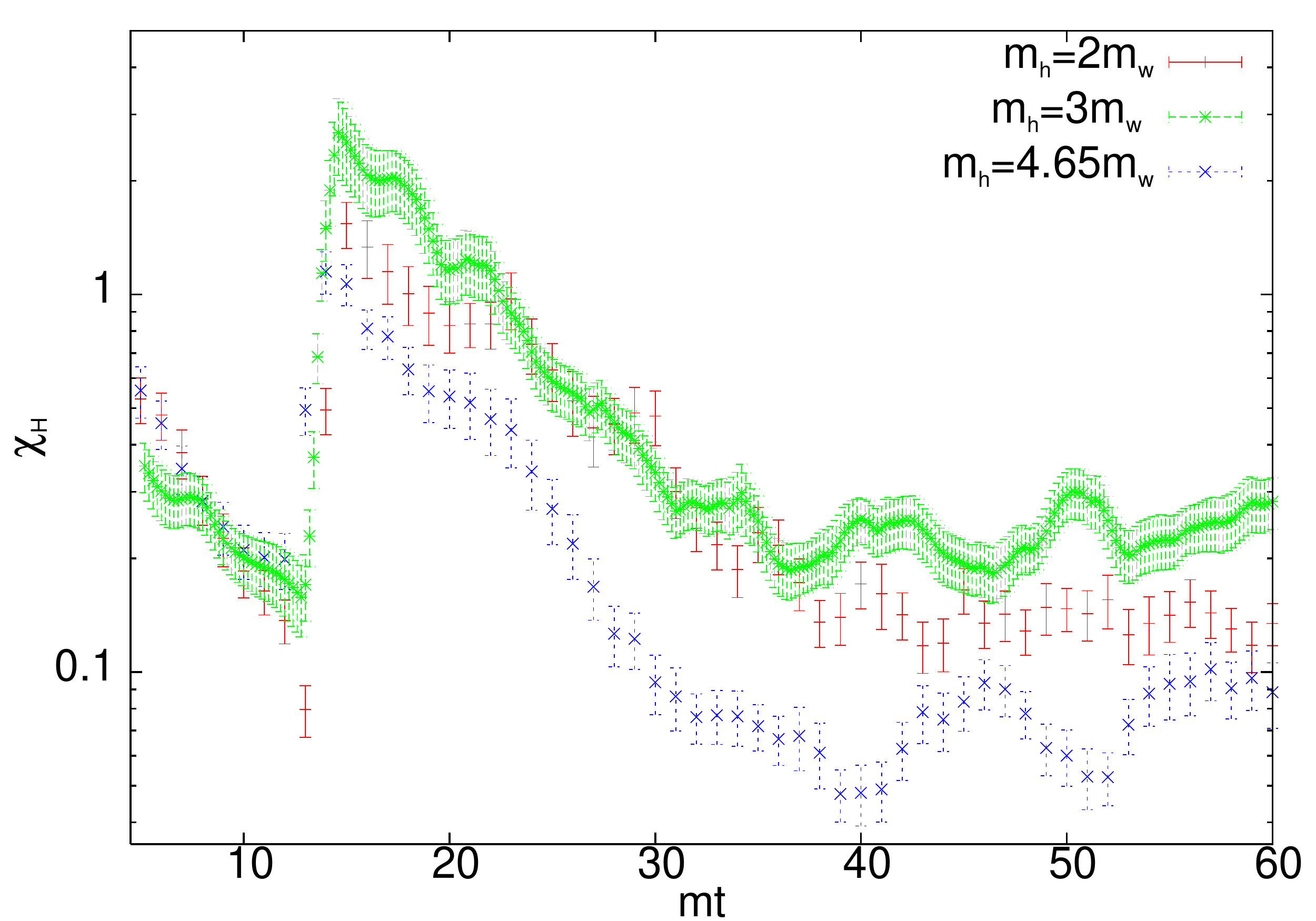,angle=-90,width=7cm}
\caption{Time evolution of the energy densities, in $m^4$ units, in
  the: Top Left: longitudinal electric field; Top Right: transverse
  electric field; Bottom Left: magnetic field.  Bottom Right: $\chi_H$
  in $m^3$ units.  Energy densities are not normalized to the total
  energy density in order to emphasize $\lambda$ independence in the
  initial stages of the evolution.  }}
\label{fig:mdep}}



\section{Conclusions}\label{conclusions}

In this paper we have analyzed the production of primordial magnetic
fields in a model of low-scale EW hybrid inflation.  Some partial
aspects of our study were anticipated in Ref.~\cite{ajmtPRL}.  For
that purpose we have studied, with the help of lattice
non-perturbative techniques, the preheating and early reheating
periods after the end of a inflationary period. Our work includes, for
the first time, the full Standard Model, SU(2) $\otimes$ U(1), gauge
degrees of freedom. The period of low-scale inflation which sets the
initial conditions of our work could be brief.  We do not need the
full 60 e-folds that are necessary to account for the CMB
anisotropies. All that is needed is a period of thermal inflation at
the EW scale which would cool down the universe during at least 10
e-folds, and set the stage for a cold (quantum) EW transition. The
metric fluctuations responsible for large scale structure could be
produced at the primordial (high energy scale) inflation.  This
secondary stage only redshifts scales by another $e^{10}$ factor, but
is irrelevant for horizon size fluctuations today, while is enough to
erase all relativistic and non-relativistic species. This scenario was
first proposed in Ref.~\cite{CEWB} and has recently been considered in
Ref.~\cite{Easther2008}.

The main results of our work can be summarized in the following
three observations. First, this set up provides a concrete realization
of the mechanism proposed by Vachaspati~\cite{Vachaspati1991} and
Cornwall~\cite{Cornwall1997}, by which inhomogeneities of the Higgs
field phases act as sources for the generation of magnetic fields
and $-$ this is essential $-$ with non-trivial helicity. To the best of our
knowledge this is the first time that this mechanism has been observed
in a fully non-perturbative set-up.  Second, the generated magnetic
field would have, when red-shifted until today, an amplitude of $\sim
0.5\, \mu$G. This is enough to explain the values of magnetic fields
observed in clusters, while those in galaxies would require a small
amount of enhancement via the usual dynamo mechanism.  Third, the
correlation length of the generated magnetic field grows linearly with
time within the time span we have analyzed. For $\mh = 3 \mw$ we find
$m \xi_B \sim 0.03\, mt$, as shown in Fig.~\ref{fig:seeds}. This linear
growth seems to be sustained by the non-trivial dynamics of the plasma
made of $W$-bosons and could be expected to hold until the decay of
the Higgs, the $W$ and the $Z$ bosons into light fermions. Our
approach does not allow us to extrapolate these results from then
onwards. Nevertheless, the helical nature of the generated magnetic
field warrants that the effect of the primordial plasma
would be that of preserving and even amplifying the magnitude of the
helicity and the magnetic field correlation
length~\cite{ChristenssonHindmarsh2005}-~\cite{Banerjee}.

We have distinguished three different stages in the evolution after
inflation ends: tachyonic growth of the Higgs-field low momentum
modes, symmetry breaking and late time evolution after SSB. In what
follows we will summarize the main features characterizing each of
these stages.

During the first tachyonic stage, non-linearities in the Higgs
potential and gauge fields can be neglected and the quantum evolution
of the system can be exactly solved. Quantum fluctuations of the
Higgs-field infrared modes are described by a multi-component Gaussian
random field. As described in detail in section~\ref{generation},
magnetic fields are already present at this stage with a non-trivial
helical susceptibility directly related to the winding number
susceptibility of the Higgs as a Gaussian random field. Although
SU(2)$\otimes$ U(1) gauge fields are very small at the end of
inflation, the magnetic fields arise through the presence of
inhomogeneities in the Higgs field phase, thus corroborating
Vachaspati's conjecture. Along this period, the spatial distribution
of the magnetic field is determined by that of the Higgs field, a
feature that is maintained and even enhanced during the second stage
of evolution corresponding to symmetry breaking.

The period of SSB arises via the formation of bubbles in the Higgs
field norm that expand with time and collide with each other.
Magnetic fields are squeezed by the expansion in string-like
structures localized in the regions between bubbles (see
Fig.~\ref{fig:higgsymag}). This stringy structure is reproduced both
in the helicity density and in the $Z$ boson magnetic field
density. We have estimated a characteristic string separation during
this period of $\mh l \sim 14$. Linked to the appearance of the
magnetic strings we find a non trivial distribution of electric fields
and $W$-boson charge and current densities. Most remarkably, we see a
very non-trivial distribution of the charge density with the formation
of extended charged clusters which track the position of the magnetic
string. This separation of unequal charges induces electric fields in
the plasma. We observe both transverse and longitudinal electric
fields also correlated with the string locations. The clusters persist
for a very long time and, as a consequence, we observe a very slow
screening of the longitudinal electric field with time, see
Fig.~\ref{fig:lenergy}. We conjecture that these electric fields will
be erased as soon as the plasma of $W$-bosons decays into light
fermions moving close to the speed of light, which will neutralize
much faster than the heavy $W$-charges.

The third stage of evolution after SSB is characterized by a very slow
approach to thermalisation. To claim a feasible mechanism for
magnetogenesis we have to guarantee that the initial helical magnetic
seed is not removed with time. We have shown in section \ref{latet}
that the magnitude of the helical susceptibility grows with time with
a power-law behaviour, $\chi_H \propto t^\alpha$, with
$\alpha=0.7(1),\, 0.8(1)$ and $0.3(1)$ for $\mh/\mw=\, 2,\, 3,\, 4.65$
respectively. At the same time the $Z$-boson helical susceptibility
decays also with a power law dependence with time. We have observed
that the magnitude of the generated magnetic susceptibility does not
depend monotonically on the Higgs- to $W$-mass ratio.  Of the values
we have analyzed, $\mh/\mw=3$ is the one that generates larger helical
fields.

In order to extract the late time behaviour of the amplitude and
correlation length of the magnetic field seed, we have performed a
detailed analysis of the magnetic field Fourier spectrum for
$\mh/\mw=3$. It shows two well differentiated and uncorrelated
components: an ultraviolet radiation sector and an infrared peak whose
amplitude increases with time (see Fig. \ref{fig:specb}). The
radiation tail is well described by a Bose-Einstein distribution of
massive photons with a non-trivial chemical potential at temperatures
$T\sim 0.23\, \mh$ slowly rising with time. The low momentum part of
the spectrum carries a fraction $f \sim 10^{-2}$ of the total energy
density.  As mentioned before, both its amplitude as its correlation
length are linearly growing with time within the analyzed time span,
showing indications of an inverse cascade towards the
infrared. However, our time scales are not long enough to demonstrate
that inverse cascade will be sustained at even later times when the
composition of the plasma changes significantly. For the moment we
can, nevertheless, rely on the results in
Refs.~\cite{ChristenssonHindmarsh2005}-~\cite{Banerjee} which show
that helical fields are optimally amplified by MHD evolution.

In summary, hybrid preheating at the EW scale could be responsible for
the observed magnetic fields associated with large scale structures
like galaxies and clusters of galaxies. Both the magnitude and
correlation length could be derived from the highly non-linear and
non-perturbative evolution after EW symmetry breaking. Our analysis
provides a concrete realization of the mechanism proposed by
Vachaspati and Cornwall many years ago. This primordial plasma enters
a regime in which helical magnetic field lines experience an inverse
cascade towards larger scales. We observe how both their energy
density and correlation length grow linearly with time.  Showing that
these magnetic fields evolve as described in the introduction until
photon decoupling would require a detailed follow up with MHD
simulations with initial conditions provided by our work. This result
would support our proposal that the helical magnetic fields produced
at the cold EW transition are responsible for the observed magnetic
fields in galaxies and clusters of galaxies.

\section*{Acknowledgements}

It is a pleasure to thank G. Aarts, A. Ach\'ucarro, J. Berges,
J.M. Cornwall, M. Hindmarsh, A. Kusenko, M. Salle, J. Smit,
J. Urrestilla and T. Vachaspati for lively discussions. JGB thanks the
Kavli Institute for Theoretical Physics, for hospitality during the
last stages of the work, and the organisers of the ``Nonequilibrium
Dynamics in Particle Physics and Cosmology 2008'' programme for the
invitation.  We acknowledge financial support from the Madrid Regional
Government (CAM) under the program HEPHACOS P-ESP-00346, and the
Spanish Research Ministry (MEC) under contracts FPA2006-05807,
FPA2006-05485, FPA2006-05423, FPA2006-26414-E. This work was supported
in part by the (US) National Science Foundation under Grant
No. PHY05-51164.  We also acknowledge use of the MareNostrum
supercomputer at the BSC-CNS and the IFT-UAM/CSIC computation cluster.
The authors participate in the Consolider-Ingenio 2010 CPAN
(CSD2007-00042) and PAU (CSD2007-00060).


\appendix

\section{The Lattice Equations of Motion}\label{app1}

To solve the classical equations of motion we discretise them on a
lattice preserving full gauge invariance. In this appendix we
introduce the lattice notation and derive the lattice equations of
motion for our particular problem.
 
As usual, the lattice points are labeled by a vector of integers
$n=(n_0,\vec n)$ in terms of which the space-time positions are given
by: $x=(n_0 \, a_t , \vec n\, a)$, with $a_t$ and $a$ the temporal and
spatial lattice spacings related by $\kappa=a_t/a$.  The adimensional
lattice scalar fields are derived from the continuum ones as: $\Phi_L
(n)=a \, \Phi(x/a)$ and $\chi_L(n)=a \, \chi(x/a)$. In what follows we
will omit the subscript $L$, since all fields will be lattice fields
unless explicitly indicated. The Higgs field is expanded as:
$\Phi(n)= \sum_\alpha \phi^{\alpha}(n)\,
\bar \sigma_\alpha$, in the basis of $2\times 2$ SU(2) matrices: 
$\bar \sigma \equiv 
(1\!{\rm l},i\vec \tau)$, with $\tau_a$ the Pauli matrices and
$\phi^{\alpha}$ real coefficients. The Standard Model Higgs doublet is
obtained trough the projection: $\varphi= \Phi \, (1, 0)^T$. Gauge
fields are given in terms of link variables: $U_\mu (n)$ and
$B_\mu(n)$ for SU(2) and hypercharge fields respectively. They are
both $2\times 2$ SU(2) matrices, with the peculiarity that $B_\mu(n)$
is diagonal.  Expanded in the $\bar \sigma$ basis, they read:
\be
U_\mu (n) = \sum_{\alpha=0,\cdots,3} \, \, u_\mu^\alpha(n)\,  \bar \sigma_\alpha\, \, , \, \, 
B_\mu (n) = \sum_{\alpha=0,3} \, \, b_\mu^\alpha(n)\,  \bar \sigma_\alpha\, \, ,
\ee
with $u_\mu^\alpha$ and $b_\mu^\alpha$ real coefficients.
The continuum limit of the gauge links is as usual:
\begin{eqnarray}
  \label{eq:links}
  U_{\mu}(n)&\sim& e^{\frac{i}{2}\, a_{\mu}g_{\rm W}{\cal A}^a_{\mu}\tau_a}\,, \\\nonumber
  B_{\mu}(n)&\sim& e^{\frac{i}{2}\, a_{\mu}g_{\rm Y}{\cal B}_{\mu}\tau_3}\,,
\end{eqnarray}
where there is no implicit sum in the $\mu$ index and where the vector
$a_{\nu}\equiv \{a_t,a,a,a \}$.

With the previous conventions, the usual U(1) hypercharge transformation is 
implemented by acting on the $\Phi$ field  with right multiplication
by a diagonal SU(2) matrix:
\be
\label{eq:doblet}
\varphi' (n)=e^{i \alpha(n)}\, \varphi(n) \, \longrightarrow\,   \Phi' (n)= \Phi (n) 
\, e^{i \alpha(n) \tau_3}\,.
\ee
The complete SU(2)$\otimes$ U(1) gauge transformation for the Higgs field then reads:
\begin{equation}
\label{eq:abelian}
\Phi(n) \rightarrow \Omega(n)\Phi(n)\Lambda(n)\,,
\end{equation}
where $\Lambda(n)= \exp(i \alpha (n) \tau_3)$ 
represents the U(1) gauge transformation and
$\Omega(n)=\sum_\alpha\Omega^{\alpha}(n)\, \bar \sigma_\alpha $
the SU(2) one. 
The corresponding  transformations of the gauge links are:
\begin{eqnarray}
\label{eq:linktransf}
U_{\mu}(n)&\rightarrow&\Omega(n)\, U_{\mu}(n)\, \Omega^{\dagger}(n+\mu)\, ,\\\nonumber
B_{\mu}(n)&\rightarrow&\Lambda(n)\, B_{\mu}(n)\, \Lambda^{\dagger}(n+\mu)\,,
\end{eqnarray}
where $\hat \mu$ is the unit vector in the $\mu$ direction.

It is useful to introduce a lattice covariant 
derivative operator defined by:
\begin{equation}
\label{eq:deriv1}
(D_{\mu}\Phi)(n)=U_{\mu}(n)\, \Phi(n+\hat \mu)\, B_{\mu}(n)-\Phi(n)\,,
\end{equation}
and its adjoint:
\be
\label{eq:aderiv1}
(\bar D_{\mu}\Phi)(n)=U_{\mu}^\dagger(n-\hat \mu)\, \Phi(n-\hat \mu)\, 
B_{\mu}^\dagger(n-\hat \mu)-\Phi(n)\,.
\ee
In addition we introduce forward and backward ordinary lattice derivatives given by:
\bea
(\Delta_\mu f) (n) &=&  f(n+ \hat \mu) - f(n) \label{eq:derivlat}\,,\\
(\bar \Delta_\mu f)(n) &=&  f(n- \hat \mu) - f(n) \label{eq:derivback}\,.
\eea

The discretization of the pure gauge part of the  Lagrangian 
is done in terms of the plaquette fields:
\begin{eqnarray}
\label{eq:plaqueta}
P_{\mu\nu}(n)&=&U_{\mu}(n)\, U_{\nu}(n+\hat \mu)\, U^{\dagger}(n+\hat \nu)\, U^{\dagger}(n)\, ,\\\nonumber
P^{ab}_{\mu\nu}(n)&=&B_{\mu}(n)\, B_{\nu}(n+\hat \mu)\, B^{\dagger}(n+\hat \nu)\, B^{\dagger}(n)\,,
\end{eqnarray}
with the transformation properties:
\begin{eqnarray}
 P_{\mu\nu}(n)&&\rightarrow
 \Omega(n)\, P_{\mu\nu}(n)\,  \Omega^{\dagger}(n)\, ,\\\nonumber
 P^{ab}_{\mu\nu}(n)&&\rightarrow\Lambda(n)\, P^{ab}_{\mu\nu}(n)\,  \Lambda^{\dagger}(n)=P^{ab}_{\mu\nu}(n)\,.
\end{eqnarray}
The pure gauge discretized Lagrangian then reads:
\begin{eqnarray}
  \label{eq:lagrangiangauge}
{L}_{\rm Y}(n)&=&\frac{2}{\kappa g_Y^{2}}\sum_{i}{\rm Tr}\,  [ 1-P^{ab}_{0i}(n)  ]
-\frac{\kappa}{g_Y^{2}}\sum_{i\neq j}{\rm Tr}\,  [ 1-P^{ab}_{ij}(n)  ]\,,\\
 { L}_{\rm SU(2)}(n)&=&\frac{2}{\kappa g_W^{2}}\sum_{i}{\rm Tr}\,  [ 1-P_{0i}(n)  ]
-\frac{\kappa}{g_W^{2}}\sum_{i\neq j}{\rm Tr}\,  [1-P_{ij}(n) ]\,.
\end{eqnarray}
And the complete lattice Lagrangian is:
\begin{eqnarray}
  {L}_{\rm L}(n) &=&{L}_{\rm Y}(n)+ {L}_{\rm SU(2)}(n)+ \rm{Tr}\Big\{(D_{\mu}\Phi)^{\dagger}(n) \, (D^{\mu}\Phi)(n)\Big\}\\\nonumber
&+&\frac{1}{2}\Delta_{\mu}\chi(n) \, \Delta^{\mu} \chi (n)
-\kappa \, V(\Phi(n),\chi(n))\,,    
\end{eqnarray}
where all the derivatives are lattice derivatives and all matter fields 
are adimensional lattice fields. To simplify notation we have introduced 
the lattice metric tensor $\eta^{\mu \nu}$ with non-zero elements:
$\eta^{00}=1/\kappa$ and $\eta^{ii}=-\kappa$, $i=1,2,3$. This allows to raise 
four-dimensional indices in the usual way. The potential, $V(\Phi(n),\chi(n))$, 
has the explicit form:
\begin{eqnarray}
  V(\Phi(n),\chi(n))=&-&M^2_L\, \rm{Tr}\{\Phi^{\dagger}(n)\Phi(n)\}+\lambda\, \Big (\rm{Tr}\{\Phi^{\dagger}(n)\Phi(n)\}\Big )^{2} \\ \nonumber
&+&\frac{\mu_L^2}{2}\, \chi^2(n) +g^2\chi^2(n)\, \rm{Tr}\{\Phi^{\dagger}(n)\Phi(n)\}\,,
\end{eqnarray}
with $M_L=a\, m$, $\mu_L=a\, \mu$, and where $g$ and $\lambda$ are the same coupling 
constants appearing in the continuum Lagrangian.

We have now all the necessary ingredients to write the lattice equations of motion.
They are derived by imposing that the variation of the lattice action with respect
to each of the fields in the Lagrangian vanishes. We obtain: 
\begin{eqnarray}
\label{eq:movlatt}  
  (\Delta_{\mu} \bar \Delta^{\mu} \chi)(n)&=&\kappa \, \Big\{\mu^2_{L}+2g^2\, \rm{Tr}[
\Phi^{\dagger}(n)\Phi(n) ] \Big\}\, \chi(n)\, ,\\\nonumber
  ( D_{\mu}\bar D^{\mu}\Phi )(n)&=&\kappa \, \Big\{-M^2_{L}+g^2\, \chi^2 (n)+2\lambda
\, \rm{Tr}[ \Phi^{\dagger}(n)\Phi(n) ] \Big\}\,  \Phi(n),\\\nonumber
(\bar D^{\rm A}_{\nu}\, {G}^{\mu\nu})(n)&=& \kappa \, J^{\mu}(n)\,,\\\nonumber
(\bar D^{Y}_{\nu}\, F^{\mu\nu})(n)&=& \kappa \, J_{\rm Y}^{\mu}(n)\,,\nonumber
\eea
with the currents  given by:
\bea
  J^{\mu}(n)&=&\frac{i g_W}{2}\, \Big [\Phi(n)\, (D^{\mu}\Phi)^{\dagger}(n)-(D^{\mu}\Phi)(n)\, \Phi^{\dagger}(n)\Big ]\,,\\\nonumber
  J_{Y}^{\mu}(n)&=&\frac{i g_Y}{2}\, \, \Big [(D^{\mu}\Phi)^{\dagger}(n)\, \Phi(n)-\Phi^{\dagger}(n)\, (D^{\mu}\Phi)(n)\Big ]_3 \bar \sigma_3 /2\,,
\end{eqnarray}
where the sub-index 3 in the second equation denotes the component, of the
term between brackets, along $\bar \sigma_3 /2$. The covariant derivatives, 
$(D_{\mu}\Phi)(n)$ and $(\bar D_{\mu}\Phi)(n)$,  are given by 
Eqs. (\ref{eq:deriv1}), (\ref{eq:aderiv1}).
We have also introduced two additional covariant derivative operators: $D^{A}_{\mu}$
and $D^{Y}_{\mu}$,  obtained from the standard one by setting either the hypercharge
or the SU(2) gauge links to the identity, i.e.:
\begin{eqnarray}
  (D^{\rm A}_{\mu}\Phi)(n)&=&U_{\mu}(n)\, \Phi(n+\hat \mu)-\Phi(n)\,,\\\nonumber
  (D^{\rm Y}_{\mu}\Phi)(n)&=&\Phi(n+\hat \mu)\, B_{\mu}(n)-\Phi(n)\,.  
\end{eqnarray}
The corresponding expressions for the plaquette fields are:
\begin{eqnarray}
  (D^{\rm A}_{\mu}P_{\rho\nu})(n)&=&U_{\mu}(n)\, P_{\rho\nu}(n+\hat \mu)\, U^{\dagger}_{\mu}(n)-P_{\rho\nu}(n)\,,\\\nonumber
  (D^{\rm Y}_{\mu}P^{ab}_{\rho\nu})(n)&=&P^{ab}_{\rho\nu}(n+\hat \mu)-P^{ab}_{\rho\nu}(n)\,.
\end{eqnarray}
The tensors ${G}_{\mu\nu}$ and ${F}_{\mu\nu}$, appearing in the equations of motion,
are  defined from the traceless part of the plaquettes by:
\begin{eqnarray}
  {F}_{\mu\nu}&=&\frac{i}{2g_Y}[P^{ab}_{\mu\nu}(n)-P^{ab}_{\nu\mu}(n)]\,,\\\nonumber
  {G}_{\mu\nu}&=&\frac{i}{2g_W}[P_{\mu\nu}(n)-P_{\nu\mu}(n)]\,.
\end{eqnarray}

In order to simplify the problem of solving
the lattice equations of motion  it is convenient to fix
the temporal gauge, realized on the lattice by fixing the temporal component
of the hypercharge and SU(2) links to unity: $B_{0}(n)=1\!{\rm l}$ , 
$U_{0}(n)=1\!{\rm l}$. In this gauge, the lattice equations of motion can be used
to solve for the fields at time $n_0+2$ in terms of the fields at times $n_0$ 
and $n_0+1$.
The lattice equations associated to the gauge fixed degrees of freedom 
become constraint equations analogous to the continuum Gauss law:
\begin{eqnarray}
  \label{eq:gauss}
  (\bar D^{\rm A}_{k}{G}^{0k})(n)=\kappa \, J^{0}(n)\,,\\\nonumber
  (\bar D^{\rm Y}_{k}{F}^{0k})(n)=\kappa \, J_{Y}^{0}(n)\,.
\end{eqnarray}
As proved in Ref.~\cite{jmt2}, these constraints are preserved by the
lattice evolution. It is hence sufficient to impose them on the
initial conditions.  The way this is done for our numerical
simulations follows exactly the procedure described in
Ref.~\cite{jmt2} for SU(2) where we refer the reader for further
details.

\section{Lattice version of the Maxwell equations}\label{app2}

In this appendix we present the derivation of the lattice version of
the Maxwell equations used in order to define the $W$ charge and
current densities.  Starting from the continuum expressions:
\bea
&&\vec \nabla \vec E (x) = \rho (x)\ \ \ , \ \ \ \vec \nabla 
\vec {\jmath} \,(x) + \partial_0 \,\rho (x) = 0\,  \\
&&\vec \nabla \vec B (x) = 0\,\nonumber\\
&&\vec \nabla \times \vec E (x) +  \partial_0\,\vec B (x) = 0\,\nonumber\\
&&\vec \nabla \times \vec B (x) - \partial_0\, \vec E (x) = \vec{\jmath}\, (x)\nonumber\,.
\eea
we look for a discretization that preserves the Bianchi identities.

In section \ref{electro}, we have defined the electromagnetic lattice
field strength, $F^{\gamma}_{\mu \nu}(n)$, in terms of clover averaged
$Z$ and $B$ field strengths.  The clover average of a space-time
tensor, like $F_{0i} (n)$, is given by:
\be
\langle F_{0i}(n) \rangle_{\rm clov} \equiv {1  \over 2} \, \Big ( F_{0i} (n) + F_{0i} (n-\hat 0) \Big )\,,
\label{eq:clover}
\ee
while for a spatial tensor we have:
\be
\langle F_{ij} (n) \rangle_{\rm clov} \equiv {1  \over 4} \, \Big ( F_{ij} (n) + 
F_{ij} (n-\hat {\imath}) + F_{ij} (n- \hat {\jmath}) + F_{ij} (n-\hat {\imath} -\hat {\jmath}) \Big )\,.
\label{eq:clovers}
\ee
From them we extract the lattice electric and magnetic fields: 
\be
E_i (n) = \frac{1}{e\,  a a_t} \, \,  \langle F_{i0} (n) \rangle_{\rm clov}  \ , \ \ \  
B_i (n) = \frac{1}{2\, e\, a^2} \,  \, \epsilon_{ijk} \, \langle F_{jk} (n) \rangle_{\rm clov} \,.
\ee
The electromagnetic  $\vec E$ and $\vec B$ fields, defined above, verify the 
following Bianchi identities:
\be
 \vec \Delta^I \cdot \vec B (n) = 0 \  , \  \  \    \vec \Delta^I \times \vec E (n)\, 
+  {1 \over \kappa} \  \Delta_0\vec B (n) = 0 \, ,
\ee
where we have introduced an improved lattice derivative given by:
\be
(\Delta_\mu^I f)(n) = {1\over 2 } \, \Big(f(n+ \hat \mu) - f(n- \hat \mu) \Big)\,.
\ee
We now define accordingly the longitudinal and transverse components of 
the electromagnetic fields. The projection is done in momentum space
with  Fourier transformed fields:
\be
\vec E (\vec k ) = \int d^3 \vec x \  \vec E(\vec x) \, e^{-i \vec k \cdot \vec x}\,\, ,
\,\,\vec B (\vec k ) = \int d^3 \vec x \  \vec B(\vec x) \, e^{-i \vec k \cdot \vec x} \, ,
\ee
with lattice momenta: $k_i = 2 \pi n_i / (N_s \,a)$, $n_i \in Z\!\!\!Z$.
Transverse components, $\vec A_t$,  of a vector $\vec A$, 
are defined  such that  $\hat q \cdot \vec A_t=0$, where:
\be
\vec q = \half \, (\vec v - \vec v^*), \ \ {\rm with} \ \  v_i = {1 \over a}\, (e^{-ik_i a}-1)\,, 
\ee
and with $\hat q$  the unit vector in the direction of $\vec q$.

The electromagnetic, Fourier transformed,  charge  and  current densities are 
computed through:
\bea
\rho \,(\vec k) &=& \hat q \cdot \vec E (\vec k) \, ,\\
\vec {\jmath} \,(\vec k)&=& \hat q \times \vec B (\vec k) - {1\over a_t} \, \bar \Delta_0 \ \vec E (\vec k) \,.
\eea

\section{Thermal radiation}\label{app3}

\newcommand{\TR}{{\rm Tr}}
In the present appendix, we prove the relation:
\begin{equation}
\langle |\vec B(\vec x)|^{2n} \rangle= \langle :|\vec B(\vec
x)|^{2n}:\rangle_{Q(T)}\ ,
\end{equation}
where the left side of the equality is calculated using the Maxwellian classical distribution:
\begin{equation}
\label{class_expec}
\langle |\vec B|^{2n} \rangle= \sqrt{\frac{2}{\pi}}\bigg(
\frac{3}{\langle B^2\rangle}\bigg)^{3/2}\int_0^{\infty}dB
B^{2n+2}e^{-\frac{B^2}{\frac{2}{3}\langle B^2 \rangle }}\ ,
\end{equation}
whereas the right hand side is calculated using the thermal quantum distribution 
in the canonical formalism.  Thus,
\begin {equation}
\label{eq:average}
  \langle :|\vec B(\vec x)|^{2n}: \rangle_{Q(T)}\equiv \frac{\TR(:|\vec B(\vec x)|^{2n}:\rho)}{\TR(\rho)}.
\end {equation}
 where $\rho$ is the  canonical distribution  density matrix:
\begin{equation}
  \rho= e^{-\frac{H}{T}}.
\end{equation}
and  $:O:$ denotes normal ordering of the operator $O$.
By performing the integral in Eq.~\ref{class_expec} we obtain the
classical thermal averages:
\begin{equation}
\langle |\vec B(\vec x)|^{2n}\rangle= \frac{(2n+1)!!}{3^{n}}(\langle |\vec B(\vec x)|^2\rangle)^{n}.
\label{eq:maxwell2}
\end{equation}

Our goal is then to compute the thermal quantum averages 
$$ \langle : \big(\vec{B}(\vec{x})\cdot \vec{B}(\vec{x})\big)^n :
\rangle_{Q(T)} $$
on the canonical  distribution at temperature T. The  only terms  of the 
normal-ordered operator that contribute to the expectation values must be 
diagonal in momentum space. If we single out that part we obtain
\begin{equation} \label{eq:b^n}
: \big(\vec{B}(\vec{x})\cdot \vec{B}(\vec{x})\big)^n : =
\prod_{i=1}^n \left( \sum_{a_i} \int d\vec{k_i}
\mathbf{a}^\dagger_{a_i}(\vec{k_i}) \mathbf{a}_{a_i}(\vec{k_i})\right) G_{a_1
\ldots a_n}(\vec{k_1}, \ldots \vec{k_n}) + X \ ,
\end{equation}
where $X$ denotes the part that does not contribute to the expectation
value and G is a a coefficient to be specified later.  

Next  we can evaluate the thermal average of the operator part, which
can be expressed as a product of $ \overline{n}(k_i,
a_i) $, the mean number of photons of momenta $\vec{k_i}$ and
polarization $a_i$. Hence, we arrive at 
$$
\label{eq:G} \prod_{i=1}^n \left( \sum_{a_i} \int d\vec{k_i} \, \overline{n}(k_i,
a_i) \right) \,  G_{a_1 \ldots a_n}(\vec{k_1}, \ldots \vec{k_n}) \ .$$
Now we should unfold the form of the coefficient $G$. It is given by
$$
\label{eq:Gunf}
 \frac{1}{n!} \prod_i \left(v_{l_{2i-1}}(\vec{k_i},a_i) v_{l_{2i}}(\vec{k_i},a_i)\right) \sum_{\sigma
\in S_{2 n}} \delta_{l_{\sigma(1)} l_{\sigma(2)}}\cdots
\delta_{l_{\sigma(2n-1)}
l_{\sigma(2n)}}\ ,$$
where the sum is over all the permutations of the 2n indices, and
$$v_{i}(\vec{k},a)=\frac{1}{(2\pi)^{3/2} \sqrt{2 k}} (\vec{k}\times
\vec{\epsilon}_a(\vec{k}))_i \ .$$ The sum over all permutations follows
from taking all creation annihilation operators as distinguishable and  
assigning them  to each of the $2n$ magnetic fields. Nonetheless,
since we are integrating over all values of momenta one has to divide
by $n!$ to eliminate double-counting. 

Now we will  introduce the matrix $M$, given by
\begin{equation}
\label{eq:M}
M_{i j}\equiv  \sum_{a} \int d\vec{k} \, \overline{n}(k,a)
v_{i}(\vec{k},a) v_{j}(\vec{k},a)= \lambda \delta_{ i j} \ .
\end{equation}
The left-hand side is a consequence of rotational invariance.
Substituting in the previous formulas  we get 
$$ \frac{1}{n!} M_{l_1 l_2} \cdots M_{l_{2n-1} l_{2n}} 
\sum_{\sigma \in S_{2 n}}
\delta_{l_{\sigma(1)} l_{\sigma(2)}} \cdots \delta_{l_{\sigma(2n-1)}
l_{\sigma(2n)}}\ .$$

The sum over permutations can be factored as follows 
$$ \sum_{\sigma \in S_{2 n}} \delta_{l_{\sigma(1)} l_{\sigma(2)}}
\cdots \delta_{l_{\sigma(2n-1)}
l_{\sigma(2n)}} = 2^n n! \sum_{\rm pairings} \prod_{\rm 
pair} \delta({\rm pair})\ , $$
where a pairing is an arrangement of the $2n$ indices into pairs
(equivalently a permutation made entirely of 2-cycles). The rest of
the calculation is very much like a calculation to $n$th order in
perturbation theory in a model with 2-leg vertices given by the $M$
matrix and a propagator given by the identity matrix. All diagrams are
now characterized by $n_l$, the number of l-cycles (loops), where l
runs from 1 to n.  Applying the standard Feynman rules one arrives at
$$ 4^n n! \prod_i \left(\sum_{n_l} \frac{\TR{M^{n_l}}}{(2l)^{n_l}
n_l!}\right)  \ .$$
The factors $2l$ and $n_l!$ provide the order of the symmetry group of
the diagram. The $2l$ term is associated with cyclic permutations of
the vertices and to a change in orientation. In the previous formula, 
the sum over $n_l$ runs over all possible integers subject to the
constraint $\sum_l l n_l=n$. One can actually perform this sum. Setting 
$M=\lambda \mathbf{I}$ our expression  becomes proportional to $\lambda^n$. 
Thus, we can eliminate the constraint on the $n_l$ by summing over $n$.
The constrained sum can be obtained from the unconstrained one by
selecting the term proportional to  $\lambda^n$.
Hence, 
$$ 4^n n! \prod_l \left(\sum_{n_l} \left(\frac{D\lambda^l}{2l}\right)^{n_l}\
\frac{1}{n_l!} \right) = 4^n n! \exp \{ D/2\sum_l \frac{\lambda^l}{l} \} =
4^n n! (1-\lambda)^{-D/2}$$
where $D$ is the space dimension, which is 3 in our case. This
quantity is the generating function of all the quantum averages. 
Differentiating $n$ times with  respect to $\lambda$ we extract the
n-th term that we were looking for:
\begin{equation}
\label{eq:final}
\langle : \big(\vec{B}(\vec{x})\cdot \vec{B}(\vec{x})\big)^n :
\rangle_{Q(T)} = (2\lambda)^n (2n+1)!! 
\end{equation}
The result for $D=\lambda=1$, given by $\frac{(2n)!}{n!}$, serves to
crosscheck the result. From the previous equation (\ref{eq:final}) we get
$\lambda=(1/6)\langle :|B|^2(v):\rangle_{Q(T)}$ allowing to re-express eq. (\ref{eq:final})
in the form of eq. (\ref{eq:maxwell2}).

To conclude we give the expression of $\langle:|B(\vec x)|^2:\rangle_{Q(T)}$ in terms of 
the temperature. Taking the trace of eq. (\ref{eq:M}) we obtain:
$$
\label{eq:b^2}
\langle :|B|^2(x):\rangle_{Q(T)}=
\int d\vec k \frac{1}{(2 \pi )^3} 2\,k\,\overline{n}(k,a)\,.
$$
Taking into account $\overline{n}(k,a)=(e^{k/T}-1)^{-1},$
we can perform the integration:
$$\frac{1}{\pi^2}\int dk k^3\frac{1}{(e^{k/T}-1)}=\frac{1}{\pi^2}\int dk k^3 \sum_{n=1}^\infty-(e^{-k/T})^n=\frac{6}{\pi^2}\,T^4\sum_{n=1}^\infty\frac{1}{n^4}.
$$
The sum over $n$ is the known $\zeta(4)=\pi^4/90$, leading to:
$$\frac{1}{2}\langle :|B|^2(x):\rangle_{Q(T)}=\frac{\pi^2}{30}T^4$$

\section{Gaussian Random fields}\label{app4}

In this appendix we revisit the predictions of the Gaussian random
field model. As explained in Section 3 of the paper, the initial
conditions produced by the quantum evolution shortly after inflation
ends are of this type. Furthermore, this distribution  seeds the 
generation of magnetic fields  and Chern-Simons number. There is an 
extensive literature 
(see Refs.~\cite{BBKS}-\cite{Pogosian:1998ar}) 
on Gaussian random fields and some of the analytic 
predictions have been included in our previous papers. However, 
here we are dealing with multicomponent fields and some of the 
predictions and methodology do not hold in this case. Besides, there
are many more relevant observables directly related to the Physics
issues addressed in this paper. To explore these matters within this
paper, we have  felt satisfied with its numerical study. Since gauge fields 
and non-linearities do not play a role at this stage, we  have profited 
to increase statistics and test systematic errors at a low computational
cost. These results can then be used as a reference to  compare with our
full-theoretical ones. 

Our Gaussian random field is homogeneous and isotropic.  The power
spectrum was set to match the one produced by the quantum evolution of
the Higgs field coupled to a linearly time-dependent inflaton and
neglecting the Higgs self-interaction. The details and nomenclature
are explained in our previous paper \cite{jmt}. We recall that the
Higgs field has 4 real components which are independent random
variables with identical power spectrum which, for simplicity, is
fitted to a simple form which reproduces nicely its shape:
\begin{equation}
\label{power_spect}
P(k,t) = \frac{1}{2 m^2 \pi^2} k^2 ( A(t) e^{-B(t)\,  k^2/m^2} + 1)
\Theta(\sqrt{ 2 V t m }- k)
\end{equation}
where $V$ is the inflaton velocity at the end of inflation, $A$ and
$B$ are time-dependent parameters and $\Theta$ is the Heaviside step 
function.

It is interesting to be able to trace the dependence of our results 
on the different parameters that enter our model. Fortunately, this 
dependence is greatly encoded in two scales that characterize the
Gaussian random field.  One scale fixes  the magnitude of the Higgs
field. We  choose this scale to be the dispersion $\sigma$ of the field
at one spatial point. Notice that the physical scale $v$, giving the
expectation value of the Higgs field in the true vacuum, has not yet
entered the scene, since the Gaussian random field is generated before
the self-interaction of the Higgs field affects the evolution. It is
precisely  the comparison between $\sigma(t)$ and $v$ that must be
taken into account in fixing the range of values of the initial 
times $t_i$ for the subsequent non-linear classical evolution of the
system. 

In addition, the other scale of the problem  is a  length scale $\xi_0$ 
associated to the Gaussian random field as follows:
\begin{equation}
\frac{1}{\xi_0^2} \equiv \frac{\int \frac{dk}{k}\,  P(k,t) k^2}{\int
\frac{dk}{k}\,  P(k,t)}
\end{equation}
With our choice of velocity $V=0.024$ at $mt_i=5$ we obtain
$\sigma=0.139\, v$, for $\mh=3\mw$. 
Thus, we are safely in the region where
non-linearities are still small. On the other hand $m \xi_0=3.09$,
which determines the adequate ranges of the ultraviolet and infrared
cut-off of our numerical procedure.  At $mt_i=6.5$ these numbers have
changed to $\sigma=0.204\, v$ and $m \xi_0=2.95$ respectively. This
observation allows us to give results in a way that are valid for all
the values of initial times employed in this work.

In line with previous analysis, we will present our results for the
density and distribution of local maxima of $|\phi|$.  The density of
maxima is given by $0.0140(4)\, \xi_0^{-3}$.  The distribution of
minimum distances among maxima can be studied directly and displays an
approximate Gaussian distribution with mean $3.1(1)\, \xi_0$ and
dispersion $ 0.62(2)\,\xi_0$.  We have also studied the distribution
of values of $|\phi|$ at the maxima, $\varphi$. The average height of
a peak being $1.52(4) \sigma$.  The histogram is much narrower than
the one obtained for a single component Gaussian random field, and is
well-fitted to the following expression
$$ \varphi^a \exp\left\{-{\varphi^2\over2\tilde\sigma^2}\right\} $$ 
with $a=10.4(5)$ and $\tilde\sigma= 0.44(1)\,\sigma$.  Nicely enough the
results presented are robust as one changes the ultraviolet, infrared
cut-offs and time within their safe windows (See
Fig.~\ref{fig:initialthings}). Errors quoted are both
statistical and systematic.

\FIGURE{
\centerline{
\psfig{file=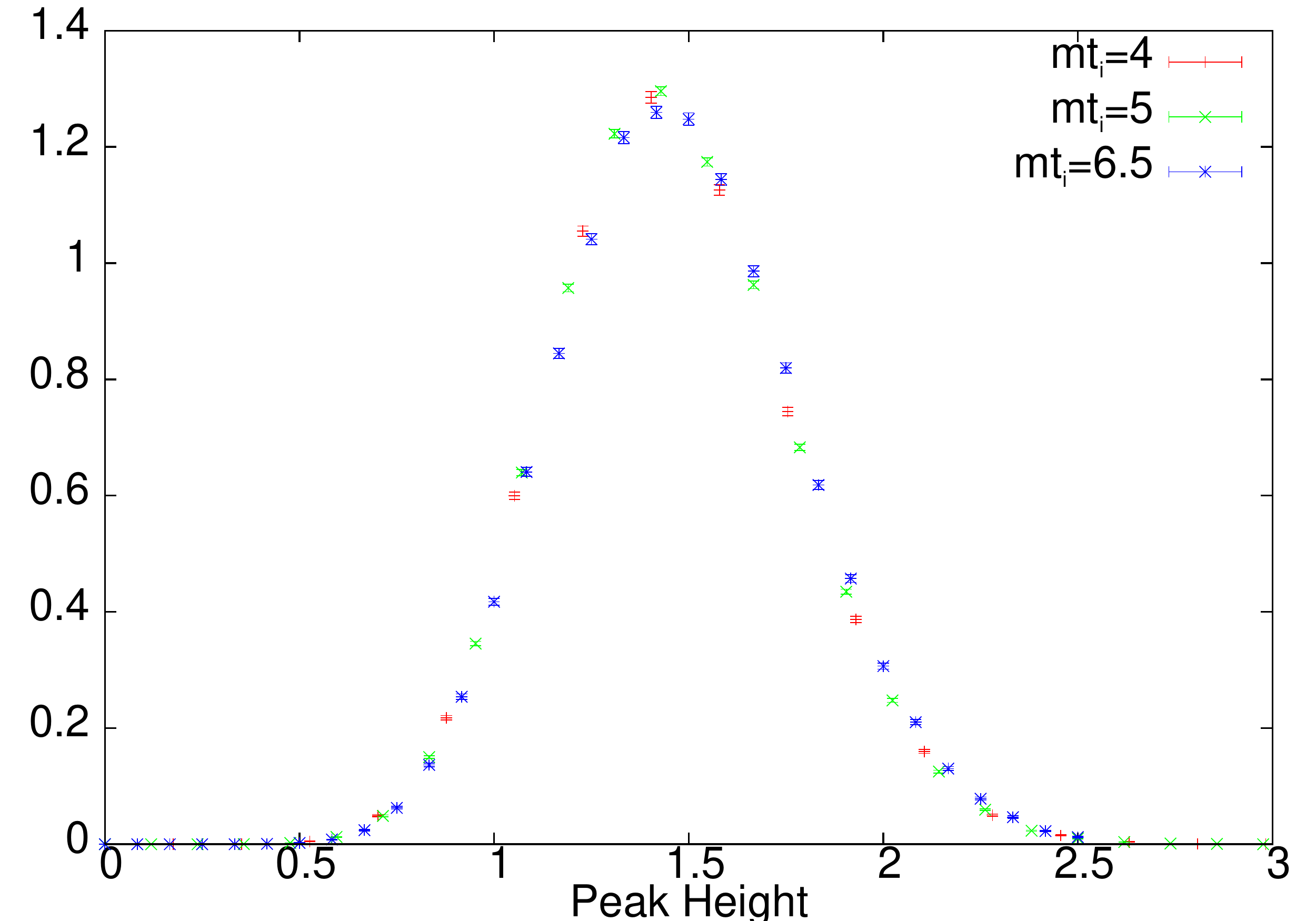,angle=-90,width=8cm}
\psfig{file=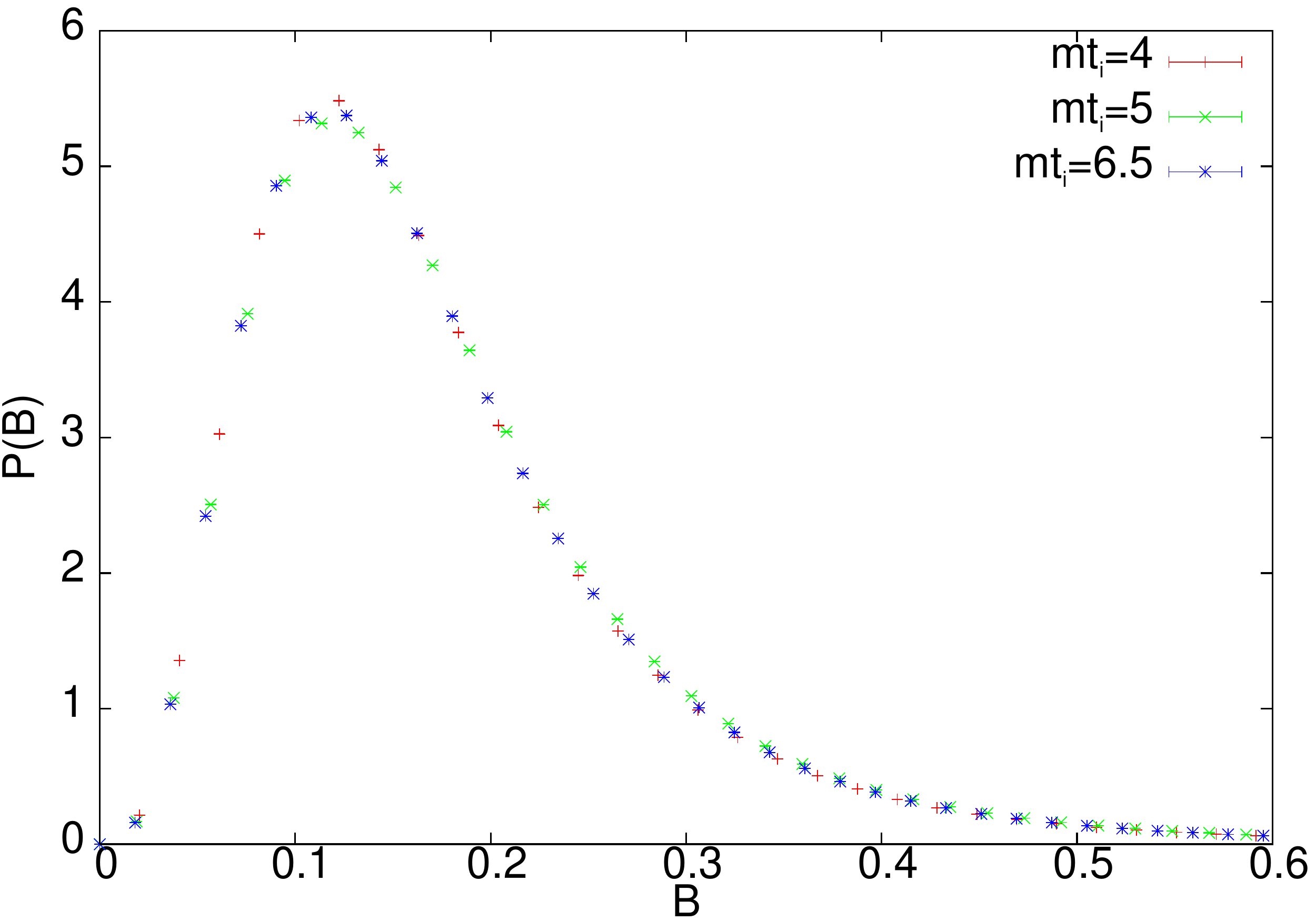,angle=-90,width=8cm}
\caption{Left: Histogram of  peak (local maxima)
  heights, expressed in $\sigma(t_i)$ units. Right: Distribution of the 
  the local magnetic field intensity $B=|\vec B(x)|$  in $\xi_0(t_i)$
  units.}}
\label{fig:initialthings}}

Now we turn to observables which are characteristic of multi-component
Gaussian random fields. A crucial role is played by the topological
susceptibility $\chi$ which is obtained by dividing the mean value of
the winding number square by the volume. We obtain $1.55(10)\times
10^{-3} \xi_0^{-3}$.  We can also compute the initial magnetic field
distribution. Notice that, as explained in the paper, despite the fact
that SU(2)$\times$U(1) gauge fields are zero at this stage, our
formulas induce a non-zero $Z$ field and a non-zero magnetic field
which is proportional to it. Computing this magnetic field at each
point of space we obtain a distribution which is well fitted by a
formula 
$$P(B) = B^b\exp\left\{-\left({B\over d_1}\right)^{h_1}\right\}+
A\,\exp\left\{-\left({B\over d_2}\right)^{h_2}\right\}\,,$$ 
with $B=|\vec B(x)|$, see Fig.~\ref{fig:maxw}.  
Our best fit values of the parameters are
$b=1.89(3),$  $d_1\,\xi_0^2=3.0(1)\, 10^{-3}, h_1=0.368(3), \, 
d_2\,\xi_0^2=2.61(2), h_2=1.34(3)$ , $A=1.0(5)
\, 10^{-7}$.  The initial magnetic field
distribution has a slower decrease at large values than the Maxwellian
distribution obtained at later times.  The aforementioned universality
can be tested here. In particular, it follows that results obtained at
different initial times $t_i$ should fall in the same curve once
normalised by the scales of $\sigma$ and $\xi_0$. This is clearly seen
in Fig.~\ref{fig:initialthings}.

We have also studied the spectrum of the magnetic field to compare it 
with the one obtained once non-linearities set in. In our case the
high momentum profile differs from the thermal tail displayed at
later times. Instead, the  high momentum tail  is well fitted by a function 
$$ \exp\left\{-\left({k\over b}\right)^c\right\}$$ 
where $b\xi_0=0.01(1)$ and $c=0.36(4)$.


\end{document}